\documentclass[12pt]{article}
\usepackage{epsfig,cite}
\usepackage{amsmath}
\usepackage{amssymb}
\usepackage{color}
\usepackage{graphicx}
\usepackage{verbatim,mathrsfs,subfigure,feynmf}
\usepackage{url}
\usepackage{ytableau}
\include{epsf}
\newcount\timecount
\newcount\hours \newcount\minutes  \newcount\temp \newcount\pmhours
\hours = \time
\divide\hours by 60
\temp = \hours
\multiply\temp by 60
\minutes = \time
\advance\minutes by -\temp
\def\hour{\the\hours}
\def\minute{\ifnum\minutes<10 0\the\minutes
            \else\the\minutes\fi}
\def\clock{
\ifnum\hours=0 12:\minute\ AM
\else\ifnum\hours<12 \hour:\minute\ AM
      \else\ifnum\hours=12 12:\minute\ PM
            \else\ifnum\hours>12
                 \pmhours=\hours
                 \advance\pmhours by -12
                 \the\pmhours:\minute\ PM
                 \fi
            \fi
      \fi
\fi
}

\def\monthname{\relax\ifcase\month 0/\or January\or February\or
   March\or April\or May\or June\or July\or August\or September\or
   October\or November\or December\else\number\month/\fi}

\def\bold#1{\setbox0=\hbox{$#1$}%
     \kern-.025em\copy0\kern-\wd0
     \kern.05em\copy0\kern-\wd0
     \kern-.025em\raise.0433em\box0 }

\def\beq{\begin{equation}}
\def\eeq{\end{equation}}

\def\ga{\mathrel{\raise.3ex\hbox{$>$\kern-.75em\lower1ex\hbox{$\sim$}}}}
\def\la{\mathrel{\raise.3ex\hbox{$<$\kern-.75em\lower1ex\hbox{$\sim$}}}}
\def\gev{{\rm \, Ge\kern-0.125em V}}
\def\tev{{\rm \, Te\kern-0.125em V}}
\def\gyr{{\rm \, G\kern-0.125em yr}}

\def\gappeq{\mathrel{\rlap {\raise.5ex\hbox{$>$}}
{\lower.5ex\hbox{$\sim$}}}}
\def\lappeq{\mathrel{\rlap{\raise.5ex\hbox{$<$}}
{\lower.5ex\hbox{$\sim$}}}}
\def\Toprel#1\over#2{\mathrel{\mathop{#2}\limits^{#1}}}

\def\m12{m_{1\!/2}}

\def\bea{\begin{eqnarray}}
\def\eea{\end{eqnarray}}

\usepackage{lscape}

\def\beq{\begin{equation}}
\def\eeq{\end{equation}}

\begin{document}\begin{titlepage}
\pagestyle{empty}
\baselineskip=30pt
\begin{flushright}
{\tt KCL-PH-TH/2021-14}, {\tt CERN-TH-2021-038}  \\
{\tt UMN-TH-4009/21, FTPI-MINN-21/02} \\
\end{flushright}

\vspace{0.2cm}
\begin{center}
{\large {\bf A Minimal Supersymmetric SU(5) Missing-Partner Model }}
\end{center}
\begin{center}
{\bf John~Ellis}$^{a,b,c}$,
{\bf Jason L. Evans}$^{d}$,
{\bf Natsumi Nagata}$^{e}$,
and
{\bf Keith A. Olive}$^{f}$
\vskip 0.2in
{\small
{\it $^a$Theoretical Physics and Cosmology Group, Department of Physics, \\ King's College London, Strand,
London~WC2R 2LS, UK}\\[5pt]
{\it $^b$Theoretical Physics Department, CERN, CH-1211 Geneva 23, Switzerland} \\[5pt]
{\it $^c$National Institute of Chemical Physics \& Biophysics, R{\" a}vala 10, 10143 Tallinn, Estonia}\\[5pt]
{\it $^d$Tsung-Dao Lee Institute, Shanghai Jiao Tong University, Shanghai 200240, China}
\\[5pt]
{\it $^e$Department of Physics, University of Tokyo, Bunkyo-ku, Tokyo
 113--0033, Japan} \\[5pt]
{\it $^f$William I. Fine Theoretical Physics Institute, School of
 Physics and Astronomy, \\ [-15pt] University of Minnesota, Minneapolis,
 Minnesota 55455, USA} }\\
\vspace{1cm}
{\bf Abstract}
\end{center}
\baselineskip=18pt \noindent
{\small
We explore a missing-partner model based on the minimal SU(5) gauge group
with $\bf{75}$, $\bf{50}$ and $\bf{\overline{50}}$ Higgs representations,
assuming a super-GUT CMSSM scenario in which soft supersymmetry-breaking 
parameters are universal at some high scale $M_{\rm in}$ above the GUT scale $M_{\rm GUT}$.
We identify regions of parameter space that are consistent with the
cosmological dark matter density, the measured Higgs mass and the experimental
lower limit on $\tau(p \to K^+ \nu)$. These constraints can be satisfied simultaneously
along stop coannihilation strips in the super-GUT CMSSM with $\tan \beta \sim 3.5 - 5$
where the input gaugino mass $m_{1/2} \sim 15 - 25$~TeV, corresponding after strong
renormalization by the large GUT Higgs representations between $M_{\rm in}$ and $M_{\rm GUT}$ to
$m_{\rm LSP}, m_{\tilde t_1} \sim 2.5 - 5$~TeV and
$m_{\tilde g} \sim 13 - 20$~TeV, with the light-flavor squarks significantly 
heavier. We find that $\tau(p \to K^+ \nu) \lesssim 3 \times 10^{34}$~yrs 
throughout the allowed range of parameter space,
within the range of the next generation of searches
with the JUNO, DUNE and Hyper-Kamiokande experiments.}


\vfill
\leftline{March 2021}
\end{titlepage}

\section{Introduction}
\label{sec:intro}

The fine-tuning of the hierarchy of the electroweak and 
grand unification scales is the bane of Grand Unified Theories (GUTs).
One aspect is how to establish the hierarchy, and a separate issue
is how to stabilize it against the depredations of radiative
corrections. A favoured resolution of the second issue is to
postulate supersymmetry that persists down to (near) the
electroweak scale~\cite{Maiani:1979cx}. 
However, supersymmetry {\it per se} does not
provide a mechanism for generating the hierarchy in the first place.

Within GUTs, the key to establishing the hierarchy of mass scales
is splitting GUT multiplets of Higgs fields so that their
electroweak components are light whereas the colored components
are heavy \cite{dg}. An elegant way to achieve this is the missing-partner
mechanism, in which the color-triplet Higgs fields combine with 
other colored fields to acquire large masses, whereas the doublet 
Higgs fields lack such partners \cite{Dimopoulos:1981xm,Masiero:1982fe,Kounnas:1983gm,Hubsch:1985em}. 
One of the most economical realizations of this
possibility is provided by the flipped SU(5)$\times$U(1) GUT,
which does not require adjoint or larger Higgs representations \cite{flipped}.
However, the missing-partner mechanism can also be
realized within the minimal SU(5) GUT model, though at the price
of introducing $\bf{75}$, $\bf{50}$ and $\bf{\overline{50}}$
Higgs representations~\cite{Masiero:1982fe}.~\footnote{In lieu of introducing a {\bf 75}, one can include non-renormalizable terms involving products of two {\bf 24} representations (which contain a {\bf 75}) to achieve the same goal \cite{Kounnas:1983gm}. We note also that examples of the missing-partner mechanism in the context of SO(10) were proposed in \cite{Dimopoulos:1981xm,Babu:2006nf}.}

As we discuss in this paper, there are challenges in formulating
this minimal supersymmetric SU(5) missing-partner model, which
originate from the relatively large sizes of the Higgs
representations it requires. In particular, the SU(5) GUT coupling runs
rapidly above the mass scales of these Higgs fields, threatening
the applicability of a perturbative treatment of the SU(5)
coupling. As we show here, requiring
perturbativity up to the input scale, $M_{\rm in}$, imposes a strong lower limit on the possible
masses of states in the $\bf{50}$ and $\bf{\overline{50}}$
Higgs representations, $M_\Theta > 2 \times 10^{17}$ GeV. However, this requirement in turn
suppresses the mass of the color-triplet Higgs field that mediates
nucleon decay though dimension-5 operators. Avoiding rapid
nucleon decay is in principle possible for  sufficiently large values of the
supersymmetry-breaking masses that enter the coefficients
of the interactions violating baryon and lepton numbers \cite{Hisano:2013exa, evno,
Ellis:2015rya, Evans:2019oyw}.
However, larger supersymmetry-breaking masses are linked in
general to larger values of the supersymmetric dark matter
relic density \cite{interplay,Ellis:2015rya}, though the relic density may be kept within the range allowed by cosmology by
invoking a coannihilation mechanism \cite{gs}. One must also verify that the predicted value of the lightest Higgs mass 
is compatible with the experimental measurement \cite{Aad:2012tfa}. 

Here we investigate how these phenomenological obstacles can be circumvented in a super-GUT version \cite{super-GUT,emo,Ellis:2010jb,Ellis:2016qra,
Ellis:2016tjc,Ellis:2017djk,Ellis:2019fwf} of the constrained minimal supersymmetric extension of the Standard Model (CMSSM)  \cite{cmssm,interplay,Ellis:2015rya,Ellis:2018jyl}, in which universality of the soft supersymmetry-breaking scalar masses is
postulated at some high scale $M_{\rm in} > M_{\rm GUT}$, the GUT scale. 
We find that in this case there is a limited range of parameters where
coannihilation \cite{gs} of the lightest supersymmetric particle (LSP) brings the relic LSP density into the range
required by Planck~\cite{Planck} and other data,~\footnote{This density constraint would be 
relaxed if there is some source of entropy that we do not take into account.} while being consistent with the lower limit
on $\tau(p \to K^+ \nu)$ \cite{Abe:2014mwa} and the measured mass of the Higgs boson as calculated
using {\tt FeynHiggs~2.18.0} \cite{FH}.

While the CMSSM needs only four free parameters - a gaugino mass, $m_{1/2}$,
a scalar mass, $m_0$, a trilinear mass term, $A_0$, and the ratio of Higgs vacuum expectation values (vevs), 
$\tan \beta$~\footnote{To which should be added the ambiguity in the sign of the light Higgs mixing parameter, $\mu$.} - the super-GUT CMSSM 
based on the missing partner model (MPM) requires several additional parameters - the 
universality input scale, $M_{\rm in} \ge M_{\rm GUT}$, and three extra trilinear couplings, 
$\lambda_{\Theta, \overline{\Theta}}$ and $\lambda^\prime$ corresponding to the $\mathbf{\bar{5}}\cdot \mathbf{75} \cdot \mathbf{50}$,
$\mathbf{5}\cdot \mathbf{75} \cdot \mathbf{\overline{50}}$,   and {\bf 75$^3$}
superpotential terms.~\footnote{We assume here equal values for $\lambda_{\Theta}$ and $\lambda_{ \overline{\Theta}}$,
so as to maximize the color-triplet Higgs mass and thereby minimize the impact of the proton decay constraint, as discussed below.}
In addition, there are two supplementary bilinear parameters, namely a  {\bf 50} $\mathbf{\overline{50}}$ mass
term, $M_{\Theta}$, 
and a {\bf 75$^2$} coupling, $\mu_\Sigma$. Whilst the former is a free parameter, the latter determines the GUT
symmetry-breaking vev and is determined by the conditions for gauge coupling unification. 
There are also two associated soft supersymmetry-breaking bilinear mass terms, $B_{\Theta}$ and $B_\Sigma$,
which are taken to be equal at the input scale. 
Thus the model is determined by the following parameters:
\begin{equation}
 m_0,\ m_{1/2},\ A_0,\ B_0,\ M_{\rm in},\ M_\Theta, \ \lambda_{\Theta, \overline{\Theta}},\ \lambda',\ \tan \beta,\  {\rm
  sign}(\mu) \, .
\end{equation}
All the parameters except $\lambda_{\Theta,{\bar \Theta}}$ and $\lambda^\prime$
are specified by their values at $M_{\rm in}$, while the Yukawa couplings are specified by their values at $M_{\rm GUT}$~\footnote{Since $\lambda_{\Theta,\bar \Theta}$ do not run below $M_\Theta$, this amounts to their running values being set at $M_{\Theta}$}.

The structure of this paper is as follows. In Section~\ref{sec:setup}
we set up the missing-partner model, describing the superpotential, the
pattern of symmetry breaking, the renormalization-group running of
model parameters, and their matching conditions at the GUT scale.
Section~\ref{sec:susyb} discusses supersymmetry breaking, including
the renormalization-group running of supersymmetry-breaking parameters
and their GUT-scale matching conditions. Section~\ref{sec:superGUT} presents the search for viable regions of parameter space in the
super-GUT CMSSM, which we find along stop coannihilation \cite{stopco,eoz,interplay} strips with restricted values of the model parameters. 

The relic density constraint fixes the value of the MSSM
soft supersymmetry-breaking scalar mass $m_0$ as a function of the gaugino mass $m_{1/2}$, and allows only a restricted
range of $m_{1/2} \lesssim 25$~TeV. Reconciling the Higgs mass prediction with the relic density
constraint requires that the MSSM Higgs mixing parameter $\mu$ be negative, and the proton decay constraint
sets a lower limit on $m_{1/2}$ that is compatible with the relic density constraint for only a limited range  of $\tan \beta \sim 3.5 - 5$.
Moreover, we also
find that the strong renormalization effects associated with the large GUT Higgs representations restrict the possible ranges
of $M_{\rm in}$ and $M_\Theta$. Typical ranges of the MSSM sparticle masses in the allowed range of parameter space are
$m_{\rm LSP}, m_{\tilde t_1} \sim 2.5-5$ TeV, $m_{\tilde g} \sim 10-20$ TeV, $m_{\tilde q} \sim 15-30$ TeV and $m_{\tilde \ell} \sim 10-25$ TeV.
We find that throughout the allowed range of parameter space $\tau(p \to K^+ \nu) \lesssim 3 \times 10^{34}$~yrs,
within the discovery reaches of the next generation of searches with the JUNO, DUNE 
and Hyper-Kamiokande experiments, which are estimated to be $1.9 \times 10^{34}$~yrs~\cite{An:2015jdp}, $1.3 \times 10^{34}$~yrs~\cite{Abi:2020kei}
and $3.2 \times 10^{34}$~yrs~\cite{HK}, respectively.

\section{Setting up the Model}
\label{sec:setup}

In this Section we outline the minimal supersymmetric SU(5) MPM we study, and specify our notation.
The representations containing the SM matter content in this model are the same as in conventional SU(5): 
the right-handed down-type quark and the left-handed lepton chiral superfields, $\overline{D}_i$
and ${L}_i$, respectively, reside in ${\bf \overline{5}}_i$
representations, ${\Phi}_i$, and the left-handed quark doublets, right-handed
up-type quarks and right-handed charged leptons,
${Q}_i$, $\overline{U}_i$ and $\overline{E}_i$, respectively, are contained in
${\bf 10}_i$ representations, $\Psi_i$. Here and subsequently, Roman letters from the middle of the alphabet 
are flavor indices. Also as in conventional SU(5),
the MSSM Higgs fields, $H_u$ and $H_d$, are combined with colored Higgs fields, 
$H_C$ and $\overline H_C$, to form a ${\bf 5}$ representation of SU(5), $H$, and a
${\bf \overline 5}$ representation, $\overline H$, respectively.

The difference from conventional SU(5) is that the SU(5) symmetry is broken down to the Standard Model (SM) gauge 
SU(3)$\times$SU(2)$\times$U(1) symmetry by a ${\bf 75}$-dimensional representation of SU(5), 
denoted by $\Sigma$.  Unlike minimal SU(5), the $H_u$ and $H_d$ have small masses without the
need for any fine-tuning. This is because, as described below,
the Higgs multiplets $H$ and $\overline H$ are coupled 
via the ${\bf 75}$ representation, $\Sigma$, to a ${\bf 50}$ representation, $\Theta$, and a
${\bf \overline{50}}$ representation, $\bar \Theta$, respectively. 
The $\Theta$ and $\bar \Theta$ contain $({\bf 3},{\bf 1}, -1/3)$ and $(\bar{\bf 3},{\bf 1}, 1/3)$ states that combine 
with the colored Higgs fields to give them large masses. However, since none of these representations contain states that transform as 
$({\bf 1}, {\bf 2}, \pm 1/2)$ under the SM SU(3)$\times$SU(2)$\times$U(1) gauge symmetry, the  $H_u$ and $H_d$ remain massless.

\subsection{The Superpotential and Symmetry Breaking}
\label{sec:superpotential}

The superpotential for this minimal supersymmetric SU(5) MPM is
\begin{align}
 W_5 &=  \frac{\mu_\Sigma}{2} \,  \Sigma^{AB}_{CD}\Sigma^{CD}_{AB} - \frac{1}{3} \lambda^\prime \,  \Sigma^{AB}_{CD}\Sigma^{CD}_{EF}\Sigma^{EF}_{AB} +\lambda_\Theta \overline{H}_A\Sigma^{DE}_{BC} \Theta^{ABC}_{DE} + \lambda_{\bar \Theta} H^A\Sigma^{BC}_{DE}\bar \Theta_{ABC}^{DE}
\nonumber \\[3pt]
&+M_\Theta \Theta^{ABC}_{DE} \bar{\Theta}_{ABC}^{DE} + \left(h_{\bf 10}\right)_{ij} \epsilon_{ABCDE}
 \Psi_i^{AB} \Psi^{CD}_j H^E +
 \left(h_{\overline{\bf 5}}\right)_{ij} \Psi_i^{AB} \Phi_{j A}
 \overline{H}_B \, ,
\label{W5}
\end{align}
where the upper-case Roman letters are SU(5) gauge indices and $\epsilon_{ABCDE}$ is the totally antisymmetric tensor. The upper and lower indices of $\Sigma^{AB}_{CD}$, $\Theta^{ABC}_{DE}$, and $\bar{\Theta}_{ABC}^{DE}$ are also totally antisymmetric, and these fields satisfy the following traceless conditions:~\footnote{These representations
may be described by the following Young tableaux:
\begin{equation}
   \Theta = 
    \ytableausetup{smalltableaux}
    \begin{ytableau}
        ~ & ~ \\ ~ & ~ \\ ~ & ~ 
    \end{ytableau}
    ~, \qquad 
    \bar{\Theta} = 
    \ytableausetup{smalltableaux}
    \begin{ytableau}
        ~ & ~ \\ ~ & ~ 
    \end{ytableau}
    ~, \qquad 
   \Sigma = 
    \ytableausetup{smalltableaux}
    \begin{ytableau}
        ~ & ~ \\ ~ & ~ \\ ~
    \end{ytableau}
    ~.
    \nonumber
\end{equation}
}
\begin{equation}
    \Sigma^{AB}_{AC} = \Theta^{ABC}_{AD} = \bar{\Theta}_{ABC}^{AD} = 0 ~.
\end{equation}
We assume a supergravity framework with the following minimal canonical form for the K\"{a}hler potential: 
\begin{align}
    K = \left| \Psi_i^{AB}\right|^2 +  \left| \Phi_{iA}\right|^2
    +  \left| H^A\right|^2 +  \left| \overline{H}_A\right|^2
    +  \left| \Theta^{ABC}_{DE}\right|^2
    +  \left| \bar{\Theta}_{ABC}^{DE}\right|^2
    +  \left| \Sigma_{CD}^{AB}\right|^2 ~,
\end{align}
where summations over the indices are understood.
The superpotential~\eqref{W5} consists of all the renormalizable terms that are allowed by the gauge symmetry and $R$-parity, except for the term bilinear in $H$ and $\overline{H}$. It is possible to suppress this term through an additional symmetry if the matter content is extended; for instance, models with extra global \cite{Hisano:1994fn, Altarelli:2000fu} or gauged \cite{Berezhiani:1996nu} U(1) symmetries have been discussed in the literature, and U(1)$_R$ symmetry may also be useful for this purpose, as used in a flipped $\text{SU}(5) \times \text{U}(1)$ model in Ref.~\cite{Hamaguchi:2020tet}. In the present work, however, we focus on the minimal matter content and assume that this bilinear term is absent.

Successful electroweak symmetry breaking requires both a SM $\mu$ term and a Higgs $B$ term
with magnitudes suitable for electroweak symmetry breaking, i.e., $\mathcal{O}(m_{3/2})$.
However, as we see below, the SM $\mu$-term does not arise from the breaking of SU(5),
so we use other ways to generate $\mu$ and the Higgs $B$ term
with the appropriate magnitudes.  One contribution comes from a Giudice-Masiero term in the K\"ahler potential~\cite{gm,Inoue:1991rk, Casas:1992mk, dlmmo}:
\begin{eqnarray}
\Delta K = c_KH\overline{H} +({\rm h.c.})~.
\label{gm1}
\end{eqnarray}
This generates $\mu$- and $B$-terms that are automatically of the correct magnitudes. However, the
magnitude of the $B$-term is fixed to be $2 c_K m_{3/2}$ while that of the $\mu$-term is $c_K m_{3/2}$. This means that we must use $m_{3/2}$ or $\tan \beta$ to satisfy 
the electroweak symmetry breaking conditions, which makes electroweak symmetry breaking much harder to realize.  
This issue can be resolved by remembering that the superpotential can also have a term of the form\cite{Evans:2020fmh}
\begin{eqnarray}
\Delta W =\frac{c_W\langle W_h \rangle }{M_P^2}H\overline{H}\ ,
\label{gm2}
\end{eqnarray}
where $M_P$ is the reduced Planck mass and $\langle W_h \rangle$ is the vacuum expectation value of the superpotential of the hidden sector that is responsible for the dominant contribution to supersymmetry breaking, i.e., the gravitino mass, $m_{3/2}$. This term provides an additional contribution to both $\mu$ and the $B$ term. If both contributions are included, the following expressions are found: 
\begin{align}
\mu&=(c_W +c_K)m_{3/2}\ , \nonumber \\
B\mu&=(-c_W+2c_K)m_{3/2}^2 \ .
\label{Bmu}
\end{align}
Clearly, the Higgs $\mu$ and  $B$ terms are now no longer directly proportional to each other,
and the freedom in $c_W$ and $c_K$ can be used to satisfy the electroweak symmetry-breaking conditions,
with $m_{3/2}$ and/or $\tan \beta$ being free parameters.

The {\bf 50} representation  $\Theta$ may be decomposed as follows in terms of SM 
representations~\footnote{We use the same conventions as~\cite{Slansky:1981yr}, except for the normalization of hypercharge.}
\begin{align}
 \Theta = ({\bf 1}, {\bf 1}, -2) \oplus ({\bf 3}, {\bf 1}, -1/3) \oplus (\overline{\bf 3}, {\bf 2}, -7/6) \oplus (\overline{\bf 6}, {\bf 3},- 1/3)
\oplus ({\bf 6}, {\bf 1}, {4}/{3}) \oplus ({\bf 8}, {\bf 2},
 {1}/{2}) ~,
 \label{eq:theta_decomp}
\end{align}
and the {\bf 75} representation $\Sigma$ as  
\begin{align}
    \Sigma &= (\mathbf{1}, \mathbf{1}, 0) \oplus (\mathbf{3}, \mathbf{1}, 5/3)  \oplus  (\bar{\mathbf{3}}, \mathbf{1}, -5/3) \oplus  (\mathbf{3}, \mathbf{2}, -5/6)  \oplus  (\bar{\mathbf{3}}, \mathbf{2}, 5/6)    \nonumber \\
    &\oplus  (\bar{\mathbf{6}}, \mathbf{2}, -5/6)\oplus  (\mathbf{6}, \mathbf{2}, 5/6) \oplus  (\mathbf{8}, \mathbf{1}, 0)   \oplus  (\mathbf{8}, \mathbf{3}, 0) ~.
    \label{eq:sigma_decomp}
\end{align}
As seen in Eq.~\eqref{eq:sigma_decomp}, the $\Sigma$ field contains an SM singlet component. We assume that $\Sigma$ develops a vev in this direction,
thereby breaking SU(5) without breaking the SM gauge symmetry.~\footnote{As discussed in Ref.~\cite{Hubsch:1985em}, there are numerous degenerate minima in the potential of the {\bf 75} field, which lead to different breaking patterns of SU(5). A discussion of the cosmological selection between these possible vacua is beyond the scope of this paper.} 
The vev of $\Sigma$ is thus of the form 
\begin{eqnarray}
&&\langle \Sigma^{\alpha\beta}_{\gamma\delta}\rangle_0=\frac{3}{2}V\left(\delta^\alpha_\gamma \delta^{\beta}_{\delta} - \delta^\alpha_\delta \delta^{\beta}_{\gamma}\right) \, , \nonumber \\
&&\langle \Sigma^{ab}_{cd}\rangle_0=\frac{1}{2}V\left(\delta^a_c \delta^{b}_{d} - \delta^a_d \delta^b_c\right) \, , \\
&&\langle \Sigma^{\alpha b}_{\gamma d}\rangle_0=\langle \Sigma^{b\alpha}_{d\gamma}\rangle_0=-\langle \Sigma^{b\alpha}_{\gamma d}\rangle_0=-\langle \Sigma^{\alpha b}_{d\gamma}\rangle_0=- \frac{1}{2}V\delta^\alpha_\gamma\delta^a_d \, , \nonumber
\end{eqnarray}
where the Greek letters $\alpha,\beta, ...$ are SU(2) indices, early Roman letters $a,b, ...$ are SU(3) indices, 
the $0$ subscript denotes, for later convenience, the vev with no supersymmetry breaking, and~\footnote{This result is consistent with those given in Refs.~\cite{Hisano:1994fn, Pokorski:2019ete}. }
\begin{eqnarray}
V=\frac{3}{4} \frac{\mu_\Sigma}{\lambda'} \, .
\end{eqnarray}
This breaks SU(5) down to the SM SU(3)$\times$SU(2)$\times$U(1) gauge symmetry, 
and gives masses to the GUT gauge bosons:~\footnote{This agrees with the result in Ref.~\cite{Hisano:1997nu}.  }
\begin{equation}
    M_X = \sqrt{24}g_5V ~,
\end{equation}
where $g_5$ is the SU(5) gauge coupling constant.

The following are the irreducible representations of SU(3)$\times$SU(2)$\times$U(1) that are contained within the ${\bf 75}$:~\footnote{These results are consistent with those in Ref.~\cite{Yamada:1993nqv}, up to a difference in overall normalization by a factor of 2 that originates from the difference in the normalization of the kinetic term. } 
\begin{align}
     \Sigma^{\alpha\beta}_{\gamma\delta} &= \frac{1}{2\sqrt{2}} \epsilon^{\alpha\beta} \epsilon_{\gamma\delta}  \Sigma_{(\mathbf{1}, \mathbf{1}, 0)} ~, \nonumber \\ 
     \Sigma^{\alpha\beta}_{cd} &= \frac{1}{2} \epsilon^{\alpha\beta} \epsilon_{cde} \Sigma^e_{(\mathbf{3}, \mathbf{1}, 5/3)} ~,
    \qquad \Sigma_{\gamma\delta}^{ab}= \frac{1}{2} \epsilon_{\gamma \delta} \epsilon^{abc} \Sigma_{c (\bar{\mathbf{3}}, \mathbf{1}, -5/3)} 
    ~, \nonumber \\ 
    \Sigma_{\gamma \delta}^{\alpha b} &= 
     \frac{1}{\sqrt{6}} \epsilon^{\alpha\beta}  \epsilon_{\gamma\delta} \Sigma^{b}_{\beta ({\mathbf{3}}, \mathbf{2}, -5/6)} 
    ~, \qquad
     \Sigma^{\alpha\beta}_{\gamma d} =  \frac{1}{\sqrt{6}} \epsilon^{\alpha\beta} \epsilon_{\gamma\delta} \Sigma^\delta_{d (\bar{\mathbf{3}}, \mathbf{2}, 5/6)} ~, 
    \nonumber  \\
     \Sigma^{a b}_{c\alpha} &= 
     \frac{1}{2} \epsilon^{abd} \Sigma^{}_{cd, \delta (\bar{\mathbf{6}}, \mathbf{2}, -5/6)} +
     \frac{1}{2\sqrt{6}} \bigl[
     \delta^b_c \Sigma^{a}_{\delta ({\mathbf{3}}, \mathbf{2}, -5/6)} -
     \delta^a_c \Sigma^b_{\delta ({\mathbf{3}}, \mathbf{2}, -5/6)}  \bigr] ~, \nonumber  \\ 
      \Sigma^{a \beta}_{cd} &= 
      \frac{1}{2} \epsilon_{bcd} \Sigma^{ab, \beta}_{(\mathbf{6}, \mathbf{2}, 5/6)} + 
      \frac{1}{2\sqrt{6}} \bigl[\delta^a_d \Sigma^{\beta}_{c (\bar{\mathbf{3}}, \mathbf{2}, 5/6)} -\delta^a_c \Sigma^{\beta}_{d (\bar{\mathbf{3}}, \mathbf{2}, 5/6)}  \bigr] ~, \nonumber \\ 
      \Sigma^{ab}_{cd} &= \frac{1}{\sqrt{6}}  \epsilon^{abe} \epsilon_{cd f}
      \Sigma_{e(\mathbf{8}, \mathbf{1}, 0) }^f + 
      \frac{1}{6 \sqrt{2}} \epsilon^{abe} \epsilon_{cd e}
      \Sigma_{(\mathbf{1}, \mathbf{1}, 0)} 
    ~, \nonumber \\ 
    \Sigma^{a\beta}_{c \delta} &= \frac{1}{2}
    \Sigma^{a \beta}_{c \delta (\mathbf{8}, \mathbf{3}, 0) }
    +\frac{1}{2\sqrt{6}}  \delta^\beta_\delta \Sigma^a_{c(\mathbf{8}, \mathbf{1}, 0) }  
    - \frac{1}{6\sqrt{2}} \delta^a_c \delta^\beta_\delta \Sigma_{(\mathbf{1}, \mathbf{1}, 0)} 
    ~,
\end{align}
where $\epsilon^{\alpha \beta} = \epsilon_{\alpha \beta}$ and $\epsilon^{abc} = \epsilon_{abc}$ are the totally antisymmetric tensors of rank 2 and 3, respectively, and each field component is labelled by its {SU}(3)$\times$SU(2)$\times$U(1) quantum numbers. 
The irreducible representations of the SM gauge symmetries contained in the ${\bf 50}$ are
\begin{align}
    \Theta^{abc}_{\delta \epsilon} &=  \frac{1}{2\sqrt{3}} 
\epsilon^{abc} \epsilon_{\delta \epsilon} \Theta_{({\bf 1}, {\bf 1}, -2)} ~, \qquad 
\Theta^{\alpha \beta c}_{de} = \frac{1}{2\sqrt{3}}
\epsilon^{\alpha\beta} \epsilon_{ade}\Theta^{ac}_{({\bf 6}, {\bf 1}, {4}/{3}) } ~,
\nonumber \\
 \Theta^{abc}_{de} &= \frac{1}{6} \epsilon^{abc} \epsilon_{def} \Theta^f_{({\bf 3}, {\bf 1}, -1/3)} ~,  \qquad 
    \Theta^{\alpha \beta c}_{ \delta \epsilon } = \frac{1}{6} \epsilon^{\alpha\beta} \epsilon_{\delta\epsilon}  \Theta^c_{({\bf 3}, {\bf 1}, -1/3)}
    ~, \nonumber \\
    \Theta^{abc}_{\delta e} &= \frac{1}{2\sqrt{6}} \epsilon^{abc} \Theta_{\delta e (\overline{\bf 3}, {\bf 2}, -7/6)} ~, \qquad 
\Theta^{ab\gamma}_{\delta \epsilon} =   \frac{1}{2\sqrt{6}} \epsilon^{abc} \epsilon^{\gamma \alpha}\epsilon_{\delta \epsilon} \Theta_{\alpha c (\overline{\bf 3}, {\bf 2}, -7/6)} ~, \nonumber \\ 
\Theta^{\alpha bc}_{\delta e} &= \frac{1}{2\sqrt{3}} \epsilon^{abc} \Theta^{\alpha}_{\delta ae (\overline{\bf 6}, {\bf 3},- 1/3)}  
+ \frac{1}{12} \epsilon^{abc} \epsilon_{ade} \delta^\alpha_{\delta} \Theta^d_{({\bf 3}, {\bf 1}, -1/3)} ~, \nonumber \\ 
\Theta^{ab\gamma}_{de} &= \frac{1}{2\sqrt{6}} \epsilon^{abc}\epsilon_{def}
\Theta^{f\gamma}_{c({\bf 8}, {\bf 2},{1}/{2}) } ~, 
\qquad 
\Theta^{a \beta \gamma}_{d \epsilon} = \frac{1}{2\sqrt{6}} 
\epsilon^{\beta \gamma } \epsilon_{\epsilon\alpha} 
\Theta^{a\alpha}_{d({\bf 8}, {\bf 2},{1}/{2}) } ~,
\end{align}
and its conjugate ${\bf \overline{50}}$ field, $\bar \Theta$, decomposes into the corresponding conjugate representations.

When the SU(5) symmetry is broken to the SM gauge symmetry SU(3)$\times$SU(2)$\times$U(1), the components in $\Sigma$ obtain masses as follows:~\footnote{These results agree with those in  Refs.~\cite{Hagiwara:1992ys,Hisano:1994fn, Hisano:1997nu}. }
\begin{align}
    M_{\Sigma_{(\mathbf{1}, \mathbf{1}, 0)}} &= 
    - \frac{4}{3} \lambda^\prime V  ~, \qquad 
    M_{\Sigma_{(\mathbf{3}, \mathbf{1}, 5/3)}} = 
    - 
    \frac{8}{3} \lambda^\prime V ~, \qquad 
    M_{\Sigma_{({\mathbf{3}}, \mathbf{2}, -5/6)}} = 0 ~, \nonumber \\
    M_{\Sigma_{(\mathbf{6}, \mathbf{2}, 5/6)}} &= \frac{4}{3} \lambda^\prime V ~, \qquad 
    M_{\Sigma_{(\mathbf{8}, \mathbf{1}, 0)}} = \frac{2}{3} \lambda^\prime V ~, \qquad 
    M_{\Sigma_{(\mathbf{8}, \mathbf{3}, 0) }} = \frac{10}{3} \lambda^\prime V ~.
    \label{eq:msigma}
\end{align}
We note that the $ (\mathbf{3}, \mathbf{2}, -5/6)  \oplus  (\bar{\mathbf{3}}, \mathbf{2}, 5/6) $ components remain massless after the SU(5) symmetry is broken, as they are the Nambu-Goldstone fields associated with this symmetry breaking that are absorbed by the SU(5) gauge vector multiplets.

As mentioned above, after the $\Sigma$ field acquires a vev, the $({\bf 3},{\bf 1}, -1/3)$ and $(\bar{\bf 3},{\bf 1}, 1/3)$ components in $\Theta$ and $\bar{\Theta}$ combine with the $(\bar{\bf 3},{\bf 1}, 1/3)$ and $({\bf 3},{\bf 1}, -1/3)$ components in $\bar{H}$ and $H$ via the couplings $\lambda_\Theta$ and $\lambda_{\bar{\Theta}}$, respectively, 
acquiring the following mass terms:
\begin{equation}
W  \supset
M_\Theta \Theta^a_{({\bf 3}, {\bf 1}, -1/3)} \bar{\Theta}_{a(\overline{\bf 3}, {\bf 1}, 1/3)} +
2 \lambda_{\Theta} V  \bar{H}_{C a}
    \Theta^a_{({\bf 3}, {\bf 1}, -1/3)} 
    + 
    2 \lambda_{\bar{\Theta}} V H_C^a \bar{\Theta}_{a(\overline{\bf 3}, {\bf 1}, 1/3)} ~. 
\end{equation}
As we discuss below, in order to preserve perturbativity of the SU(5) gauge coupling, we must require $M_\Theta\gg V$: indeed,
we find that $V \sim (6 - 7) \times 10^{16}$~GeV over the interesting region of parameter space. In this limit, we can integrate out $\Theta$ and $\bar{\Theta}$ at the scale of $M_\Theta$ to construct an effective theory in which the mass term for the color-triplet Higgs multiplets is given by 
\begin{equation}
  W_{\rm eff} \supset  -\lambda_\Theta \lambda_{\bar \Theta} \frac{(2V)^2}{M_\Theta}H_C^a\bar H_{Ca} \, ,\label{eq:ColorHiggsMass}
\end{equation}
from which we see that the colored Higgs mass $M_{H_C}$ is given by 
\begin{eqnarray}
M_{H_C}=\lambda_\Theta \lambda_{\bar \Theta} \frac{(2V)^2}{M_\Theta} \, .
\label{eq:mcolored}
\end{eqnarray}
The colored Higgs mass will be an important constraint on the model that will dictate the range of allowed parameter values,
in order to avoid rapid dimension-5 proton decay.~\footnote{It may be possible to avoid this complication in non-minimal models \cite{Masiero:1982fe}, 
but their exploration lies beyond the scope of this paper.}

On the other hand, the couplings $\lambda_\Theta$ and $\lambda_{\bar{\Theta}}$ do not give masses to the doublet components in $H$ and $\bar{H}$, since there is no corresponding doublet component in $\Theta$ and $\bar{\Theta}$, as seen in Eq.~\eqref{eq:theta_decomp}.

\subsection{Renormalization-Group Running}

We provide in this Section the supersymmetric Renormalization-Group equations (RGEs) for our minimal MPM,
starting with the running of the gauge coupling. We recall that in this model the matter fields consist of three $\bf 10$'s, 
four $\bf{\bar 5}$'s, one $\bf 5$, one $\bf 75$, one $\bf  50$, and one $\bf{\overline{50}}$.  The Casimir indices for these representations are
\begin{eqnarray}
&&C({\bf 5})=C(\bar{\bf 5})=\frac{1}{2} \, , \quad \quad \quad\quad\quad \!\!\!  C({\bf 10})=\frac{3}{2} \, , \nonumber\\
&&C({\bf 50})=C(\overline{\bf 50})=\frac{35}{2} \, , \quad \quad \quad C({\bf 75})=25 ~.
\end{eqnarray}
These imply that the one-loop RGE for the SU(5) gauge coupling is
\begin{eqnarray}
\frac{d g_5^2}{dt}=\frac{b_5}{8\pi^2}g_5^4 \, , \quad {\rm where} \quad \quad b_5=52 ~,
\end{eqnarray}
where $t \equiv \ln Q$ with $Q$ the renormalization scale. 
Because of the large SU(5) representations, the one-loop beta function is almost non-perturbative: $\frac{52}{8\pi^2}\simeq 0.65$.
For this reason, one must either place a lower bound on $M_\Theta$ in order to ensure perturbativity of the SU(5) gauge coupling up to the Planck scale,
or there must be an effective cutoff for the theory where it has to be UV-completed.  
As we expect some new physics to enter at $M_{\rm in}$, we require only
that $M_\Theta$ be large enough to push the Landau pole
beyond this input scale.  This is done by solving the one-loop RGE for the gauge couplings in two regimes,
above and below $M_\Theta$, and matching them at $M_\Theta$.
We find that the Landau pole is above the input scale if
\begin{eqnarray}
M_\Theta> M_{\rm GUT} \left(\frac{M_{\rm in}}{M_{\rm GUT}}\right)^{\frac{52}{35}} \exp\left(\frac{-8\pi^2}{g^2_5(M_{\rm GUT})}\frac{1}{35}\right)\simeq 2 \times 10^{17}~{\rm GeV} \, ,
\label{thetabound}
\end{eqnarray}
where $M_{\rm GUT}$ is the scale where we match the GUT theory to the MSSM, defined to be the scale where
the Standard Model couplings, $g_1$ and $g_2$ are equal and $g_5(M_{\rm GUT})$ is the gauge coupling at $M_{\rm GUT}$.  The room for RGE running in the presence of $\Theta$
is limited, but we seek to keep $M_\Theta$ as light as possible, in order that the triplet Higgs mass
not be so light that the proton lifetime is too short.

The one-loop RGEs for the Yukawa couplings are:
\begin{align}
    \frac{d \lambda^\prime}{dt} &= \frac{\lambda'}{16\pi^2}\left(\frac{56}{3}|\lambda'|^2 +2|\lambda_\Theta|^2+2|\lambda_{\bar \Theta}|^2-48g_5^2\right) \, , \\
\frac{d \lambda_\Theta}{dt} &= \frac{\lambda_\Theta}{16\pi^2}\left(\frac{35}{3}|\lambda_\Theta|^2 +\frac{2}{3}|\lambda_{\bar \Theta}|^2+\frac{56}{9}|\lambda'|^2+2{\bf Tr}(h_5^\dagger h_5)-\frac{188}{5}g_5^2\right) \, , \\
\frac{d \lambda_{\bar{\Theta}}}{dt} &= \frac{\lambda_{\bar \Theta}}{16\pi^2}\left(\frac{35}{3}|\lambda_{\bar \Theta}|^2 +\frac{2}{3}|\lambda_{ \Theta}|^2+\frac{56}{9}|\lambda'|^2 + 48{\bf Tr}(h_{10}^\dagger h_{10}) -\frac{188}{5}g_5^2\right) \, , \\
\frac{d h_{{10}_{33}}}{dt} &= \frac{h_{{10}_{33}}}{16\pi^2}\left(144|h_{{10}_{33}}|^2+2|h_{5_{33}}|^2 +10|\lambda_{\bar \Theta}|-\frac{96}{5}g_5^2\right) \, , \\
\frac{d h_{{5}_{33}}}{dt}&=\frac{h_{5_{33}}}{16\pi^2}\left(48|h_{10_{33}}|^2+5|h_{5_{33}}|^2 +10|\lambda_\Theta|^2-\frac{84}{5}g_5^2\right) \, ,
\end{align}
where we have assumed that the Yukawa couplings of the first two fermion generations can be neglected. Notice that this set of RGEs is to be used above the mass threshold of $\Theta$ and $\bar{\Theta}$; below this mass scale, the terms including the couplings $\lambda_{\Theta}$ and $\lambda_{\bar{\Theta}}$ in the above equations are set to zero. 

\subsection{Supersymmetric Matching Conditions}

The GUT-scale matching conditions for the gauge couplings change drastically from those in minimal SU(5), 
since the $\bf 75$ has many component fields with different SM charges and masses, as seen above.  
In addition, there is the possibility of a Planck-scale-suppressed dimension-five operator constructed out of the gauge and {\bf 75} fields,
\begin{eqnarray}
W^{\Delta g}_{\rm eff}=\frac{d}{M_P}{\cal W}^C_A{\cal W}^D_B \Sigma^{AB}_{CD} \, , \label{eq:SigWW}
\end{eqnarray}
which would also affect the matching conditions~\footnote{It is stated in~\cite{Pokorski:2019ete} that the minimal missing partner model we consider is ruled out because matching conditions force a non-perturbative gauge coupling at the GUT scale. We evade this conclusion by including the dimension-five operator in Eq.~(\ref{eq:SigWW}), which alters the matching conditions and allows viable models. In the models considered, a value of  $d \sim 0.2$ is sufficient to ensure perturbative gauge couplings.}~\cite{Hisano:1997nu, Huitu:1999eh}, where ${\cal W}^A_B$ denotes the gauge field strength chiral superfields. We then have the following matching conditions for the gauge coupling constants in the $\overline{\text{DR}}$ scheme: 
\begin{align}
\frac{1}{g_1^2(Q)} &=\frac{1}{g_5^2(Q)}+\frac{1}{8\pi^2} \biggl[
 10\ln \biggl( \frac{Q}{M_{\Sigma_{(\mathbf{3}, \mathbf{1}, 5/3)}}} \biggr)
+ 10\ln \biggl(\frac{Q}{M_{\Sigma_{({\mathbf{6}}, \mathbf{2}, 5/6)}}}\biggr) 
\nonumber \\[2pt] 
& \hspace{2cm} +\frac{2}{5}\ln\left(\frac{Q}{M_{H_C}}\right)-10\ln \left(\frac{Q}{M_X}\right)
\biggr]+\frac{5}{2}\left(\frac{8dV}{M_P}\right)~, \\[4pt]
\frac{1}{g_2^2(Q)} &=\frac{1}{g_5^2(Q)}+\frac{1}{8\pi^2} 
\biggl[ 
6\ln \biggl(\frac{Q}{M_{\Sigma_{({\mathbf{6}}, \mathbf{2}, 5/6)}}}\biggr) + 16\ln \biggl(\frac{Q}{M_{\Sigma_{({\mathbf{8}}, \mathbf{3}, 0)}}}\biggr) - 6 \ln\left(\frac{Q}{M_X}\right)
\biggr]-\frac{3}{2}\left(\frac{8dV}{M_P}\right) ~, \\ 
\frac{1}{g_3^2(Q)}&=\frac{1}{g_5^2(Q)}+\frac{1}{8\pi^2}
\biggl[
10\ln \biggl( \frac{Q}{M_{\Sigma_{({\mathbf{6}}, \mathbf{2}, 5/6)}}} \biggr)
+\ln \biggl( \frac{Q}{M_{\Sigma_{(\mathbf{3}, \mathbf{1}, 5/3)}}} \biggr)
   + 3\ln \biggl( \frac{Q}{M_{\Sigma_{({\mathbf{8}}, \mathbf{1}, 0)}}} \biggr)
   \nonumber \\[2pt] 
  & \hspace{2cm} + 9\ln \biggl( \frac{Q}{M_{\Sigma_{({\mathbf{8}}, \mathbf{3}, 0)}}}\biggr)
  + \ln\left(\frac{Q}{M_{H_C}}\right)-4\ln\left(\frac{Q}{M_X}\right)
\biggr]-\frac{1}{2}\left(\frac{8dV}{M_P}\right)  ~.
\end{align}
The conditions on the effective theory below the mass threshold of $\Theta$ and $\bar{\Theta}$ when these fields are integrated out can be
simplified to~\footnote{
These simplified equations are modified if $\lambda'$ is small and there are higher-dimensional operators consisting of {\bf 75} fields, such as $\left(\Sigma^{AB}_{CD}\Sigma^{CD}_{AB}\right)^2$, that become important. 
In this case, the masses of the component fields in $\Sigma$ are shifted from those in Eq.~\eqref{eq:msigma}, altering the GUT-scale matching conditions. A general discussion of the possible impacts of this and other higher-dimensional operators lies beyond the scope of this paper.}
\begin{align}
 \frac{3}{g_2^2(Q)} - \frac{2}{g_3^2(Q)} -\frac{1}{g_1^2(Q)}
&=-\frac{3}{10\pi^2} \ln \left(\frac{Q}{M_{H_C}}N_{H_C}\right)
-\frac{48dV}{M_P}
~,\label{eq:matchmhc} \\[3pt]
 \frac{5}{g_1^2(Q)} -\frac{3}{g_2^2(Q)} -\frac{2}{g_3^2(Q)}
&= -\frac{3}{2\pi^2}\ln\biggl(\frac{Q^3}{M_X^2 M_{\Sigma_{(\mathbf{8}, \mathbf{1}, 0)}}}N_X\biggr)+\frac{144dV}{M_P} ~,
\label{eq:matchmgut}
\\[3pt]
 \frac{5}{g_1^2(Q)} +\frac{3}{g_2^2(Q)} -\frac{2}{g_3^2(Q)}&= -\frac{15}{2\pi^2} \ln\biggl(N_{g_5}\frac{M_{\Sigma_{(\mathbf{8}, \mathbf{1}, 0)}}^2}{M_XQ}\biggr) + \frac{6}{g_5^2(Q)} +\frac{72dV}{M_P} ~,\label{eq:matchg5}
\end{align}
where we have used (see Eq.~\eqref{eq:msigma}) 
\begin{equation}
    M_{\Sigma_{(\mathbf{6}, \mathbf{2}, 5/6)}} = 2 M_{\Sigma_{(\mathbf{8}, \mathbf{1}, 0)}} ~, \quad 
    M_{\Sigma_{(\mathbf{3}, \mathbf{1}, 5/3)}} = 4 M_{\Sigma_{(\mathbf{8}, \mathbf{1}, 0)}} ~, \quad 
     M_{\Sigma_{(\mathbf{8}, \mathbf{3}, 0) }} = 5 M_{\Sigma_{(\mathbf{8}, \mathbf{1}, 0)}} ~,
\end{equation}
and 
\begin{align}
    N_{H_C} &\equiv \biggl(\frac{M_{\Sigma_{({\mathbf{8}}, \mathbf{3}, 0)}}^5 }{M_{\Sigma_{(\mathbf{3}, \mathbf{1}, 5/3)}}^2 M_{\Sigma_{({\mathbf{6}}, \mathbf{2}, 5/6)}}^2  M_{\Sigma_{({\mathbf{8}}, \mathbf{1}, 0)}}}\biggr)^{\frac{5}{2}} = \frac{5^{\frac{25}{2}}}{2^{15}} ~, \nonumber \\ 
    N_X & \equiv 
    \frac{M_{\Sigma_{(\mathbf{3}, \mathbf{1}, 5/3)}}^4
    M_{\Sigma_{({\mathbf{6}}, \mathbf{2}, 5/6)}}
    M_{\Sigma_{({\mathbf{8}}, \mathbf{1}, 0)}}^{1/2}
    }{   
    M_{\Sigma_{({\mathbf{8}}, \mathbf{3}, 0)}}^{11/2}
    } = \frac{2^9}{5^{\frac{11}{2}}} ~, \nonumber \\ 
    N_{g_5} & \equiv 
     \frac{M_{\Sigma_{(\mathbf{3}, \mathbf{1}, 5/3)}}^{4/5}
    M_{\Sigma_{({\mathbf{6}}, \mathbf{2}, 5/6)}}^{4/5}
    M_{\Sigma_{({\mathbf{8}}, \mathbf{3}, 0)}}^{1/2}
    }{   M_{\Sigma_{({\mathbf{8}}, \mathbf{1}, 0)}}^{21/10}
    } = 2^{\frac{12}{5}} 5^{\frac{1}{2}} ~.
\end{align}
We note that Eq.~\eqref{eq:matchmgut} is dependent on the contribution of the dimension-five operator~\cite{Hisano:1997nu}, 
which is contrary to the case in minimal SUSY SU(5) GUT~\cite{Ellis:2016tjc, Hall:1992kq}. 
The inclusion of terms $\propto d \sim 0.2$ in (\ref{eq:matchmhc}, \ref{eq:matchmgut}, \ref{eq:matchg5})
avoids the problem of non-perturbative gauge couplings
emphasized in~\cite{Pokorski:2019ete}.

We can find a condition that is independent of $d$ and gives a constraint on the masses that must always be satisfied, namely:
\begin{eqnarray}
\frac{6}{g_2^2(Q)}-\frac{8}{g_3^2(Q)}+\frac{2}{g_1^2(Q)}=\frac {27}{5\,{\pi}^{2}}\ln  \left( 5^{-\frac{5}{9}}
\frac{M_G}{Q} \right) \, ,
\end{eqnarray}
where 
\begin{eqnarray}
M_G=\left(M_{H_C}^3M_{\Sigma_{({\mathbf{8}}, \mathbf{1}, 0)}}^{5} M_X^{10}\right)^\frac{1}{18} \, .
\end{eqnarray}
We can substitute the above expressions for $M_{H_C}$, $M_{\Sigma_{({\mathbf{8}}, \mathbf{1}, 0)}}$ and $M_X$ into this expression and solve for $V$ to get
\begin{eqnarray}
V=\frac{2^\frac{16}{21}}{4} \left(\frac{M_G^{18}M_{\Theta}^3}{g_5^{10}\lambda'^5 \lambda_\Theta^3\lambda_{\bar \Theta}^3}\right)^{\frac{1}{21}} \, .
\end{eqnarray}
We can then use this expression and those for $M_X$ and $M_{\Sigma_{({\mathbf{8}}, \mathbf{1}, 0)}}$ in Eq. (\ref{eq:matchg5}) to solve for $g_5$.

\begin{figure}[!ht]
\centering
\includegraphics[width=12cm]{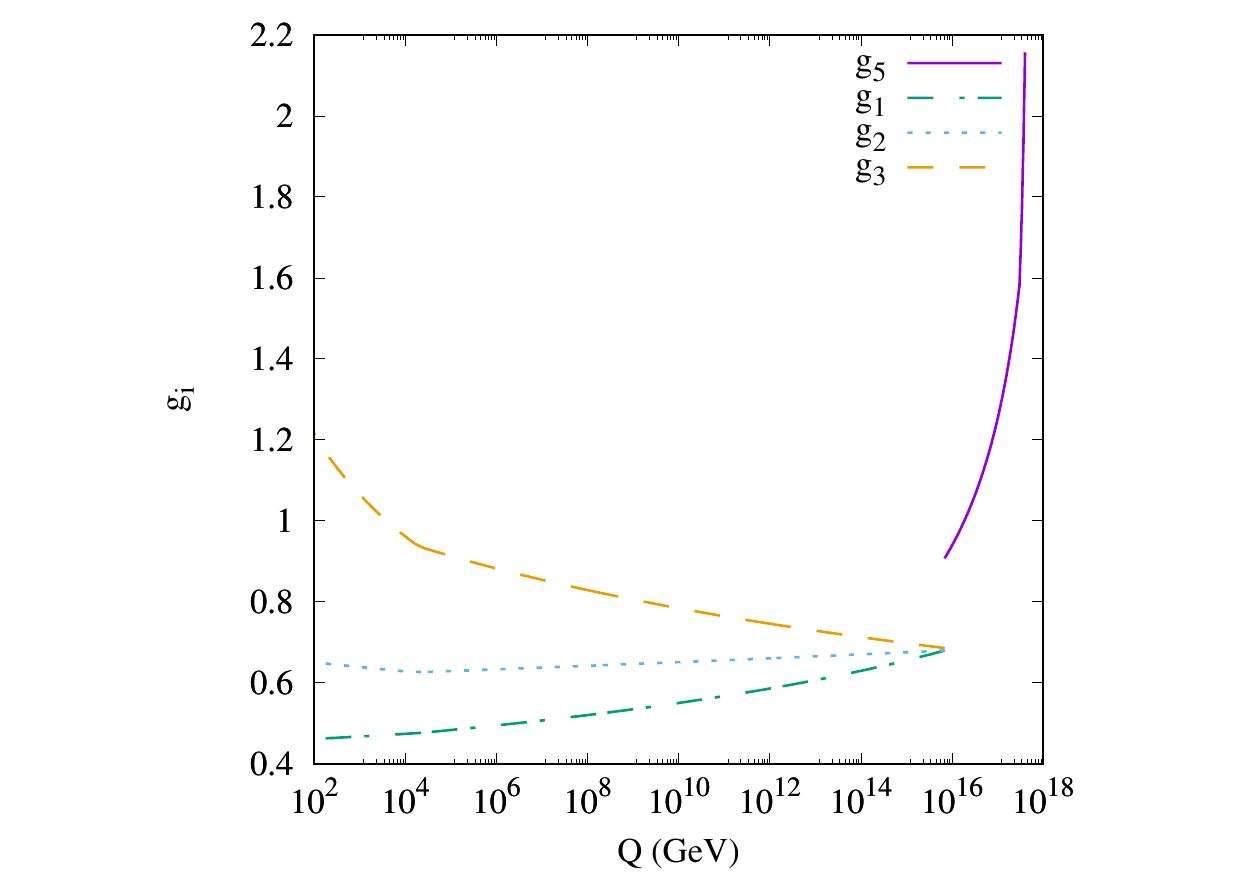}
\caption{\it Evolutions of the SM gauge couplings $g_{1,2,3}$ with the renormalization scale $\mu$
below the GUT scale, exhibiting their unification and the rapid increase of $g_5$ above the GUT scale.
We assume a benchmark point with $m_{1/2} = 20$ TeV: the values of the other benchmark parameters
are specified in the text. }
\label{fig:gaugeunification}
\end{figure}

We consider as a benchmark point a
model with $\tan \beta = 4.5$, $m_{1/2} = 20$~TeV and $A_0/m_0 = 2.25$, and assume that the input $B$-term satisfies the minimal supergravity condition $B_0 = A_0 - m_0$~\cite{bfs}.
We choose $M_{\rm in} \simeq 4\times 10^{17}$~GeV and $M_{\Theta} = 3 \times 10^{17}$~GeV, which satisfies the bound (\ref{thetabound}). We take $\lambda^\prime = 0.005$, $\lambda_{\Theta,{\bar \Theta}} = 1$, and sgn($\mu) < 0$. In order to obtain the correct relic density, we take $m_0 = 15.9$~TeV,
so that the lightest supersymmetric particle (LSP) (which is the bino) and lighter stop are nearly degenerate, with the LSP mass at 4.2 TeV and $\Delta m = 12$ GeV. 
The input parameters for this benchmark point are given in Table~\ref{tab:numbers}, together with its derived GUT-scale quantities, MSSM parameters and observables.
Using these inputs, 
we see in Fig.~\ref{fig:gaugeunification} that the SM gauge couplings $g_{1,2,3}$
nearly unify at the GUT scale, where $g_1 =g_2$ and $g_3/g_1 = 1.009$. There is a large
threshold effect on the gauge coupling at the GUT scale, where the unified SU(5) gauge coupling $g_5 = 0.907$. It then increases rapidly at higher scales because 
of the large contributions to the one-loop renormalization coefficient $\beta_5$ coming from the $\mathbf{75}$,
$\mathbf{50}$ and $\mathbf{\overline{50}}$ representations in the MPM.

\begin{table}
\small
\begin{center}
\label{tbl:MassesFV}
\begin{tabular}{|ccc|}
\hline 
\multicolumn{3}{|c|}{Inputs}\\
\hline
$m_{1/2} = 20$ TeV & $m_0 = 15.9$ TeV & $A_0/m_0 = 2.25$\\
$\tan \beta = 4.5$ & $M_{\rm in} = 10^{17.6}$ GeV & $M_\Theta = 3\times 10^{17}$ GeV \\
$\lambda^\prime = 0.005$ &  $\lambda_{\Theta,{\bar \Theta}} = 1$ & $B_0 = A_0 - m_0$ \\
\hline
\multicolumn{3}{|c|}{GUT-scale parameters (masses in units of $10^{16}$ GeV)} \\
\hline
$M_{\rm GUT} = 0.692$  & $M_{H_C} = 5.53$ &  $M_{\Sigma} = 0.0215$ \\
$M_G = 2.95$  & $M_X = 28.6$ & $V= 6.46$ \\
$g_5 = 0.907$  & $d = 0.24$  & \\
\hline
\multicolumn{3}{|c|}{MSSM parameters (masses in units of TeV)} \\
\hline
$m_\chi = 4.2$ &   $m_{\tilde t_1} = 4.2 $ & $m_{\tilde g} = 17.7 $  \\
$m_{\chi_2} = 8.5$ & $m_{\tilde H} = 24.1$ & $\mu = -23.5$  \\
$m_{\tilde l_L} = 21.5$ & $m_{\tilde l_R} = 23.2$ & $m_{\tilde \tau_1} = 20.5$ \\
$m_{\tilde q_L} = 26.6$ & $m_{\tilde d_R} = 24.4$ & $m_{\tilde t_2} = 18.1$ \\
$A_t = 32.7$ & $A_d = 80.9$ & $B = -14.6$ \\
$c_K =-0.043$ & $c_W = -1.44$ & \\
\hline
\multicolumn{3}{|c|}{Observables} \\
\hline
$\Omega_\chi h^2 = 0.125$ & $m_h = 125.3$ GeV & $\tau_p = (0.099 \pm 0.026) \times 10^{35}$~yrs \\
\hline
\end{tabular}
\end{center}
\caption{\it {Benchmark point parameters. We first give the input parameters defined at $M_{\rm in}$, with the exception of the Yukawa couplings, which are defined at $M_{\rm GUT}$, and assume $\mu < 0$. The second set of parameters are GUT scale quantities determined by the matching conditions described in the text. The third set of quantities are derived from the running of the RGEs and matching at the weak scale. 
We find that $m_{\tilde t_1} - m_\chi = 12$ GeV at this point.  The chargino masses are nearly degenerate with the wino and Higgsino masses given as $m_{\chi_2}$ and $m_{\tilde H}$, $m_{\tilde \tau_2}$ is similar to $m_{\tilde l_L}$, $m_{\tilde u_R}$ is similar to $m_{\tilde q_L}$, $m_{\tilde b_1}$ is similar to $m_{\tilde t_2}$, and $m_{\tilde b_2}$ is similar to $m_{\tilde d_R}$. Other $A$-terms take values between $A_t$ and $A_d$. The final entries correspond to the observables we concentrate on in this work: the relic density, the Higgs mass, and the proton lifetime.}  \label{tab:numbers}}
\end{table}

\section{Supersymmetry Breaking}
\label{sec:susyb}

We now look at supersymmetry breaking in the minimal MPM 
and the associated RGEs. The soft supersymmetry-breaking terms in the model are
\begin{align}
 {\cal L}_{\rm soft} = &- \left(m_{\bf 10}^2\right)_{ij}
 \widetilde{\psi}_i^* \widetilde{\psi}_j
- \left(m_{\overline{\bf 5}}^2\right)_{ij} \widetilde{\phi}^*_i
 \widetilde{\phi}_j
- m_H^2 |H|^2 -m_{\overline{H}}^2 |\overline{H}|^2 - m_\Sigma^2 |\Sigma|^2
- m_{\Theta}^2 |\Theta|^2 - m_{\bar{\Theta}}^2 |\bar{\Theta}|^2
\nonumber \\
&-\biggl[
\frac{1}{2}M_5 \widetilde{\lambda}^{A} \widetilde{\lambda}^A
+ A_{\bf 10} \left(h_{\bf 10}\right)_{ij}
 \epsilon_{\alpha\beta\gamma\delta\zeta} \widetilde{\psi}_i^{\alpha\beta}
 \widetilde{\psi}^{\gamma\delta}_j H^\zeta
+ A_{\overline{\bf 5}}\left(h_{\overline{\bf 5}}\right)_{ij}
 \widetilde{\psi}_i^{\alpha\beta} \widetilde{\phi}_{j \alpha}  \overline{H}_\beta
\nonumber \\
&+ \frac{1}{2}B_\Sigma \mu_\Sigma \Sigma^2 -\frac{1}{3} A_{\lambda^\prime
 } \lambda^\prime  \Sigma^3+A_\Theta \lambda_\Theta \bar H\Sigma\Theta +A_{\bar \Theta}\lambda_{\bar \Theta}H\Sigma \bar \Theta+B_\Theta M_\Theta \Theta\bar \Theta+{\rm h.c.}
 \biggr]~.
 \label{eq:softparam}
\end{align}

\subsection{Renormalization Group Equations}

Some of the one-loop RGEs in the MPM differ from those in minimal SU(5).  
The new one-loop RGEs for soft supersymmetry-breaking scalar masses are
\begin{align}
    \frac{dm_{\Sigma}^2}{dt}&= \frac{1}{8\pi^2}\left[\frac{56}{9}|\lambda'|^2S_{\lambda'}+ \frac{2}{3}|\lambda_\Theta|^2S_\Theta + \frac{2}{3}|\lambda_{\bar \Theta}|^2S_{\bar\Theta} -32g_5^2|M_5|^2\right] \, , \\
\frac{dm_\Theta^2}{dt}&= \frac{1}{8\pi^2}\left[|\lambda_\Theta|^2S_\Theta-\frac{168}{5}g_5^2|M_5|^2\right] \, , \\
\frac{dm_{\bar \Theta}^2}{dt}&= \frac{1}{8\pi^2}\left[|\lambda_{\bar \Theta}|^2S_{\bar \Theta}-\frac{168}{5}g_5^2|M_5|^2\right] \, , \\
\frac{dm_H^2}{dt} &= \frac{1}{8\pi^2}\left[48|h_{10}|^2S_{10} +10|\lambda_{\bar \Theta}|^2S_{\bar \Theta}-\frac{48}{5}g_5^2|M_5|^2\right] \, , \\
\frac{dm_{\bar H}^2}{dt} &= \frac{1}{8\pi^2}\left[2|h_{5}|^2S_{5} +10|\lambda_{\Theta}|^2S_{ \Theta}-\frac{48}{5}g_5^2|M_5|^2\right] \, , 
\end{align}
where $h_{10}$ and $h_{5}$ are short-hand notations for $h_{10_{33}}$ and $h_{5_{33}}$, and
\begin{align}
S_{\lambda'} &=3m_\Sigma^2+|A_{\lambda'}|^2 \, ,  \nonumber \\
S_\Theta &= m_{\bar H}^2+m_\Sigma^2+m_\Theta^2+|A_\Theta|^2 \, , \nonumber\\
S_{\bar \Theta}&= m_{H}^2+m_\Sigma^2+m_{\bar \Theta}^2+|A_{\bar \Theta}|^2 \, , \nonumber\\
S_{10} &=m_H^2+2m_{10}^2+|A_{10}|^2 \, , \nonumber\\
S_5 &=m_{\bar H}^2 +m_{10}^2+m_{5}^2+|A_5|^2 \, ,
\end{align}
where $m_{10}^2$, $m_{5}^2$, $A_{10}$, and $A_{5}$ are analogous
short-hand notations for $m_{10_{33}}^2$, $m_{5_{33}}^2$, $A_{10_{33}}$
and $A_{5_{33}}$. The RGEs for the first two generations can be found 
from these by setting $h_5$ and $h_{10}$ to zero.

The RGEs for the $A$-terms are
\begin{align}
\frac{dA_{10}}{dt} &=\frac{1}{8\pi^2}\left[144|h_{10}|^2A_{10}+2|h_5|^2A_5 +10|\lambda_{\bar \Theta}|^2A_{\bar\Theta}-\frac{96}{5}g_5^2M_5\right] \, , \\
\frac{dA_{5}}{dt} &=\frac{1}{8\pi^2}\left[48|h_{10}|^2A_{10}+5|h_5|^2A_5 +10|\lambda_{ \Theta}|^2A_{\Theta}-\frac{84}{5}g_5^2M_5\right] \, , \\
\frac{d A_{10_1}}{dt} &=\frac{1}{8\pi^2}\left[48|h_{10}|^2A_{10} +10|\lambda_{\bar \Theta}|^2A_{\bar\Theta}-\frac{96}{5}g_5^2M_5\right] \, , \\
\frac{dA_{5_1}}{dt} &=\frac{1}{8\pi^2}\left[2|h_5|^2A_5 +10|\lambda_{ \Theta}|^2A_{\Theta}-\frac{84}{5}g_5^2M_5\right] \, , \\
\frac{dA_{\lambda'}}{dt} &=\frac{1}{8\pi^2}\left[\frac{56}{3}|\lambda'|^2A_{\lambda'} +2|\lambda_{\bar \Theta}|^2A_{\bar\Theta} +2|\lambda_{\Theta}|^2A_{\Theta} -48g_5^2M_5\right] \, , \\
\frac{dA_{\lambda_{\bar \Theta}}}{dt} &=\frac{1}{8\pi^2}\left[48|h_{10}|^2A_{10} +\frac{56}{9}|\lambda'|^2A_{\lambda'} +\frac{35}{3}|\lambda_{\bar\Theta}|^2A_{\bar\Theta} +\frac{2}{3}|\lambda_{\Theta}|^2A_{\Theta} -\frac{188}{5}g_5^2M_5\right] \, , \\
\frac{dA_{\lambda_{ \Theta}}}{dt} &=\frac{1}{8\pi^2}\left[2|h_{5}|^2A_{5} +\frac{56}{9}|\lambda'|^2A_{\lambda'} +\frac{2}{3}|\lambda_{\bar\Theta}|^2A_{\bar\Theta} +\frac{35}{3}|\lambda_{\Theta}|^2A_{\Theta} -\frac{188}{5}g_5^2M_5\right] \, , \\
\frac{dB_{\Sigma}}{dt} &=\frac{1}{8\pi^2}\left[\frac{112}{9}|\lambda'|^2 A_{\lambda'}+\frac{4}{3}\left(|\lambda_\Theta|^2A_\Theta +|\lambda_{\bar\Theta}|^2A_{\bar\Theta}\right)-32g_5^2M_5\right] \, , \\
 \frac{dB_{\Theta}}{dt} &=\frac{1}{8\pi^2}\left[|\lambda_\Theta|^2A_\Theta +|\lambda_{\bar\Theta}|^2A_{\bar\Theta}-\frac{168}{5}g_5^2M_5\right] \, .
\end{align}
Again, below the mass scale of $\Theta$ and $\bar{\Theta}$, the contributions of the couplings $\lambda_{\Theta}$ and $\lambda_{\bar{\Theta}}$ are set to zero.

\subsection{Supersymmetry-Breaking Matching Conditions}
\label{sec:susybreakingmat}

We consider next the matching conditions of the supersymmetry-breaking parameters, focusing on those for the gauginos,
since those for squarks and sleptons are the same as in minimal SU(5) and expressions for $\mu$ and $B$ were given
in (\ref{Bmu}). The matching for the missing-partner model has been done in Ref.~\cite{Hisano:1993zu}, but without 
expressing explicitly, in terms of the soft parameters in the original Lagrangian, the $B$-term
needed to match the components of the $\Theta$ and $\bar \Theta$ that mix with the colored Higgs fields. Since $\Theta$ and $\bar \Theta$ must be significantly heavier than the GUT scale, there are two matching scales.  At $M_\Theta$, $\Theta$ and $\bar \Theta$ must be integrated out in an SU(5)-symmetric way, and we use the superpotential in Eq. (\ref{eq:ColorHiggsMass}) for the GUT-scale matching conditions. The $B$-term for the colored Higgs bosons in this superpotential is 
\begin{eqnarray}
{\cal L}_{\rm soft}\supset -\lambda_\Theta \lambda_{\bar \Theta}\left(B_\Theta-A_\Theta-A_{\bar \Theta}+2A_\Sigma-2B_\Sigma\right) \frac{(2V)^2}{M_\Theta}\bar H_C H_C \, .
\end{eqnarray}
Using this, we find that the SU(5) gaugino gets the following correction from integrating out $\Theta$ and $\bar \Theta$:
\begin{eqnarray}
\Delta M_5=\pm 35 \frac{g_5^2}{16\pi^2}B_\Theta \, ,
\label{ThetaCorrection}
\end{eqnarray}
where the correction is positive when running down from 
$M_{\rm in}$. 
In order to obtain the matching condition for the gauginos at the GUT scale, we need the correction to the vacuum expectation value of $\Sigma$ from the soft supersymmetry-breaking terms in Eq. (\ref{eq:softparam}), which is
\begin{eqnarray}
\langle \Sigma^{AB}_{CD}\rangle = \left[1+\frac{A_\Sigma-B_\Sigma}{\mu_\Sigma}+\left(A_\Sigma- B_\Sigma\right) \theta^2\right]\langle \Sigma^{AB}_{CD}\rangle_0 \, .
\end{eqnarray}
where the term proportional to $\theta^2$ is the $F$-term of $\Sigma$. This $F$-term leads to an additional correction to the gaugino matching conditions.  
Including this, we find that the gaugino mass matching conditions are
\begin{align}
 M_1 &= \frac{g_1^2}{g_5^2} M_5
-\frac{g_1^2}{16\pi^2}\left[10 M_5 -10(A_{\lambda^\prime} -B_\Sigma)
 -20 B_\Sigma \frac{}{}\right. \nonumber\\  &\quad \quad \quad \quad\quad \quad \quad  \left. +\frac{2}{5}\left(B_\Theta-A_\Theta -A_{\bar\Theta}+2A_\Sigma-2B_\Sigma\right)\right]+\frac{5}{4}\left(\frac{8dV}{M_P}\right)(A_{\lambda'}-B_\Sigma) \, , 
\label{eq:m1match}
\\[3pt]
M_2 &= \frac{g_2^2}{g_5^2} M_5
-\frac{g_2^2}{16\pi^2}\left[6 M_5 - 6
\left(A_{\lambda^\prime}-B_\Sigma\right)-22B_\Sigma
 \right]-\frac{3}{4}\left(\frac{8dV}{M_P}\right)(A_{\lambda'}-B_\Sigma) \, ,
\label{eq:m2match}
\\[3pt]
M_3 &= \frac{g_3^2}{g_5^2} M_5
-\frac{g_3^2}{16\pi^2}\left[4 M_5 - 4\left(A_{\lambda^\prime} - B_\Sigma\right)
-23B_\Sigma+B_\Theta-A_\Theta-A_{\bar \Theta}+2A_\Sigma-2B_\Sigma \right] \nonumber \\   &\quad \quad \quad \quad\quad \quad \quad \quad\quad \quad \quad \quad \quad \quad \quad \quad\quad \quad \quad \quad\quad \quad  -\frac{1}{4}\left(\frac{8dV}{M_P}\right)(A_{\lambda'}-B_\Sigma)
~.
\label{eq:m3match}
\end{align}
The matching conditions for the squarks and sleptons are the same as in the minimal SU(5) super-GUT, and can be found in Ref.~\cite{Ellis:2019fwf} and references therein.
As discussed earlier, there is no matching conditions for the MSSM $\mu$- and $B$-terms. 
These are fixed at the weak by the electroweak symmetry breaking minimization conditions by introducing the Giudice-Masiero terms in Eqs.~(\ref{gm1}) and (\ref{gm2}).

\begin{figure}[!ht]
\centering
\includegraphics[width=9cm]{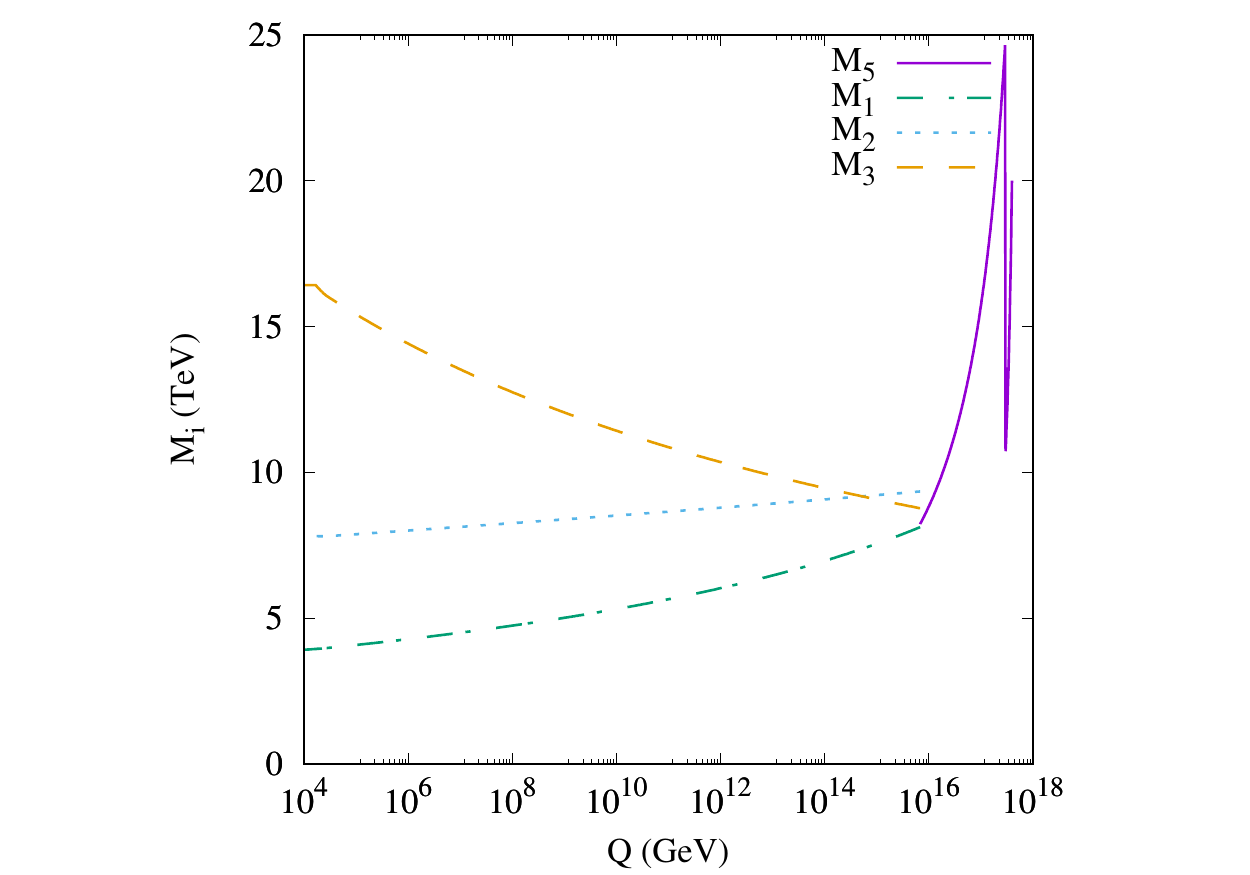}
\hskip -0.75in
\includegraphics[width=9cm]{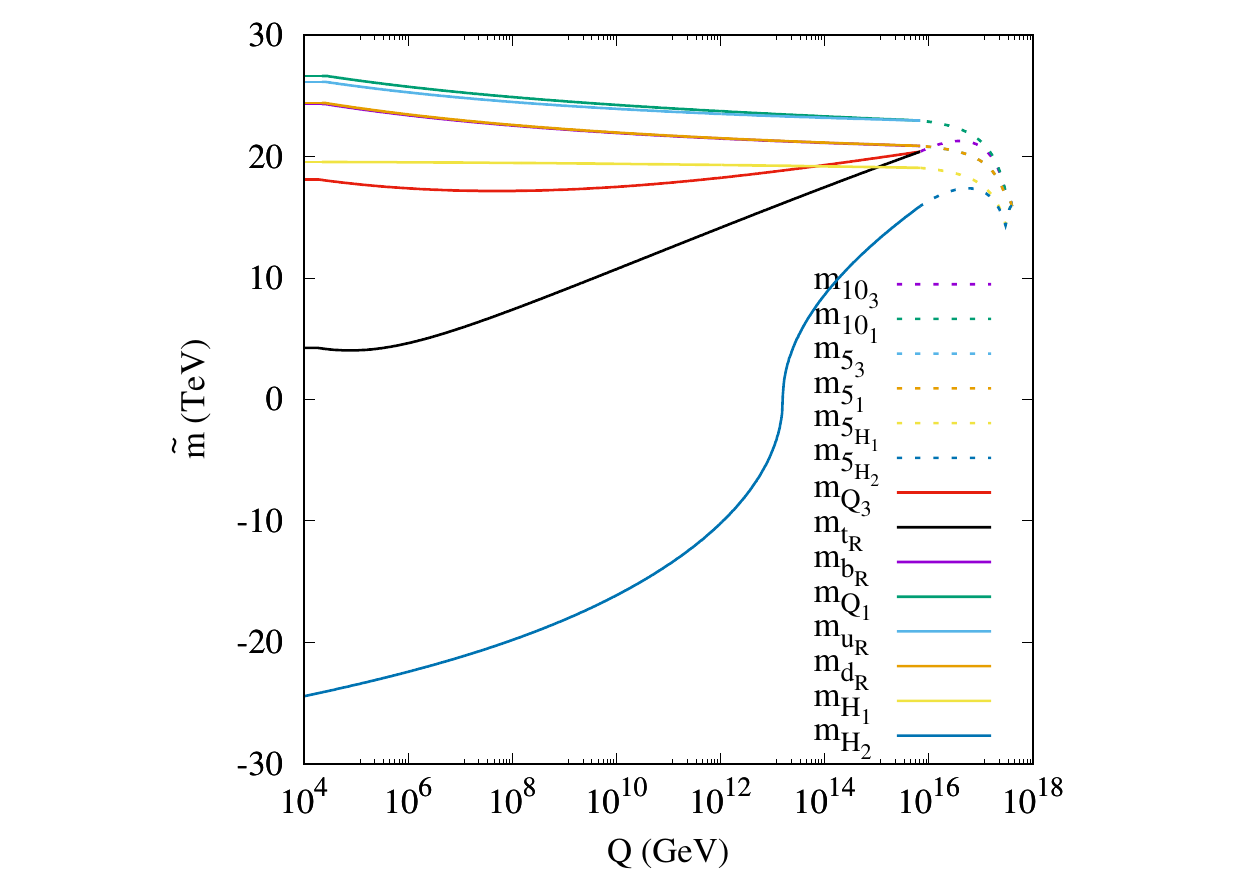}
\caption{\it Left panel: Renormalization effects on the SU(5) gaugino mass $M_5$, exhibiting its sharp decrease as $\mu \to M_X$
from above, and the subsequent evolution of the SM gaugino masses $M_{1,2,3}$ at lower scales, following threshold corrections at
the GUT scale. Right panel: Corresponding evolution of the squark and Higgs masses. Our benchmark point with $m_{1/2} = 20$ TeV is assumed,
with the other parameter values specified in the text.}
\label{fig:gaugino+sfermionunification}
\end{figure}

The renormalization effects on  the gaugino masses for our benchmark point are shown in the left
panel of Fig.~\ref{fig:gaugino+sfermionunification}. Descending from $20$~TeV at the input scale $M_{\rm in} = 4 \times 10^{17}$~GeV, 
we see that $M_5$ initially decreases rapidly to $\sim 10$~TeV at $M_\Theta$, where the large threshold
matching correction shown (\ref{ThetaCorrection}) brings $M_5$ to $\sim 25$~TeV. It then resumes its
decrease until it reaches the GUT scale, where $M_5 \sim 8.2$~TeV.
The SM gaugino masses do not unify at the GUT scale, because of
the non-trivial matching conditions described in Eqs.~(\ref{eq:m1match})-(\ref{eq:m3match}). The right panel of Fig.~\ref{fig:gaugino+sfermionunification}
shows the RGE evolution of sfermion mass parameters $\tilde m \equiv m^2/\sqrt{|m^2|}$, assuming that they are all equal to 15.9~TeV at the same input scale. We see that the physical squark masses are in the range $\sim 18 - 27$~TeV, with
the exception of the right-handed stop squark, whose physical value (after diagonalization) is $\sim 4.2$~TeV. 
This is very similar to the physical value of the $\tilde B$ mass shown 
in the left panel of Fig.~\ref{fig:gaugino+sfermionunification}, as is required along the stop coannihilation
strip that we discuss below. As usual, this reduction in the stop mass is due to renormalization by the top Yukawa 
coupling, which also drives the $H_2$ mass-squared negative at a scale $Q \sim 10^{13}$~GeV, making electroweak
symmetry breaking possible~\footnote{Although the Higgs soft mass goes tachyonic at around $Q\sim 10^{13}$~GeV, the total Higgs scalar mass $m_{H_2}^2+|\mu|^2$ does not go tachyonic until $Q\sim 10^4$~GeV, suggesting the vacuum we examined is indeed stable.}. 
The evolution of the slepton masses is similar to that shown for $H_1$. Representative squark and slepton masses at the weak scale are given in Table~\ref{tab:numbers}. 
In the following phenomenological analysis we study the regions of parameter space
where there is a consistent electroweak vacuum.~\footnote{We do not explore the possible existence of other vacua,
nor the possibility of tunnelling towards them or their possible cosmological implications. However, we do not
expect them to cause problems for the super-GUT CMSSM that we study: see the discussion of the conventional CMSSM in~\cite{EGLOS}.}

\section{Phenomenology of the Super-GUT CMSSM MPM}
\label{sec:superGUT}

The version of the super-GUT CMSSM with the missing-partner mechanism
that we study has the following parameters, in addition to the universal
soft supersymmetry-breaking gaugino mass $m_{1/2}$, scalar mass $m_0$,
trilinear and bilinear couplings $A_0$ and $B_0$, ratio of Higgs vevs $\tan \beta$ and the sign of the
Higgsino mixing parameter $\mu$: the input scale $M_{\rm in}$ at which
universality is assumed, $M_\Theta$~\footnote{Because of the rapid running of
the gauge couplings above $M_\Theta$, this must be close to $M_{\rm in}$.}
and two trilinear superpotential couplings
$\lambda_\Theta$ and $\lambda'$.

\begin{figure}[!ht]
\centering
\includegraphics[width=8.25cm]{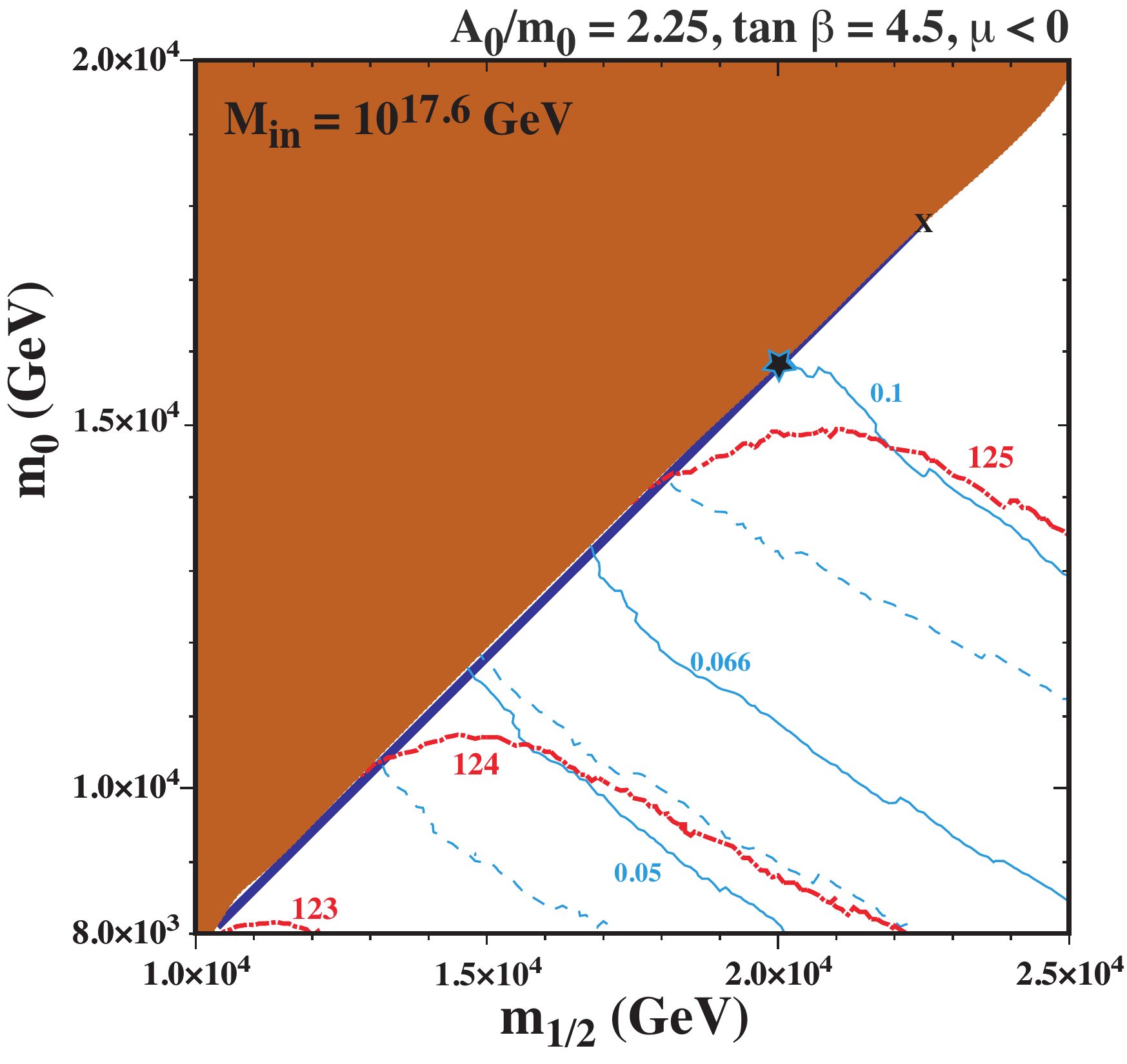}
\includegraphics[width=8cm]{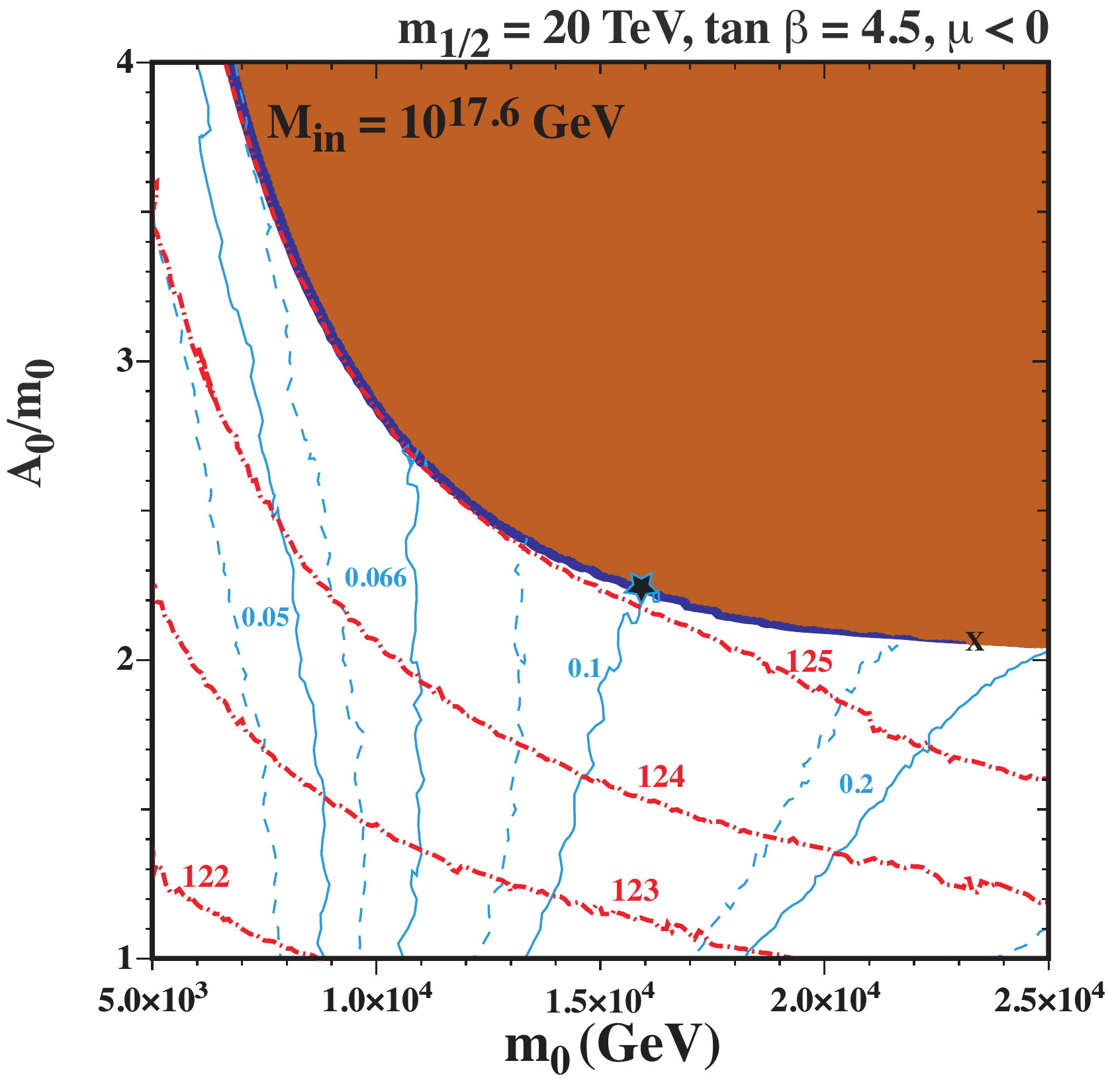}\\
\caption{\it Left panel: The $(m_{1/2}, m_0)$ plane for the parameters $\tan \beta = 4.5, A_0 = 2.25 m_0,
B_0 = 1.25 m_0, M_{\rm in} = 4 \times 10^{17}$~GeV, 
$M_\Theta = 3 \times 10^{17}$~GeV, $\lambda' = 0.005$ and $\lambda_\Theta = 1$ with $\mu < 0$.
Right panel: The $(m_0, A_0/m_0)$ plane with $B_0 = A_0 - m_0$ and unchanged values for the other
model parameters.
Here and subsequently, the brown shaded regions are excluded because
the LSP is the lighter stop, in
the narrow dark blue strips the relic LSP density $\Omega_{\rm LSP} h^2 \in (0.05, 0.2)$,
the dot-dashed red lines are contours of $m_h$ calculated using
{\tt FeynHiggs~2.18.0}, and the solid (dashed) blue lines are contours of $\tau(p \to K^+ \nu)$ 
in units of $10^{35}$~y calculated with central (- 1-$\sigma$) values of the decay matrix elements and $\alpha_s$.}
\label{fig:plane1}
\end{figure}

The left panel of Fig.~\ref{fig:plane1} shows a representative
$(m_{1/2}, m_0)$ plane for the choices $\tan \beta = 4.5, A_0/m_0 = 2.25, B_0 = A_0 - m_0,
\mu < 0, M_{\rm in} = 4 \times 10^{17}$~GeV, 
$M_\Theta = 3 \times 10^{17}$~GeV, $\lambda' = 0.005$ and
$\lambda_\Theta = 1$. In the region at smaller $m_{1/2}$ and larger $m_0$ that is shaded brown
the lighter stop is lighter than the bino, which is cosmologically unacceptable, or tachyonic, and
there is a narrow blue strip just below its boundary where the relic LSP density
falls within a factor $\sim 2$ of the cosmological range \cite{Planck} in the absence of entropy generation, i.e., $\Omega_{\rm LSP} h^2 \in (0.05, 0.2)$.
The red dash-dotted lines are contours of the value of $m_h$ calculated using
{\tt FeynHiggs~2.18.0} \cite{FH}, and the blue lines are contours of
$\tau(p \to K^+ \nu)$ in units of $10^{35}$~y. In order to obtain
conservative bounds, the coefficients of the contributing dimension-5 
operators are calculated 
using the down-quark Yukawa couplings (rather than those of the 
corresponding charged leptons) for the first two generations, 
and choosing values of the GUT phases that minimize the decay rate, as in~\cite{Ellis:2019fwf}. 
The solid lines are for the lifetime calculated using central values
of the relevant hadronic decay matrix elements and $\alpha_s$, 
and the dashed lines are calculated adding in quadrature their estimated 1-$\sigma$ reductions,
following the procedure described in~\cite{Ellis:2019fwf},
corresponding to a longer lifetime for the same values of the model parameters.

The dark matter strip in the left panel of Fig.~\ref{fig:plane1}
terminates at $m_{1/2} \simeq 22.5$~TeV, where the cosmological dark matter density
can no longer be attained in the absence of entropy generation even when the LSP and the lighter stop 
are degenerate, as indicated by the cross. We note also that the experimental lower limit $\tau(p \to K^+ \nu) = 6.6 \times 10^{33}$~y~\cite{Abe:2014mwa}
sets a lower limit
of $m_{1/2} \gtrsim 17 \, (15)$~TeV on the allowed extent of the strip if the central (+~1-$\sigma$)
estimate of the lifetime is used. 
We see that $m_h$ is within  1~GeV of the experimental value $\sim 125$~GeV along
all the allowed dark matter strip in the left panel of Fig.~\ref{fig:plane1}.
The star indicates the benchmark point that we have chosen for more detailed analysis, which has $m_0 = 15.9$~TeV. The parameter inputs and resulting GUT scale and MSSM masses as well as values for $\Omega_\chi h^2, m_h$, and $\tau_p$ are given in Table \ref{tab:numbers}. 

We now discuss how the phenomenological features of the MPM change as 
the model parameters are varied, starting with the sensitivity to $A_0$. 
The right panel of
Fig.~\ref{fig:plane1} shows the $(m_0, A_0/m_0)$ plane for the same
values of $\tan \beta, M_{\rm in}, 
M_\Theta, \lambda'$ and $\lambda_\Theta$, with 
$B_0 = A_0 - m_0$ and $\mu < 0$. We see again at large $m_0$ and $A_0/m_0$
(shaded brown) the region that is excluded because the LSP is the lighter stop and, just below it,
a stop coannihilation strip along which $m_0$ increases as $A_0/m_0$ decreases. 
We see that $\tau(p \to K^+ \nu)$ increases with $m_0$, and the experimental limit
$\tau(p \to K^+ \nu) > 6.6 \times 10^{33}$~y is satisfied for $m_0 \gtrsim 11$~TeV and $A_0/m_0 \lesssim 2.7$, or $m_0 \gtrsim 7$~TeV and $A_0/m_0 \lesssim 3.8$ when the 1-$\sigma$ uncertainty in $\tau_p$ is taken into account.
We also see that $m_h \sim 125$~GeV along all the displayed part of the strip.
The star again represents our chosen benchmark point with $(m_0, A_0/m_0) = (15.9~ {\rm TeV}, 2.25)$, and
the cross marks the tip of this strip, which is at $m_0 \simeq 23.2$~TeV and $A_0/m_0 \simeq 2.05$, corresponding to
central values of $\tau(p \to K^+ \nu) \sim 1.7 \times 10^{34}$~y and $m_h \sim 125.6$~GeV.
Our benchmark point lies
midway along the allowed part of the dark matter strip, corresponding to a representative choice within the
narrow allowed range of $A_0/m_0$.

\begin{figure}[!ht]
\centering
\includegraphics[width=8cm]{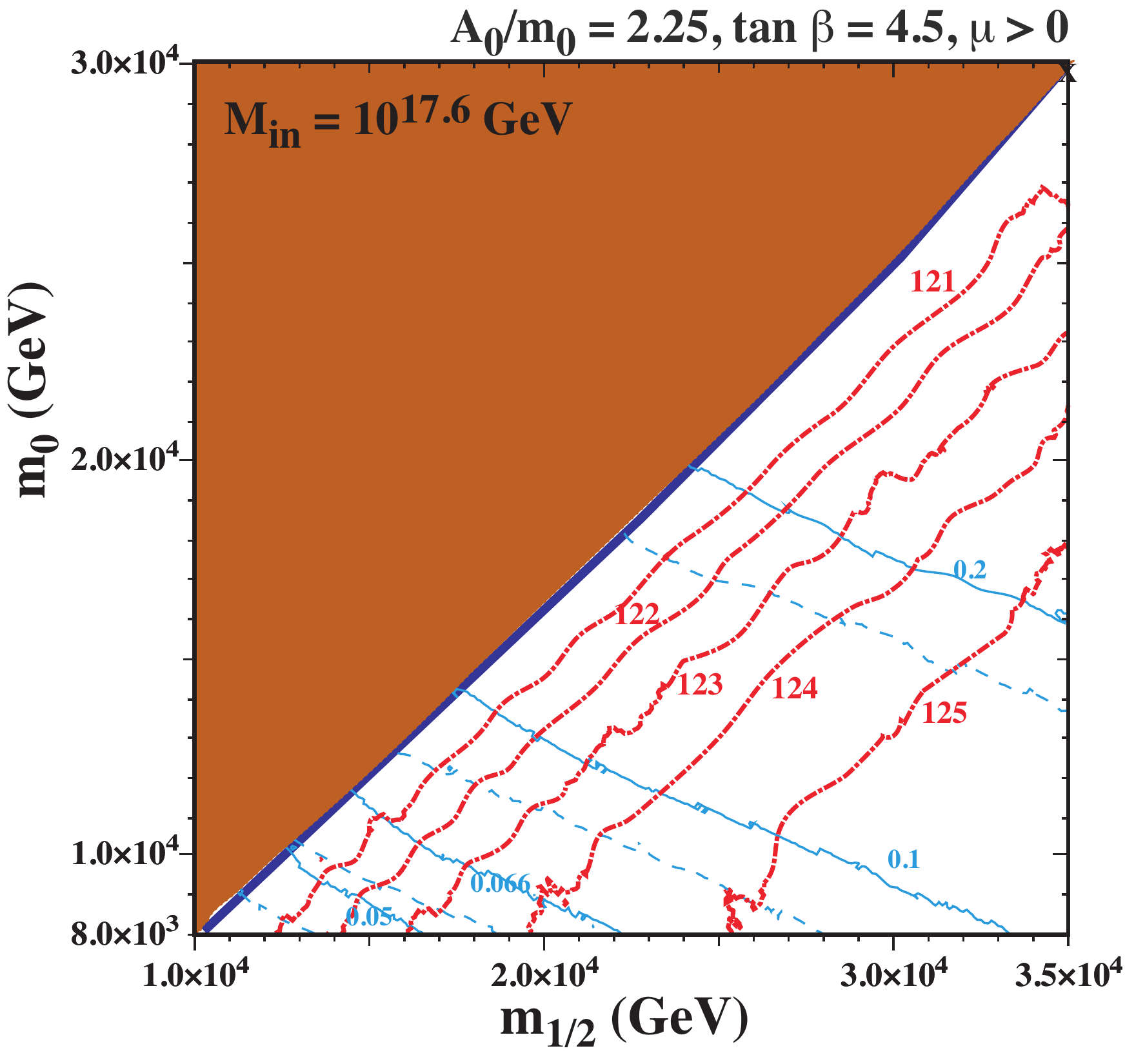}
\includegraphics[width=8cm]{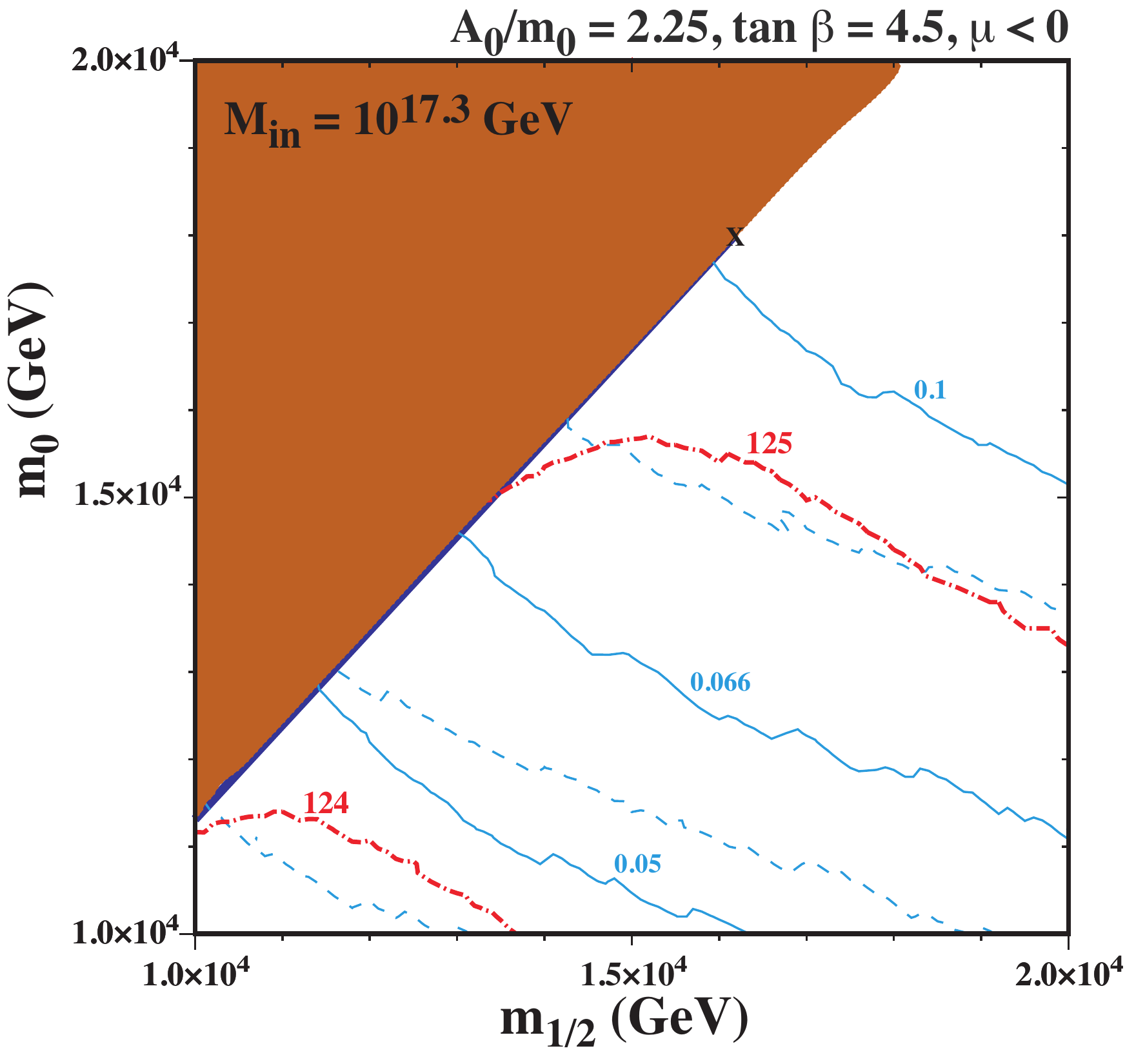}\\
\vspace{0.25in}
\includegraphics[width=8cm]{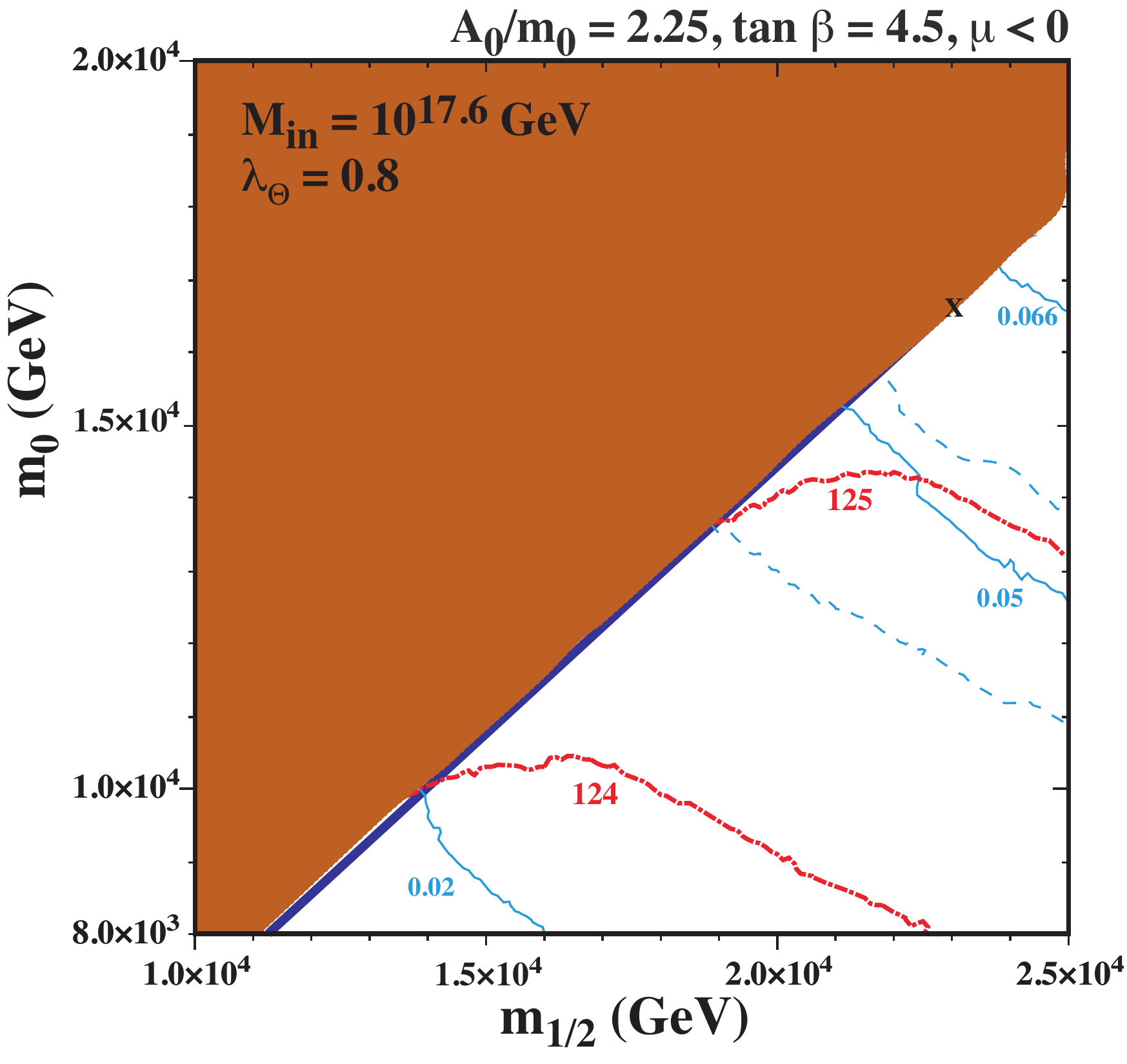}
\includegraphics[width=8cm]{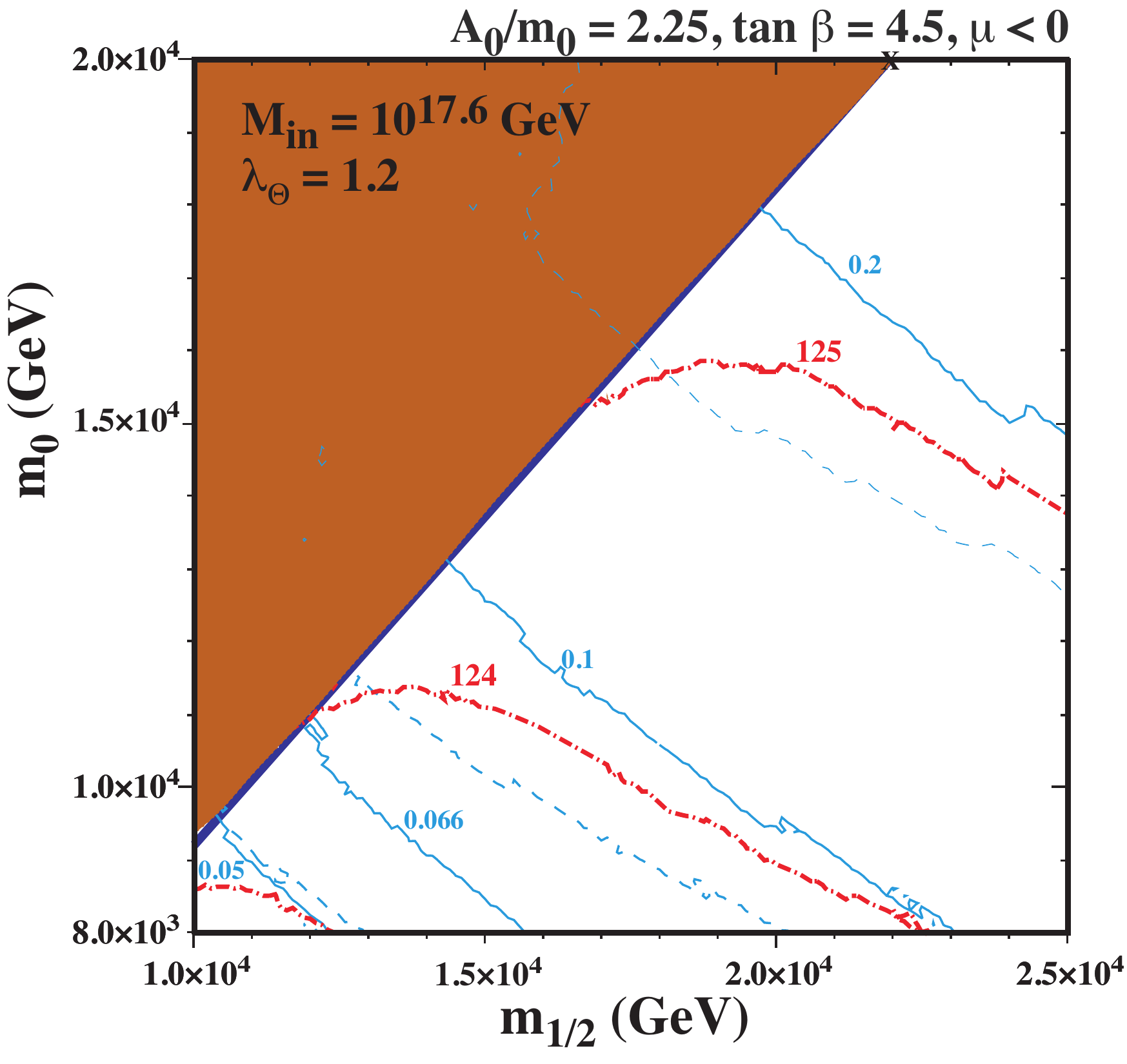}
\caption{\it Upper left panel: The $(m_{1/2}, m_0)$ plane for the parameters $\tan \beta = 4.5, A_0 = 2.25 m_0,
B_0 = 1.25 m_0, M_{\rm in} = 4 \times 10^{17}$~GeV, 
$M_\Theta = 3 \times 10^{17}$~GeV, $\lambda' = 0.005$ and $\lambda_\Theta = 1$ with $\mu > 0$. Upper right panel:
The corresponding plane with $M_{\rm in} = 2 \times 10^{17}$~GeV, $\mu < 0$ and unchanged values of the other
parameters. Lower panels: The corresponding planes with $\lambda_\Theta = 0.8$ (left) and $\lambda_\Theta = 1.2$ (right),
with $\mu < 0$, and $M_{\rm in} = 4 \times 10^{17}$~GeV and unchanged values of the other parameters. In each panel, the cross marks the endpoint of the stop coannihilation strip.}
\label{fig:plane2}
\end{figure}

Fig.~\ref{fig:plane2} illustrates the effects of changing some other MPM parameters.
In the upper left panel we see the $(m_{1/2}, m_0)$ plane with $\mu > 0$ and the 
same values of the other parameters as in Fig.~\ref{fig:plane1}, namely $\tan \beta = 4.5, A_0 = 2.25 m_0,
B_0 = 1.25 m_0, M_{\rm in} = 4 \times 10^{17}$~GeV, 
$M_\Theta = 3 \times 10^{17}$~GeV, $\lambda' = 0.005$ and $\lambda_\Theta = 1$.
We see that $m_h < 121$~GeV along all the dark matter strip. It is a general feature
of the $\mu > 0$ planes we have studied that $m_h$ is too small, whereas generic
planes with $\mu < 0$ have acceptable values of $m_h$. For this reason, we do not
discuss further any cases with $\mu > 0$.
In the upper right panel of Fig.~\ref{fig:plane2} we choose $M_{\rm in} = 2 \times 10^{17}$~GeV, i.e., below the value of $M_\Theta$, with the other parameters unchanged from those in Fig.~\ref{fig:plane1}. In this case, there is little effect of the large $\beta$-function coefficient ascribed to the {\bf 50}. We see that $\tau(p \to K^+ \nu)$ is slightly reduced at the tip of the coannihilation strip. 
In the lower left panel of Fig.~\ref{fig:plane2}, $\lambda_\Theta = 0.8$, with the other 
parameters unchanged from those in Fig.~\ref{fig:plane1}. There is a more significant reduction in 
$\tau(p \to K^+ \nu)$ at the tip of the coannihilation strip, as it falls below the
nominal experimental limit. We conclude that there is limited
scope for decreasing either $M_{\rm in}$ or $\lambda_\Theta$ below the benchmark values used
in Fig.~\ref{fig:plane1}. On the other hand, we see in the lower right panel of
Fig.~\ref{fig:plane1} that $\tau(p \to K^+ \nu)$ is increased for $\lambda_\Theta = 1.2$
and unchanged values of the other parameters. The sensitivity to $\lambda_\Theta$ is due
mainly to its effect on the colored Higgs mass: see Eq.~(\ref{eq:ColorHiggsMass}).

\begin{figure}[!ht]
\centering
\includegraphics[width=8.2cm]{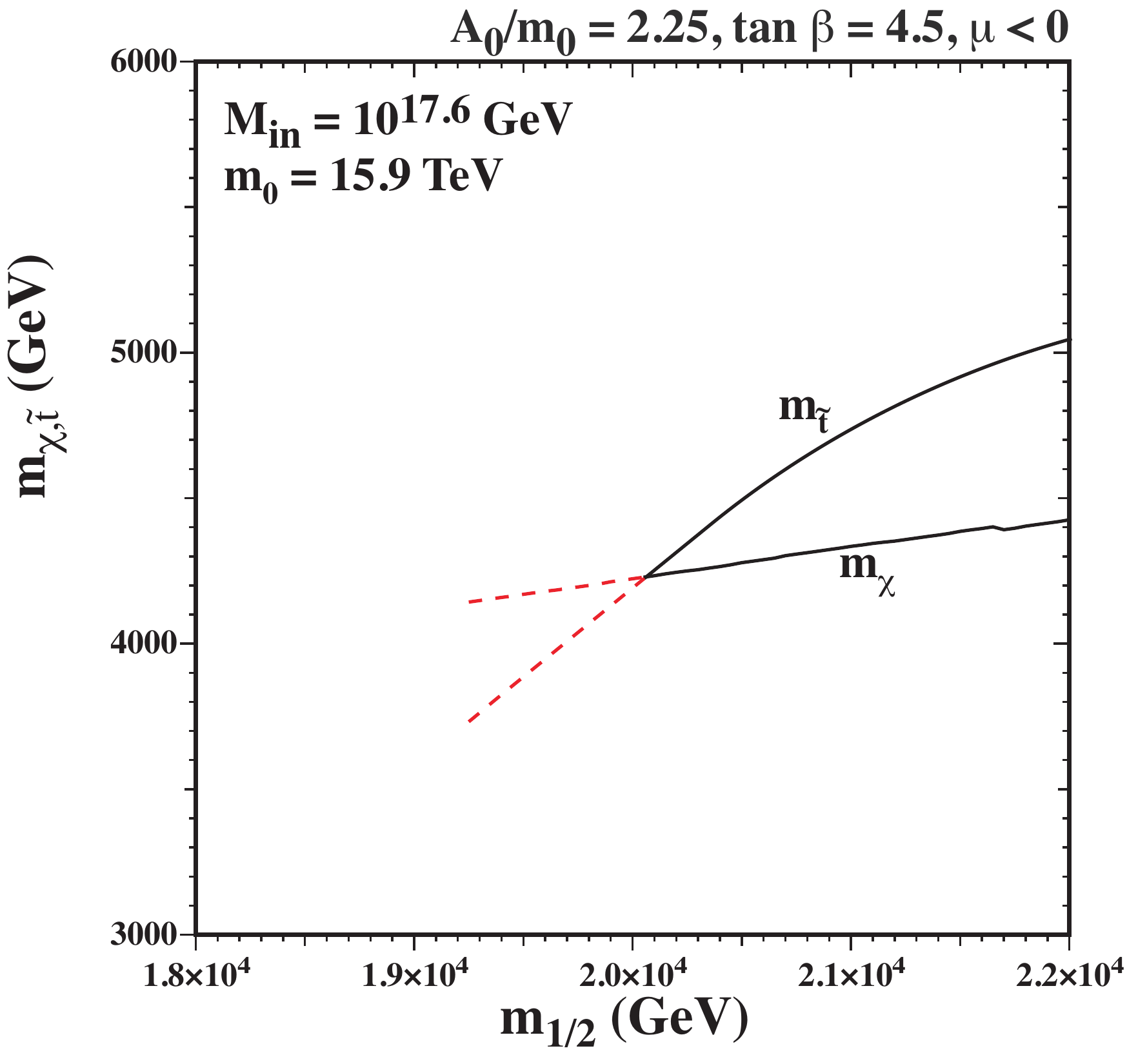}
\caption{\it The values of $m_\chi$ and $m_{\tilde t_1}$ 
(solid black lines in the bino LSP region, dashed red lines in the stop LSP region)
as functions of $m_{1/2}$
for the parameters $\tan \beta = 4.5, M_{\rm in} = 4 \times 10^{17}$~GeV, 
$M_\Theta = 3 \times 10^{17}$~GeV, $\lambda' = 0.005$ and $\lambda_\Theta = 1$, with 
$m_0$ chosen to obtain the cosmological value of $\Omega_{\rm LSP} h^2$ for $m_{1/2} = 20$~TeV.
}
\label{fig:varym12}
\end{figure}

As an aid to visualising the context of the stop coannihilation strip in more
detail, we show in Fig.~\ref{fig:varym12} a section across the dark matter strip and into
the stop LSP region for the parameters $\tan \beta = 4.5, M_{\rm in} = 4 \times 10^{17}$~GeV, 
$M_\Theta = 3 \times 10^{17}$~GeV, $\lambda' = 0.005$ and $\lambda_\Theta = 1$, with 
$m_0 = 15.9$~TeV chosen so as to obtain the cosmological value of $\Omega_{\rm LSP} h^2$ for $m_{1/2} = 20$~TeV.
We see how the role of the LSP is exchanged between the bino and the stop, 
as indicated by the switch from solid black lines to dashed red lines. The
dark matter strip is located just inside the bino LSP region, where $m_{\tilde t_1} - m_\chi \simeq 12$~GeV
for the parameter choices displayed, which is too close to the mass crossover point to be distinguishable
in this plot. The dark matter density increases rapidly for larger values of $m_{1/2}$,
and we display in the following figures the sensitivity of $\Omega_{\rm LSP} h^2$ to other model parameters.

\begin{figure}[!ht]
\centering
\includegraphics[width=5.3cm]{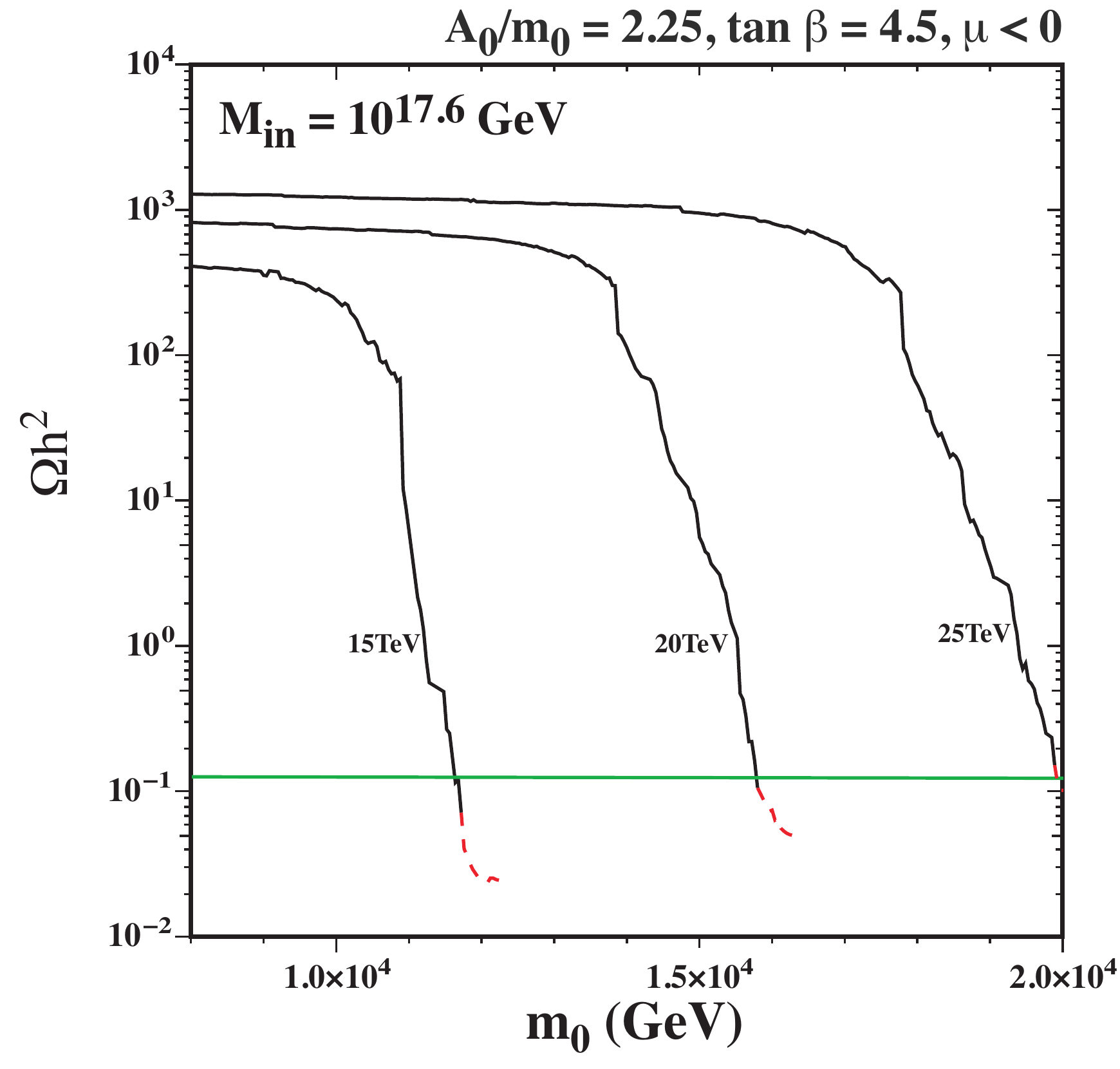}
\includegraphics[width=5.3cm]{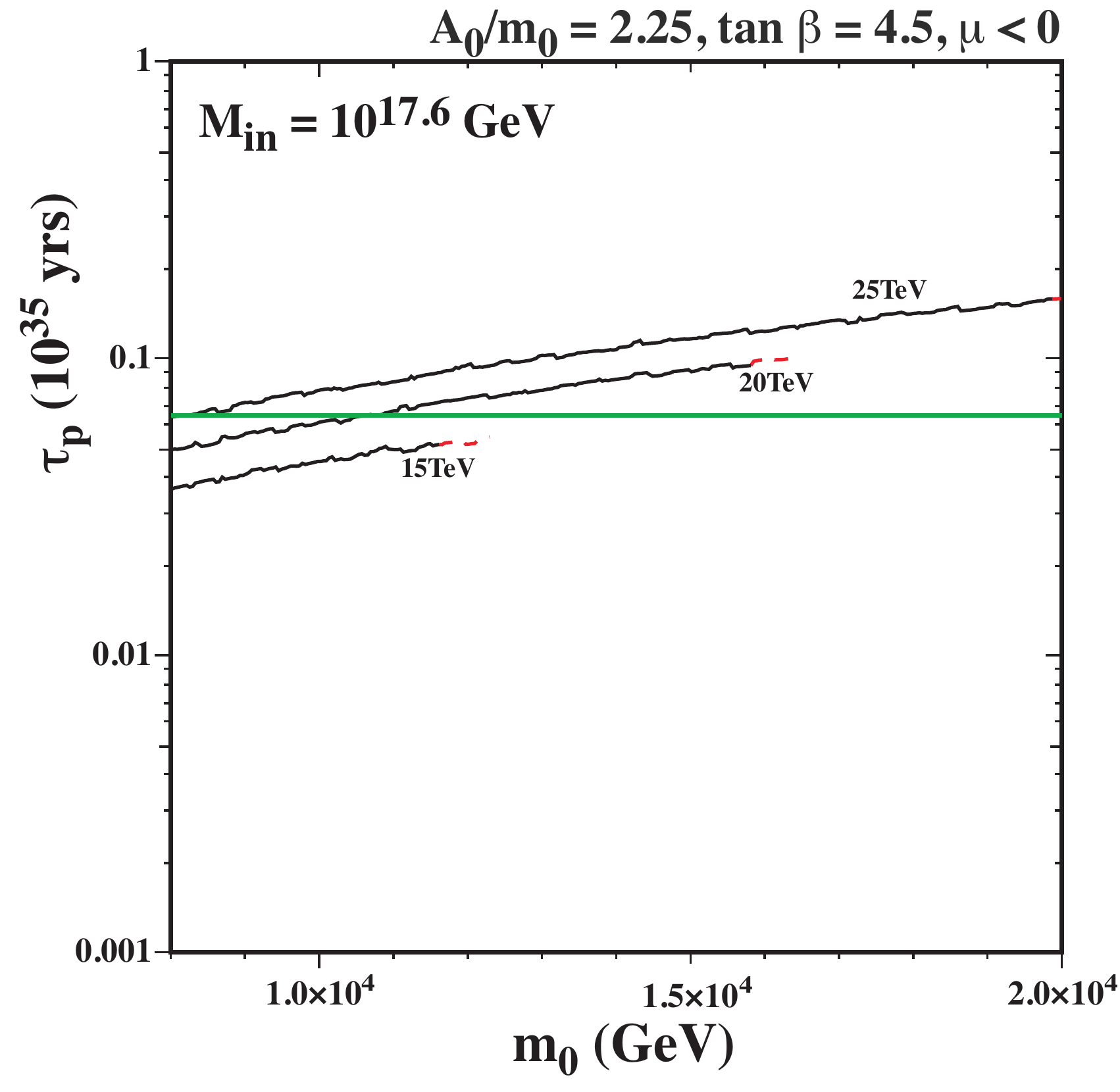}
\includegraphics[width=5.3cm]{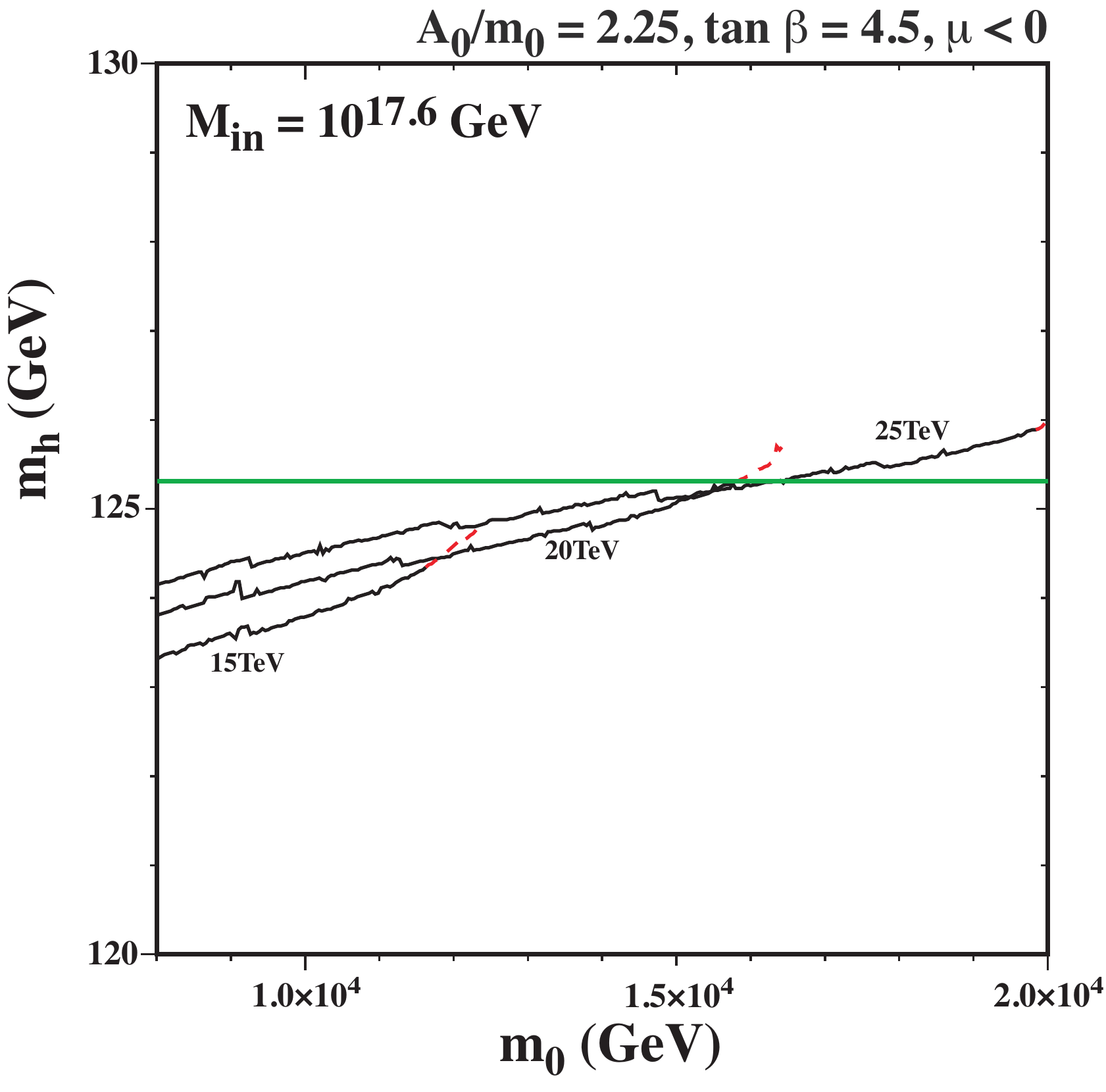}
\caption{\it Values of $\Omega_{\rm LSP} h^2$ (left panel), $\tau(p \to K^+ \nu)$ (center panel) and $m_h$ (right panel) as functions of $m_0$ for
the indicated fixed values of $m_{1/2} \in [15, 25]$~TeV and for the same input parameters as in the upper left panel
of Fig.~\ref{fig:plane1}. Values calculated when the LSP is a bino are shown as solid black lines, becoming dashed red lines when the LSP is
the lighter stop and terminating when the RGE calculations
break down. The horizontal green lines correspond to the cosmological value of $\Omega_{\rm LSP} h^2$, the experimental lower
limit on $\tau(p \to K^+ \nu)$ and the experimental value of $m_h$, respectively.}
\label{fig:stripprofile}
\end{figure}

The effects of varying $m_0$ for the indicated fixed values of $m_{1/2}$, 
for the same input parameters as in the upper left panel
of Fig.~\ref{fig:plane1}, are exhibited in Fig.~\ref{fig:stripprofile}. The values of
$\Omega_{\rm LSP} h^2$ (left panel), $\tau(p \to K^+ \nu)$ (middle panel) and $m_h$ (right panel) are plotted as functions of $m_0$. These quantities are plotted as solid black lines as long as the LSP is a bino,
changing to a dashed red line when the LSP is the lighter stop, and terminating when the RGE calculations
break down. We see in the left panel that $\Omega_{\rm LSP} h^2 \gtrsim 100$ over a large range of $m_0$,
before dropping precipitously as the LSP and the lighter stop become more degenerate and coannihilation kicks in
bringing $\Omega_{\rm LSP} h^2$ into the allowed range (indicated by the horizontal green line) at a value of
$m_0$ that increases with $m_{1/2}$. We find that $\Omega_{\rm LSP} h^2$ never falls into the allowed 
cosmological range for $m_{1/2} \gtrsim 22.5$~TeV.~\footnote{Close examination of the $m_{1/2} = 25$ TeV curve shows that the range with a bino LSP terminates before $\Omega_{\rm LSP} h^2$ drops into the measured range.} We see in the middle panel how $\tau(p \to K^+ \nu)$
varies with $m_0$ for the chosen values of $m_{1/2}$. The proton lifetime increases monotonically with $m_{1/2}$ for any fixed value of $m_0$, 
and also with $m_0$ for fixed $m_{1/2}$. It remains above the experimental lower limit
$\tau(p \to K^+ \nu) = 6.6 \times 10^{33}$~y (indicated by the horizontal green line)
for all the exhibited range of $m_0$ for $m_{1/2} = 25$~TeV, but for $m_{1/2} = 20$~TeV
the range over which it reaches the experimental limit is limited. 
When $m_{1/2} = 20$~TeV we find $m_{\rm LSP} \simeq m_{\tilde t_1} \simeq 4.2$~TeV,
and $m_{\tilde g} \sim 18$~TeV, whereas the light-flavour
squarks have masses ${\cal O}(25)$~TeV. The large negative renormalization of
$m_{\tilde g}$ arises from the large group-theoretical factors associated with the large representations appearing
in the MPM. 
Note that these curves represent central values of the computed values of $\tau_p$
and the lifetime could be increased by $\sim 20 \%$ when uncertainties in the hadronic matrix elements and $\alpha_s$ are included.  Finally, the right panel of Fig.~\ref{fig:stripprofile} shows how $m_h$ calculated using
{\tt FeynHiggs~2.18.0} \cite{FH} varies as a function of $m_0$
for the indicated fixed values of $m_{1/2}$ and the same input parameters as in the upper left panel
of Fig.~\ref{fig:plane1}. We see that $m_h \in (123, 126)$~GeV over all the ranges of $m_0$ displayed, which is
quite consistent with the experimental value (indicated by the horizontal green line) in view of the theoretical uncertainties in the calculation.

\begin{figure}[!ht]
\centering
\includegraphics[width=5.3cm]{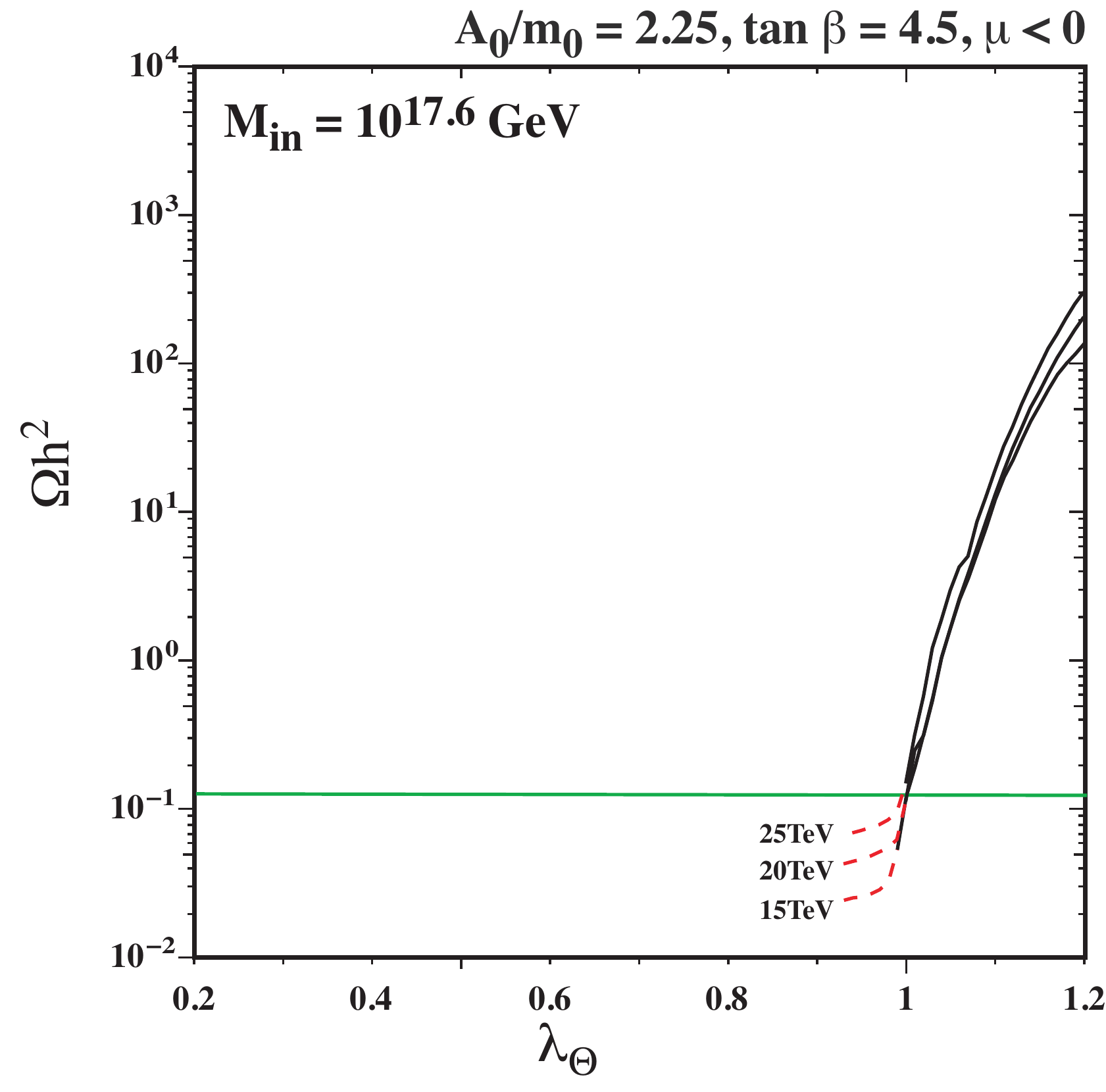}
\includegraphics[width=5.3cm]{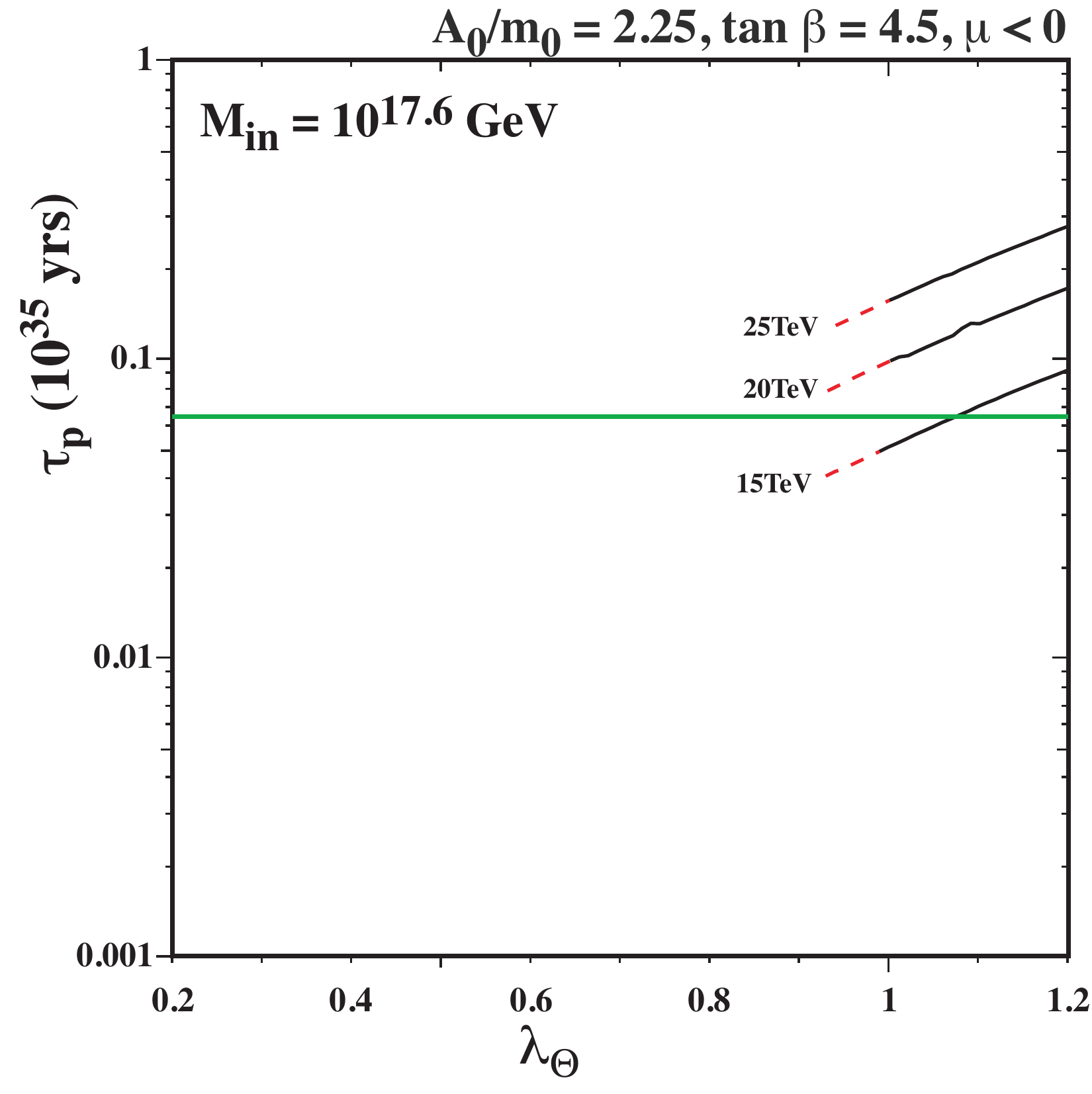}
\includegraphics[width=5.3cm]{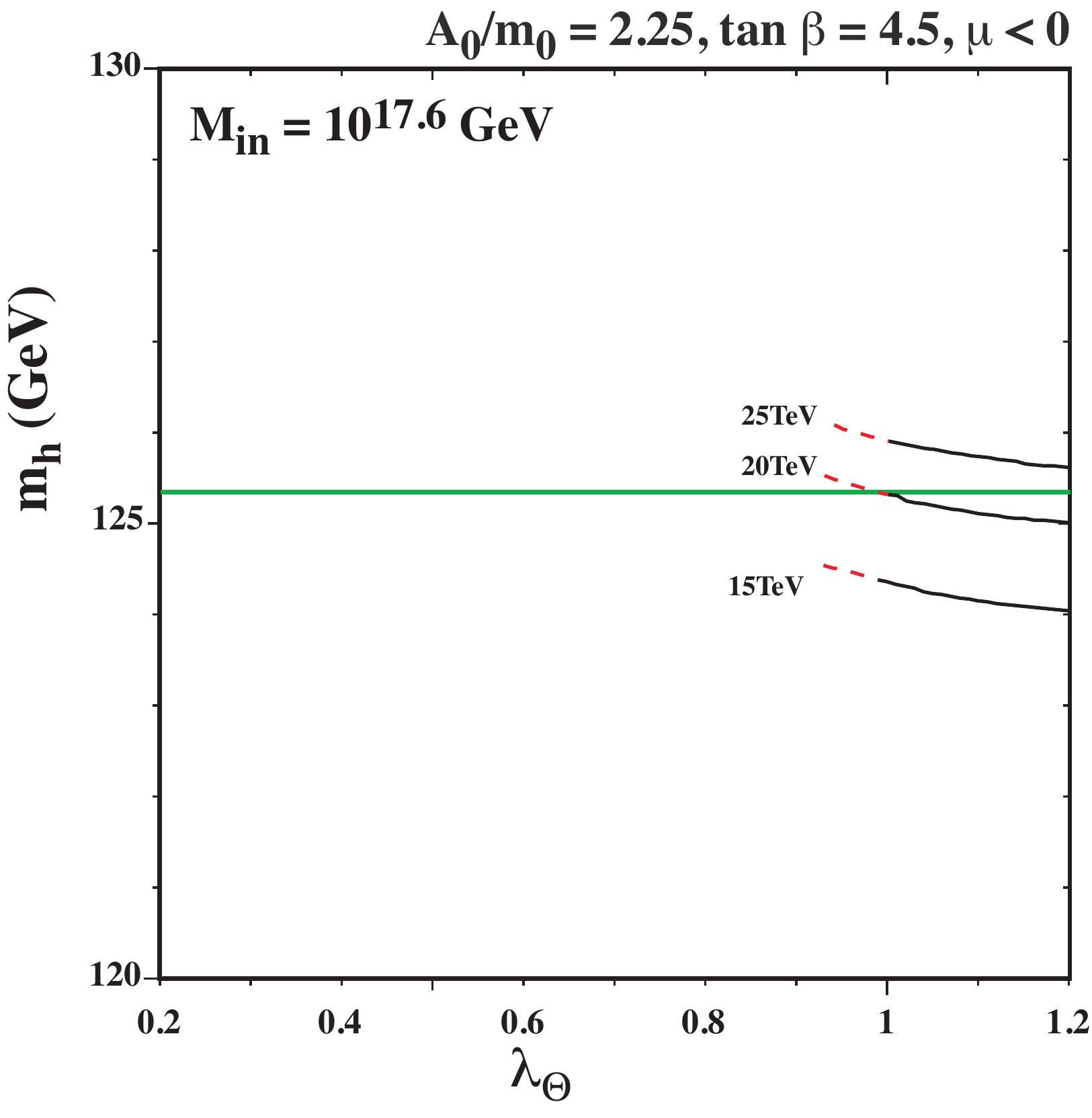}
\caption{\it As in Fig.~\ref{fig:stripprofile}, as functions of $\lambda_\Theta$ for
the indicated fixed values of $m_{1/2} \in [15, 25]$~TeV and $m_0$ chosen to obtain the cosmological value of
$\Omega_{\rm LSP} h^2$ for $\lambda_\Theta = 1$, with the same values of the other input parameters as in Fig.~\ref{fig:plane1}.}
\label{fig:stripprofile2}
\end{figure}

Fig.~\ref{fig:stripprofile2} displays $\Omega_{\rm LSP} h^2$ (left panel), $\tau(p \to K^+ \nu)$ (middle panel)
and $m_h$ (right panel) as functions of $\lambda_\Theta = \lambda_{\bar \Theta}$ for
the indicated fixed values of $m_{1/2} \in [15, 25]$~TeV and $m_0$ chosen so as to obtain the cosmological value of
$\Omega_{\rm LSP} h^2$ for $\lambda_\Theta = 1$, with the same values of the other input parameters as in the upper left panel of Fig.~\ref{fig:plane1}.
Specifically, we take $m_0 = 11.6, 15.9$ and 19.9 TeV for $m_{1/2} = 15, 20$, and 25 TeV. In the latter case, the relic density is high, $\Omega_{\rm LSP}  h^2 = 0.15$, when $m_\chi  = m_{\tilde t_1}$, i.e., it is past the stop coannihilation endpoint. 
We see that $\Omega_{\rm LSP} h^2$ is very similar for all the chosen values of $m_{1/2}$.
On the other hand, as seen in the middle panel of Fig.~\ref{fig:stripprofile2},
$\tau(p \to K^+ \nu)$ decreases when the value of $\lambda_\Theta$ is reduced. For this reason,
the allowed range of the stop coannihilation strip disappears when $\lambda_\Theta \lesssim 1$ if the other model
parameters are unchanged from those in the upper left panel of Fig.~\ref{fig:plane1} and in Fig.~\ref{fig:stripprofile}. For $\lambda_\Theta \lesssim 0.9$,
RGE running breaks down for the this parameter set, though lower values of $\lambda_\Theta$ are possible for other choices of $m_{1/2}$ and $m_0$, as seen in the lower left panel of Fig.~\ref{fig:plane2}. Moreover, 
$\tau(p \to K^+ \nu)$ is below the nominal experimental limit at the point of stop-coannihilation (when the bino and stop masses near equality, which occurs when the black solid curves become red dashed)
when $m_{1/2} = 15$~TeV, whereas for $m_{1/2} = 20$ and $25$~TeV the lifetime is above the experimental limit for all values of $\lambda_\Theta$ displayed. 
Once again, the right panel of Fig.~\ref{fig:stripprofile2} shows that $m_h$ is within 2~GeV of
the experimental value for all values of $\lambda_\Theta$ for the three indicated choices of $m_{1/2}$.

\begin{figure}[!ht]
\centering
\includegraphics[width=5.3cm]{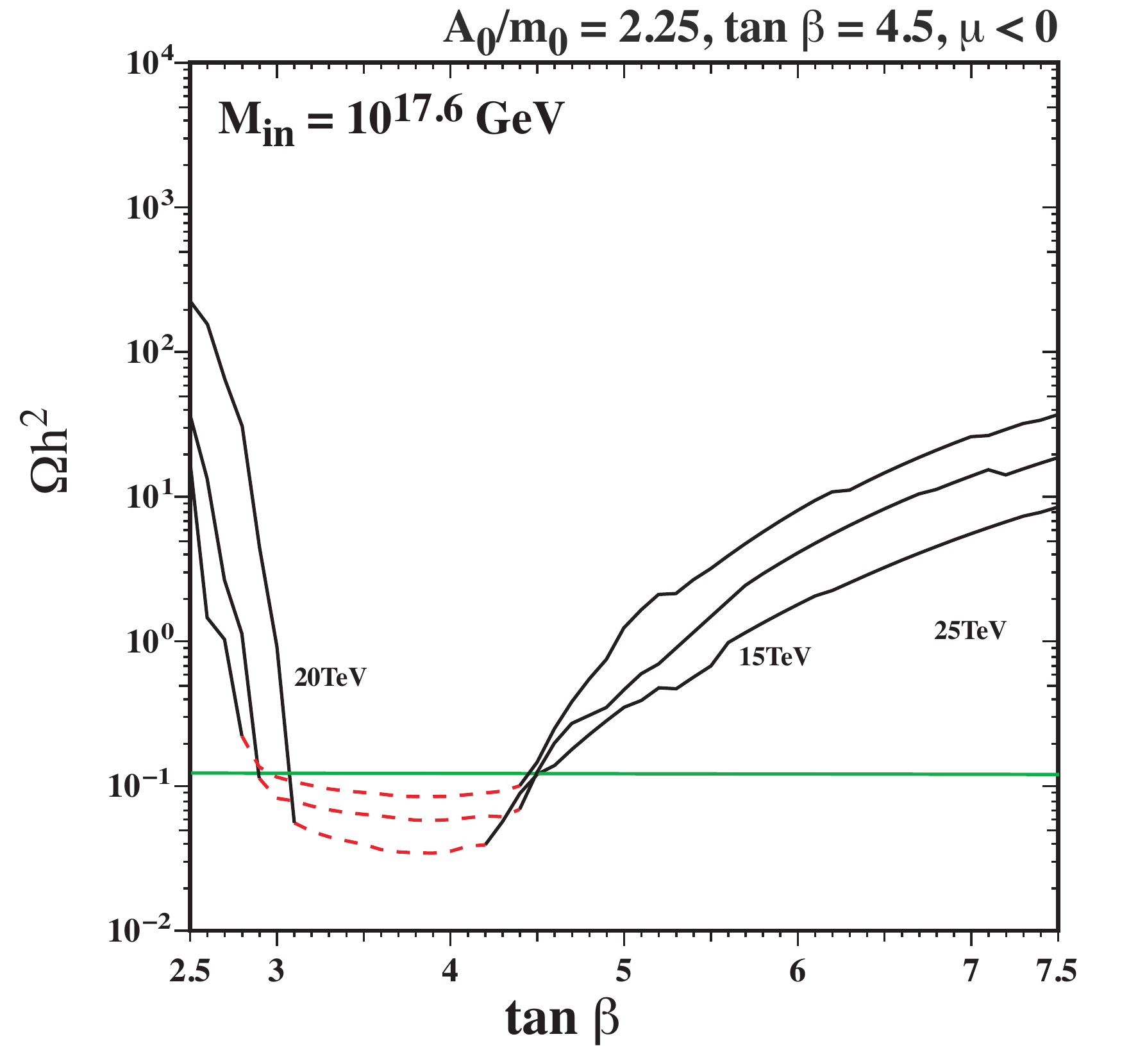}
\includegraphics[width=5.3cm]{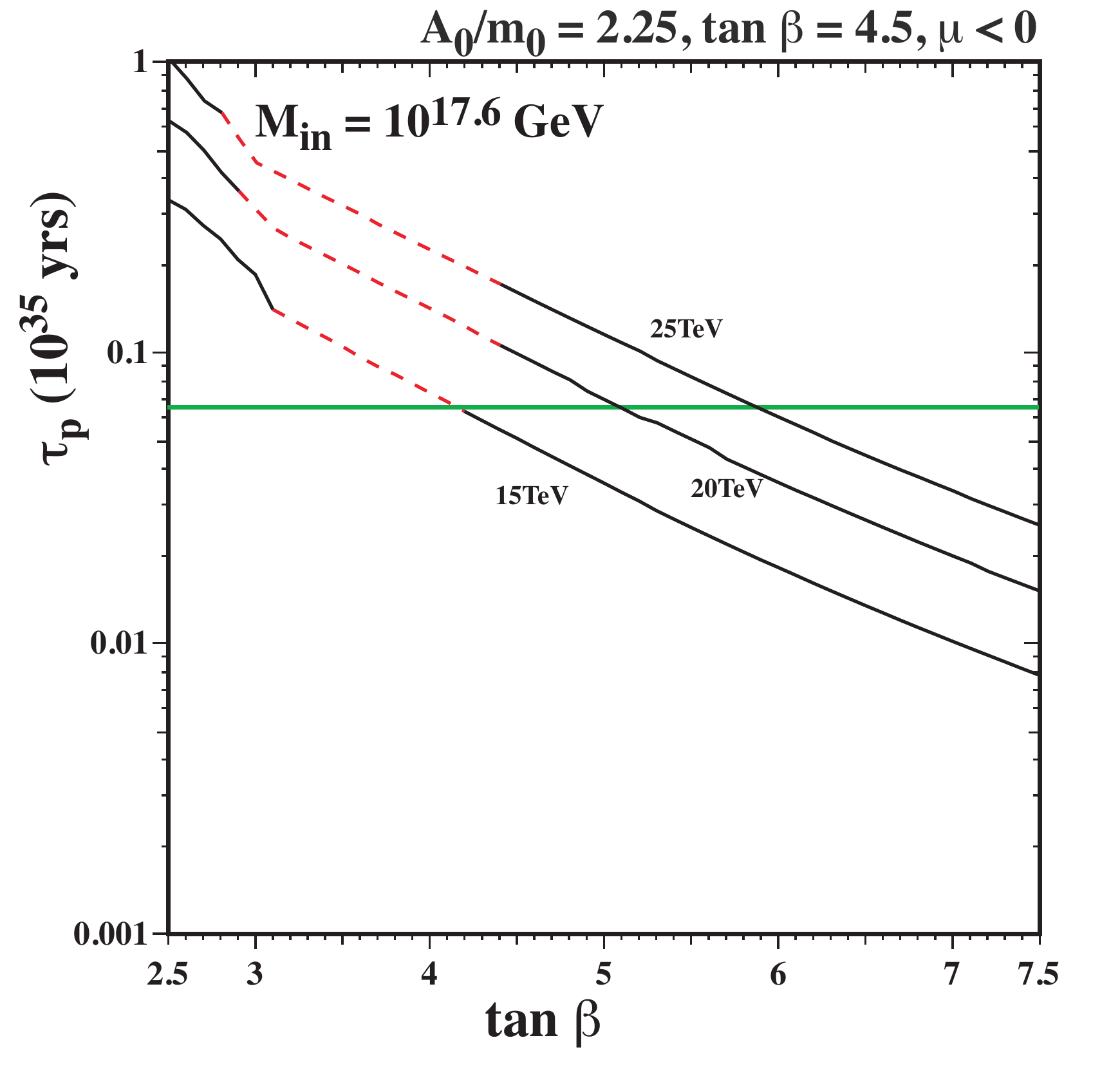}
\includegraphics[width=5.3cm]{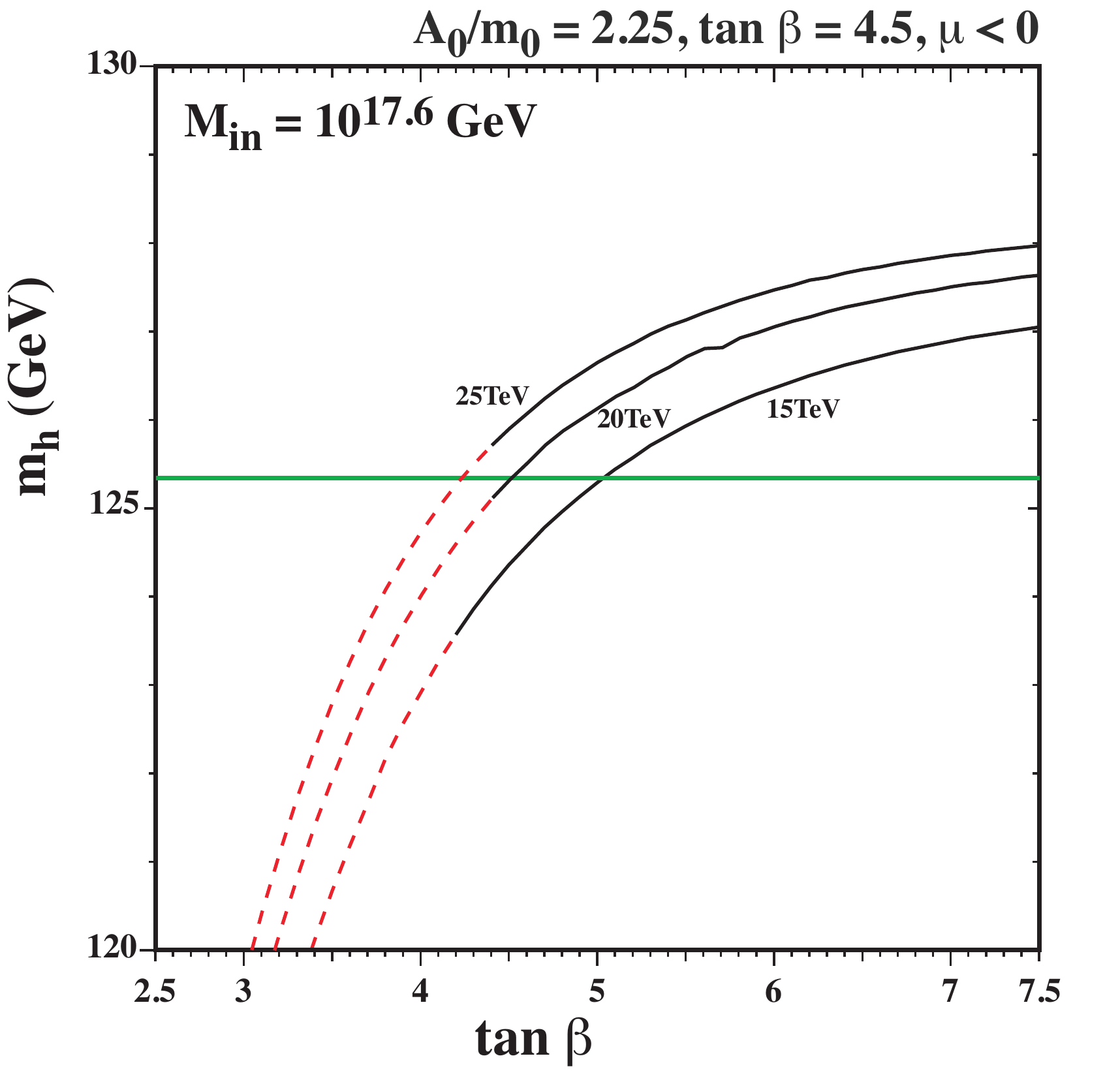}
\caption{\it As in Fig.~\ref{fig:stripprofile}, as functions of $\tan \beta$ for
the indicated fixed values of $m_{1/2} \in [15, 25]$~TeV and $m_0$ chosen to obtain the cosmological value of
$\Omega_{\rm LSP} h^2$ for $\tan \beta = 4.5$ with the same values of the other input parameters as in Fig.~\ref{fig:plane1}.}
\label{fig:stripprofile3}
\end{figure}

We study in Fig.~\ref{fig:stripprofile3} the sensitivity of the predictions to $\tan \beta$. We see in
the left panel that when $m_0$ is chosen (as in Fig.~\ref{fig:stripprofile2}) so as to obtain the cosmological value of
$\Omega_{\rm LSP} h^2$ for $\tan \beta = 4.5$, the range $m_{1/2} \in [15, 25]$~TeV
is consistent with the cosmological value of $\Omega_{\rm LSP} h^2$ only for limited ranges of $\tan \beta \sim 3$ and $\sim 4.5$.
Values of $\tan \beta$ in between are largely excluded because the LSP is the lighter stop.
The middle panel shows how $\tau(p \to K^+ \nu)$ decreases with $\tan \beta$, disfavouring values $\gtrsim 5$.
On the other hand, the right panel shows that the value of $m_h$ decreases with $\tan \beta$ and $m_{1/2}$,
becoming incompatible with the experimental value when $\tan \beta \lesssim 3.5$, where the LSP is the lighter stop.
We conclude that a range of $\tan \beta \sim 4.5$ is favoured.

\begin{figure}[!ht]
\centering
\includegraphics[width=5.3cm]{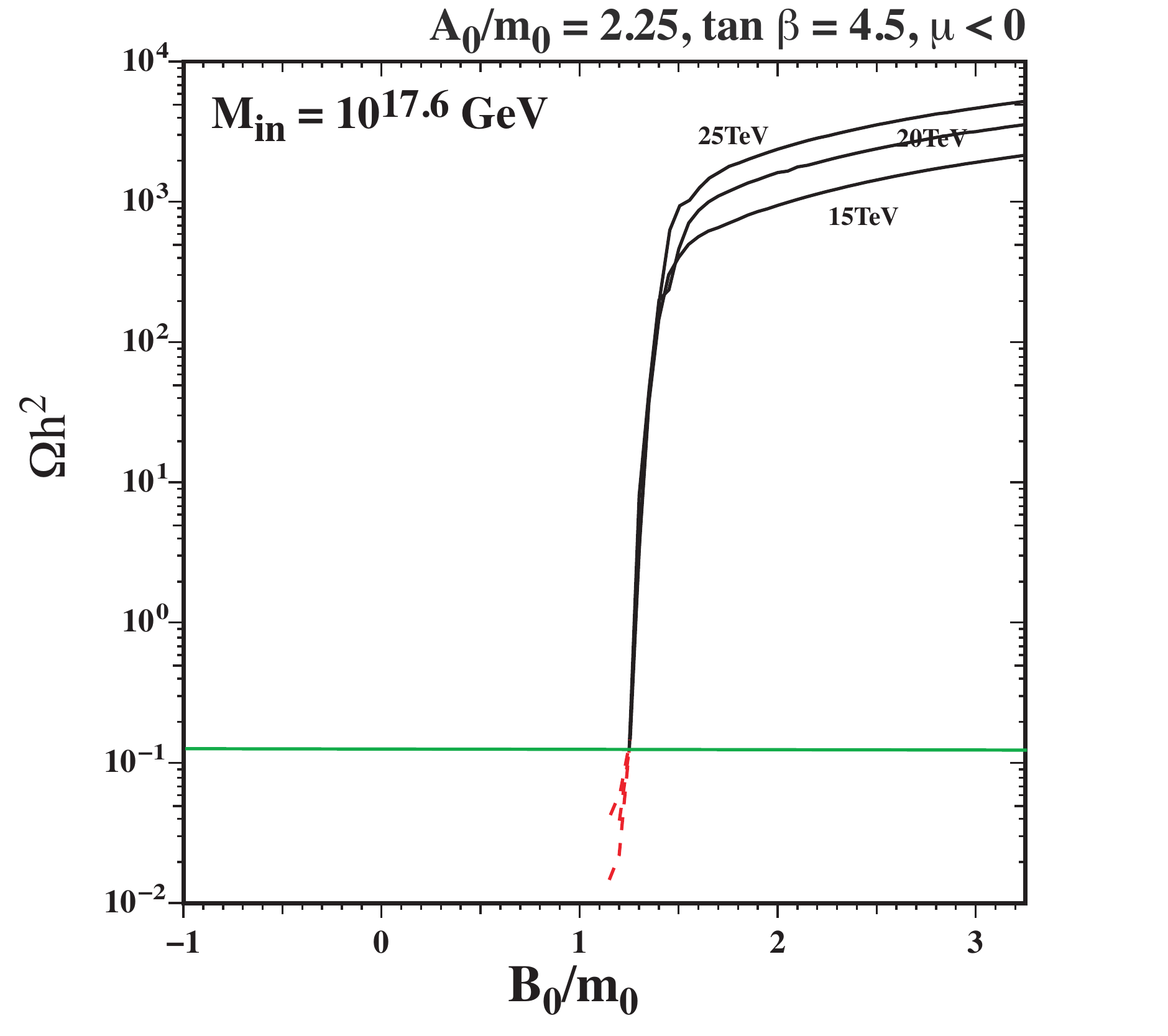}
\includegraphics[width=5.3cm]{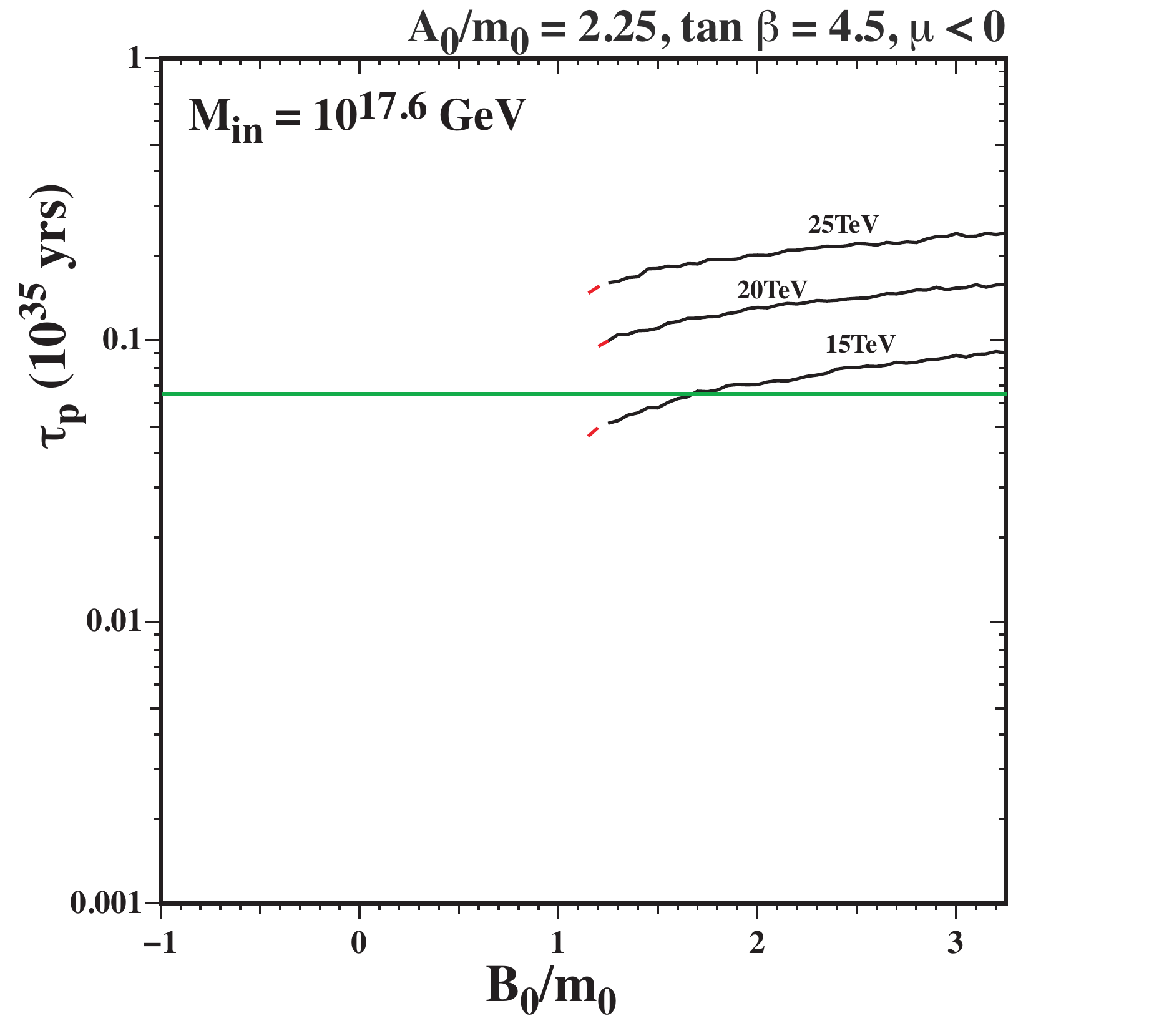}
\includegraphics[width=5.3cm]{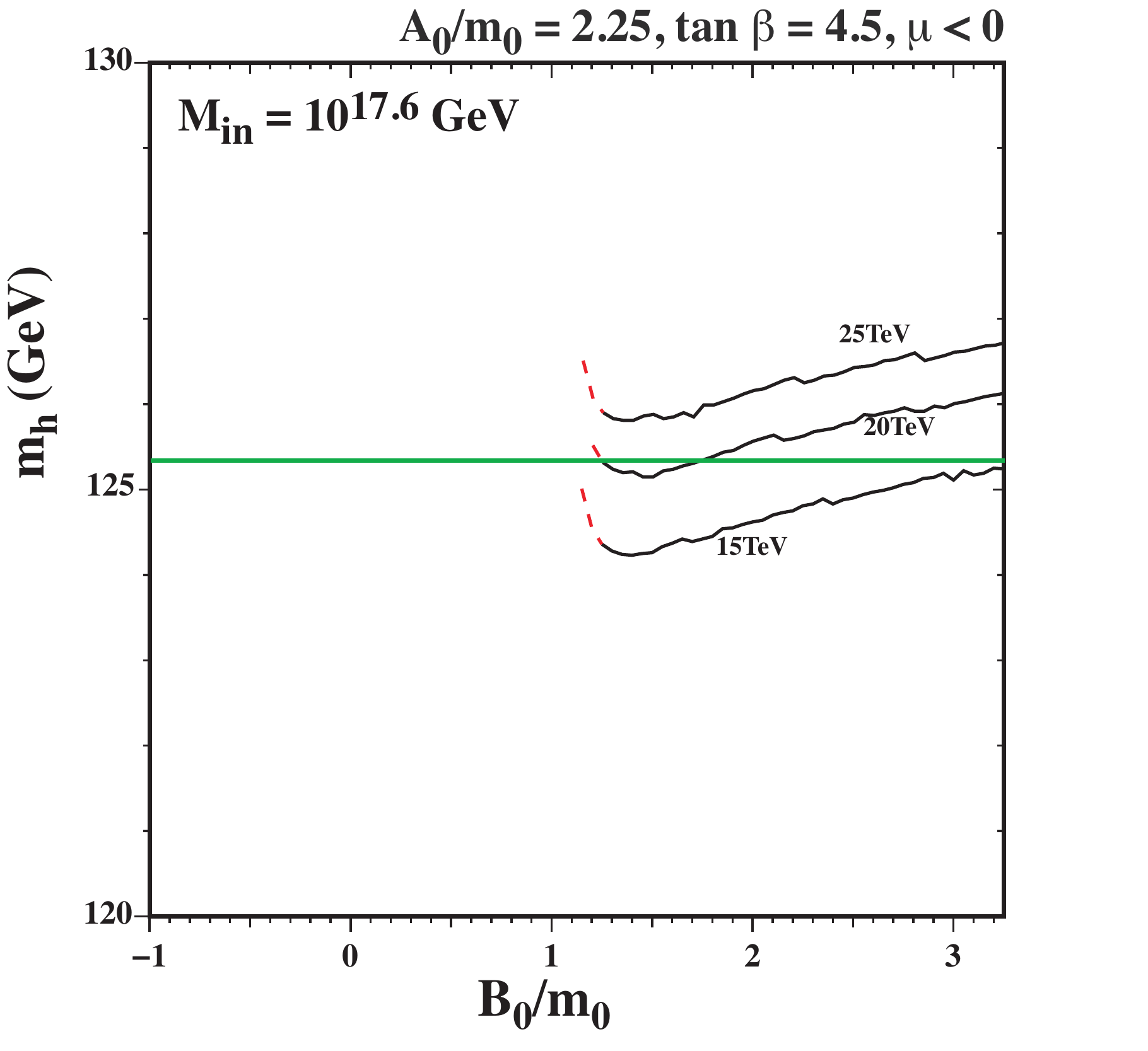}
\caption{\it As in Fig.~\ref{fig:stripprofile}, as functions of $B_0/m_0$ for
the indicated fixed values of $m_{1/2} \in [15, 25]$~TeV and $m_0$ chosen to obtain the cosmological value of
$\Omega_{\rm LSP} h^2$ for $B_0 = A_0 - m_0$, with the same values of the other input parameters as in Fig.~\ref{fig:plane1}.}
\label{fig:stripprofile4}
\end{figure}

Fig.~\ref{fig:stripprofile4} analyzes the sensitivity of $\Omega_{\rm LSP} h^2$, $\tau(p \to K^+ \nu)$ and $m_h$
to $B_0/m_0$ when $m_0$ is chosen to obtain the cosmological value of $\Omega_{\rm LSP} h^2$ for $B_0 = A_0 - m_0$. For these choices of parameters, the bino is the LSP
as long as $B_0 > A_0 - m_0$.
We see that $\Omega_{\rm LSP} h^2$ is quite insensitive to $m_{1/2}$, whereas the lower limit on $\tau(p \to K^+ \nu)$
prefers $m_{1/2} \gtrsim 15$~TeV. Finally, the right panel of Fig.~\ref{fig:stripprofile4} shows that $m_h$ is within 2~GeV of
the experimental value for all values of $B_0/m_0$ for the 3 indicated choices of $m_{1/2}$.

\begin{figure}[!ht]
\centering
\includegraphics[width=5.3cm]{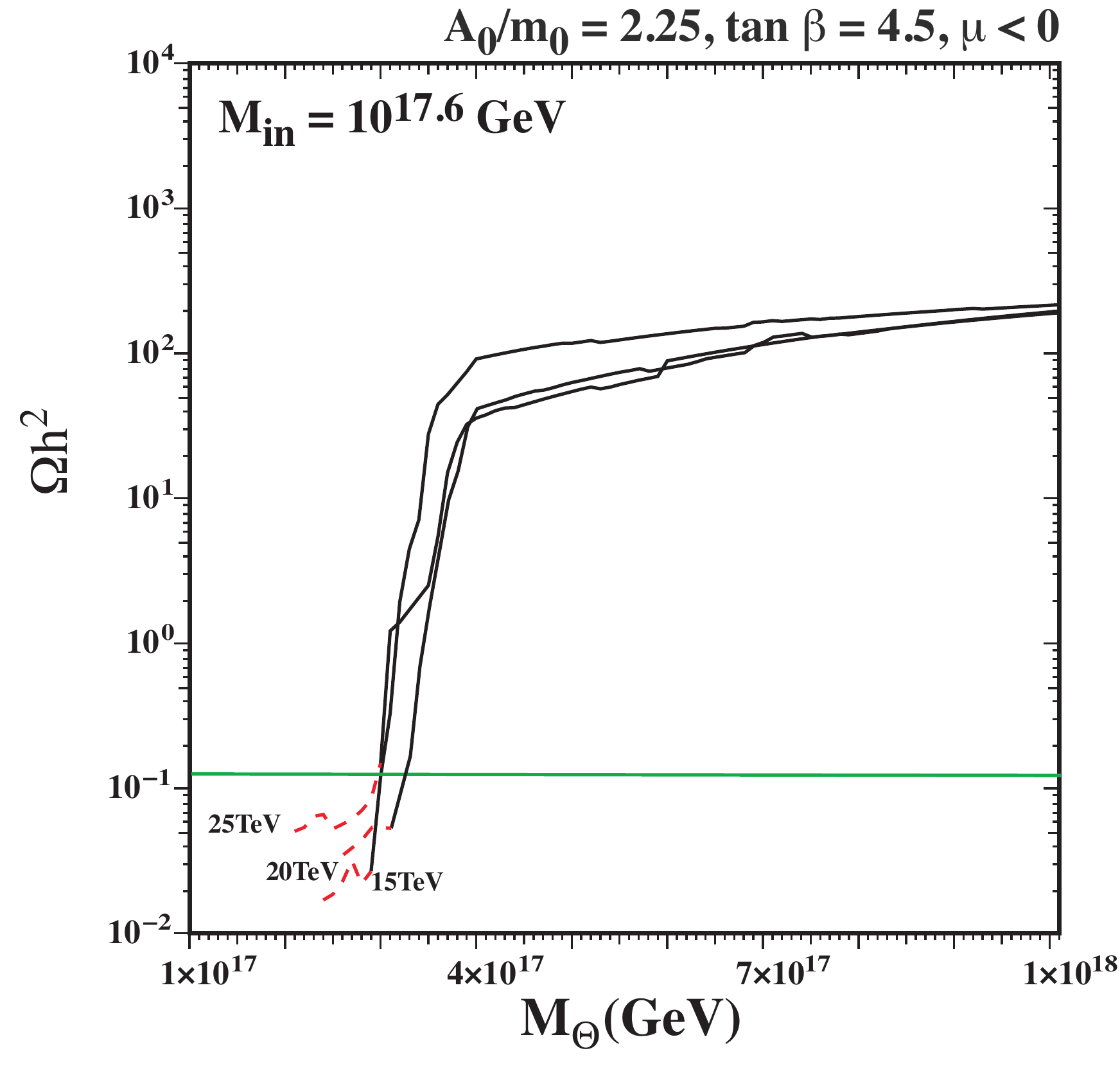}
\includegraphics[width=5.3cm]{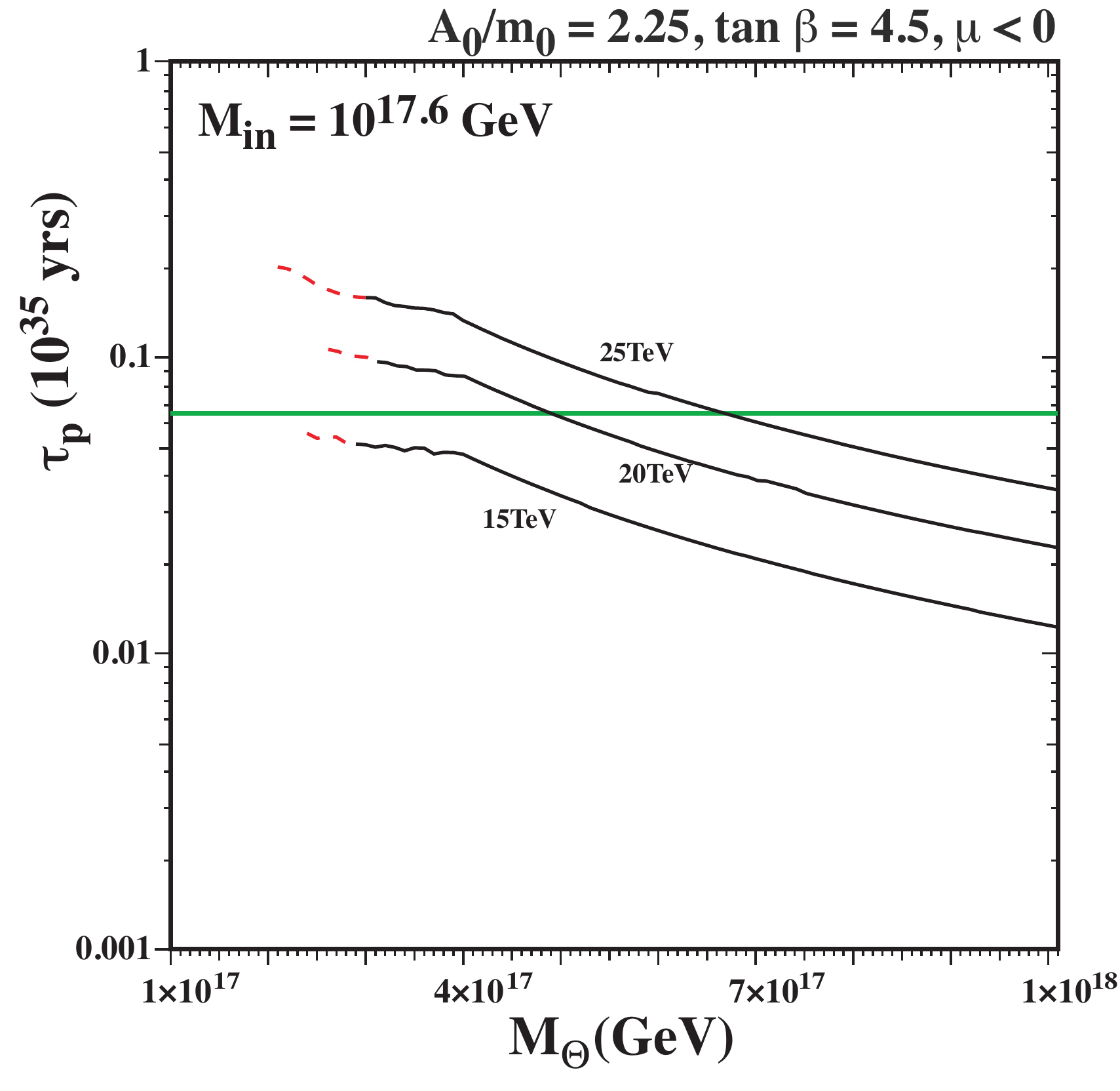}
\includegraphics[width=5.3cm]{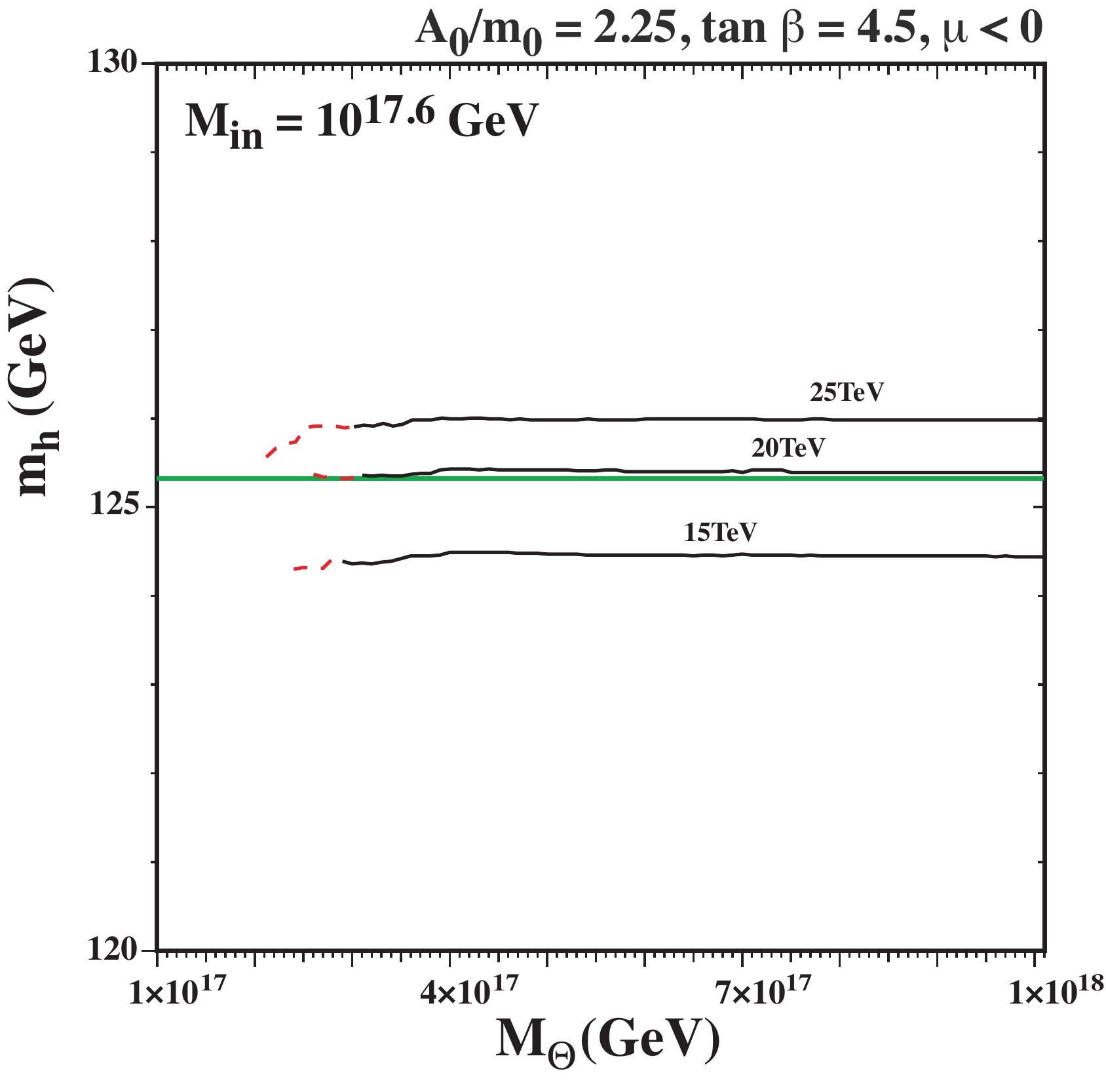}
\caption{\it As in Fig.~\ref{fig:stripprofile}, as functions of $M_\Theta$ for
the indicated fixed values of $m_{1/2} \in [15, 25]$~TeV and $m_0$ chosen to obtain the cosmological value of
$\Omega_{\rm LSP} h^2$ for $M_{\Theta} = 3 \times 10^{17}$~GeV, with the same values of the other input parameters as in Fig.~\ref{fig:plane1}.}
\label{fig:stripprofile5}
\end{figure}

Fig.~\ref{fig:stripprofile5} shows the corresponding sensitivity of $\Omega_{\rm LSP} h^2$, $\tau(p \to K^+ \nu)$ and $m_h$
to $M_\Theta$ when $m_0$ is chosen to obtain the cosmological value of $\Omega_{\rm LSP} h^2$ for $M_\Theta = 3 \times 10^{17}$~GeV.
We see that $\Omega_{\rm LSP} h^2$ is very sensitive to $M_\Theta$, and that $\tau(p \to K^+ \nu)$ disfavours large values of 
$M_\Theta$, as should be clear from Eq.~(\ref{eq:mcolored}). Here we also see that, for a given
$M_{\rm in}$, low values of $M_\Theta < M_{\rm in}$ are problematic because of rapid RGE running with 
large $\beta$-function coefficients. This problem is avoided when $M_\Theta > M_{\rm in}$. The Higgs mass, $m_h$, is quite insensitive to $M_\Theta$ and always close to the experimental value.

\begin{figure}[!ht]
\centering
\includegraphics[width=5.3cm]{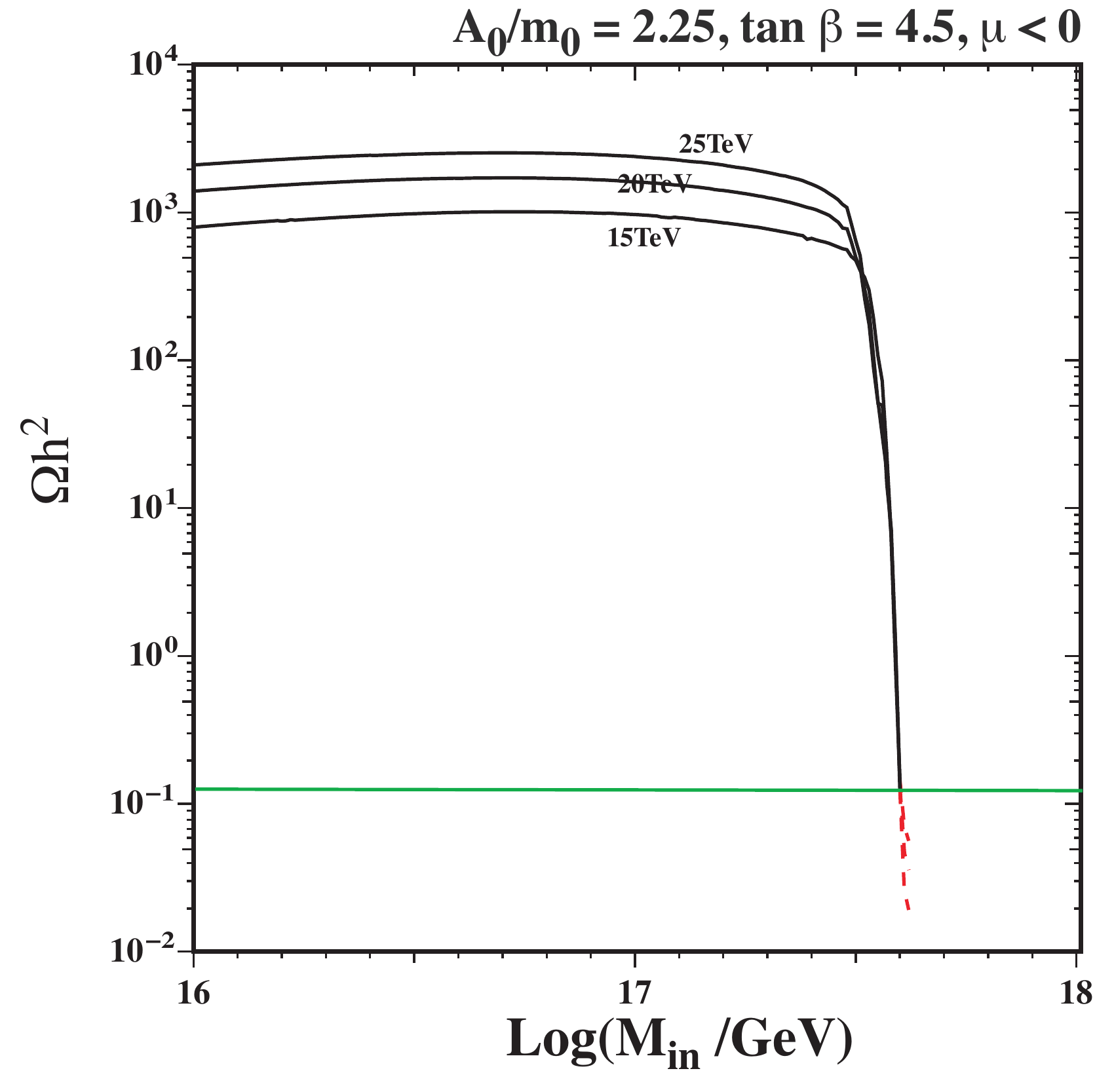}
\includegraphics[width=5.3cm]{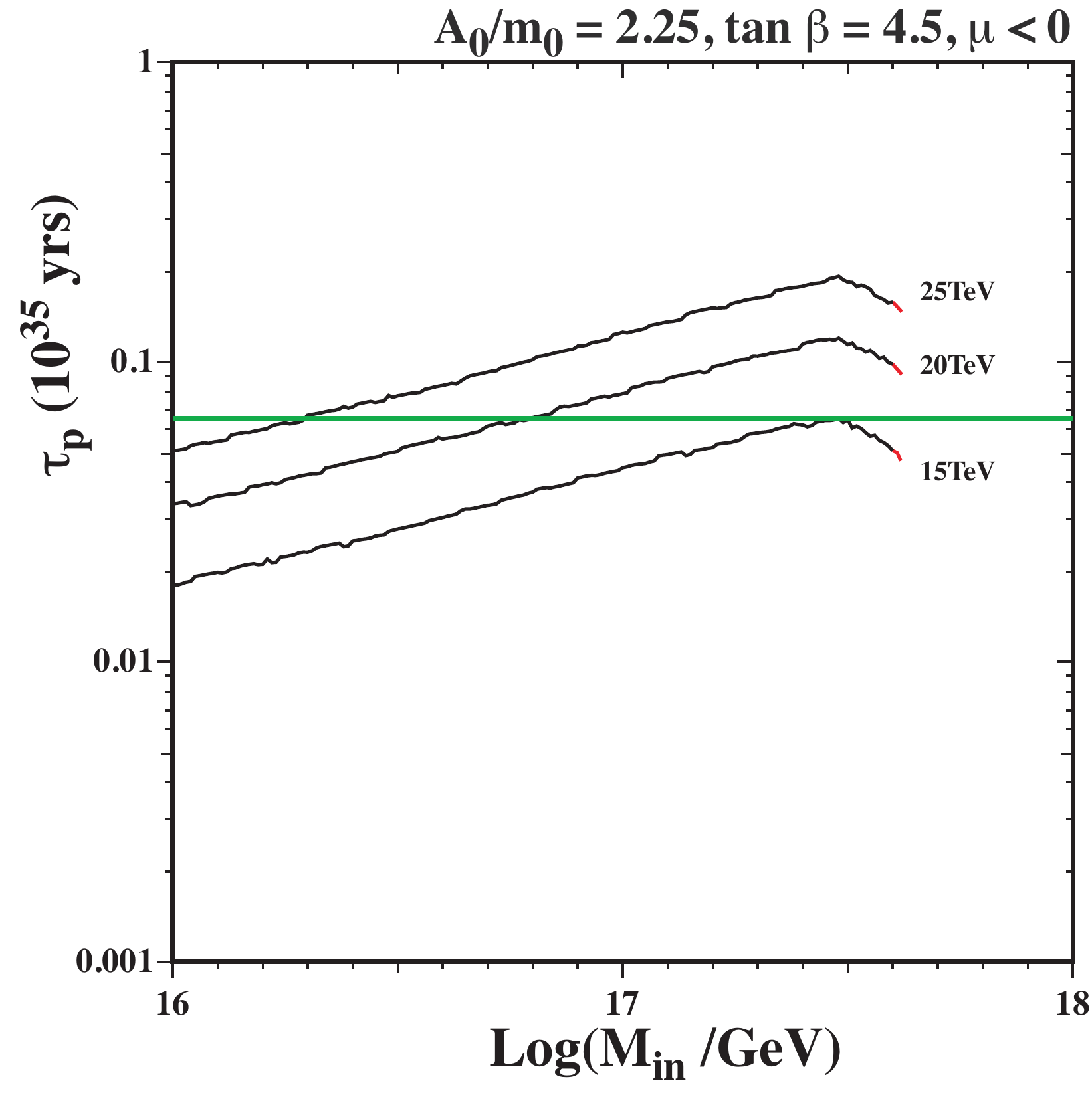}
\includegraphics[width=5.3cm]{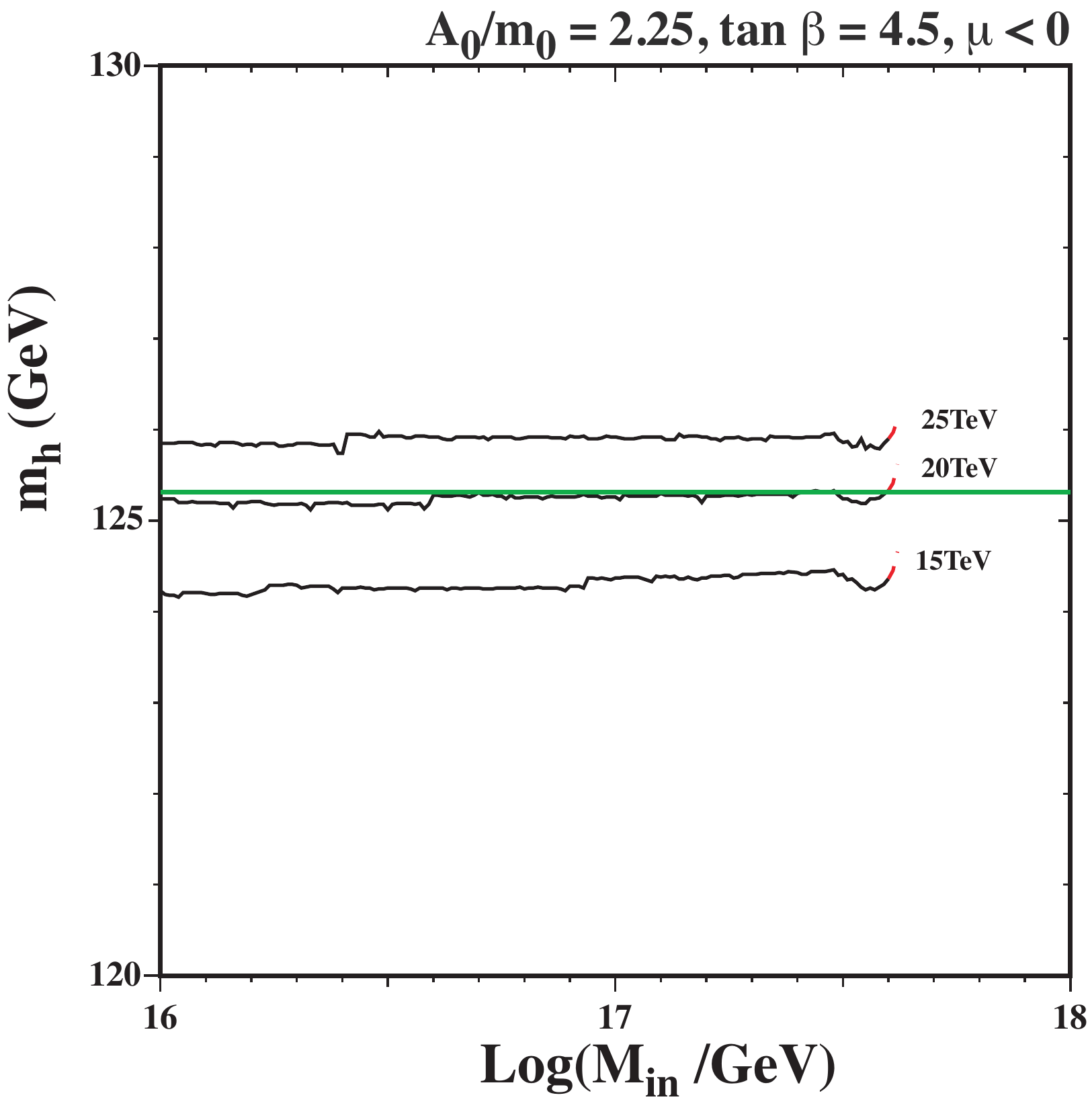}
\caption{\it As in Fig.~\ref{fig:stripprofile}, as functions of $M_{\rm in}$ for
the indicated fixed values of $m_{1/2} \in [15, 25]$~TeV and $m_0$ chosen to obtain the cosmological value of
$\Omega_{\rm LSP} h^2$ for $M_{\rm in} = 4 \times 10^{17}$~GeV, with the same values of the other input parameters as in Fig.~\ref{fig:plane1}.}
\label{fig:stripprofile6}
\end{figure}

We show in Fig.~\ref{fig:stripprofile6} the corresponding results for $\Omega_{\rm LSP} h^2$, $\tau(p \to K^+ \nu)$ and $m_h$
as $M_{\rm in}$ varies. 
Once again, we see that when the difference between $M_\Theta$ and $M_{\rm in}$ is large, RGE running becomes problematic. 
In this case $\Omega_{\rm LSP} h^2$ increases rapidly for $M_{\rm in} < 4 \times 10^{17}$~GeV, and
$\tau(p \to K^+ \nu)$ decreases gradually towards small $M_{\rm in}$, falling below the experimental lower limit for
$M_{\rm in} \lesssim {\rm a~few} \times 10^{16}$~GeV, even for $m_{1/2} = 20$ or 25~TeV. On the other hand, $m_h$ is insensitive to $M_{\rm in}$
and always compatible with the experimental value.

Finally, Fig.~\ref{fig:stripprofile6} the sensitivity of $\Omega_{\rm LSP} h^2$, $\tau(p \to K^+ \nu)$ and $m_h$
to $\lambda'$. In this case $\Omega_{\rm LSP} h^2$ increases rapidly for $\lambda' > 0.005$, while
$\tau(p \to K^+ \nu)$, decreases falling below the experimental lower limit for
$\lambda' \gtrsim 0.015$, even for $m_{1/2} = 25$~TeV. Once again, $m_h$ is insensitive to $M_{\rm in}$
and always compatible with the experimental value, so long as the bino is the LSP.

\begin{figure}[!ht]
\centering
\includegraphics[width=5.3cm]{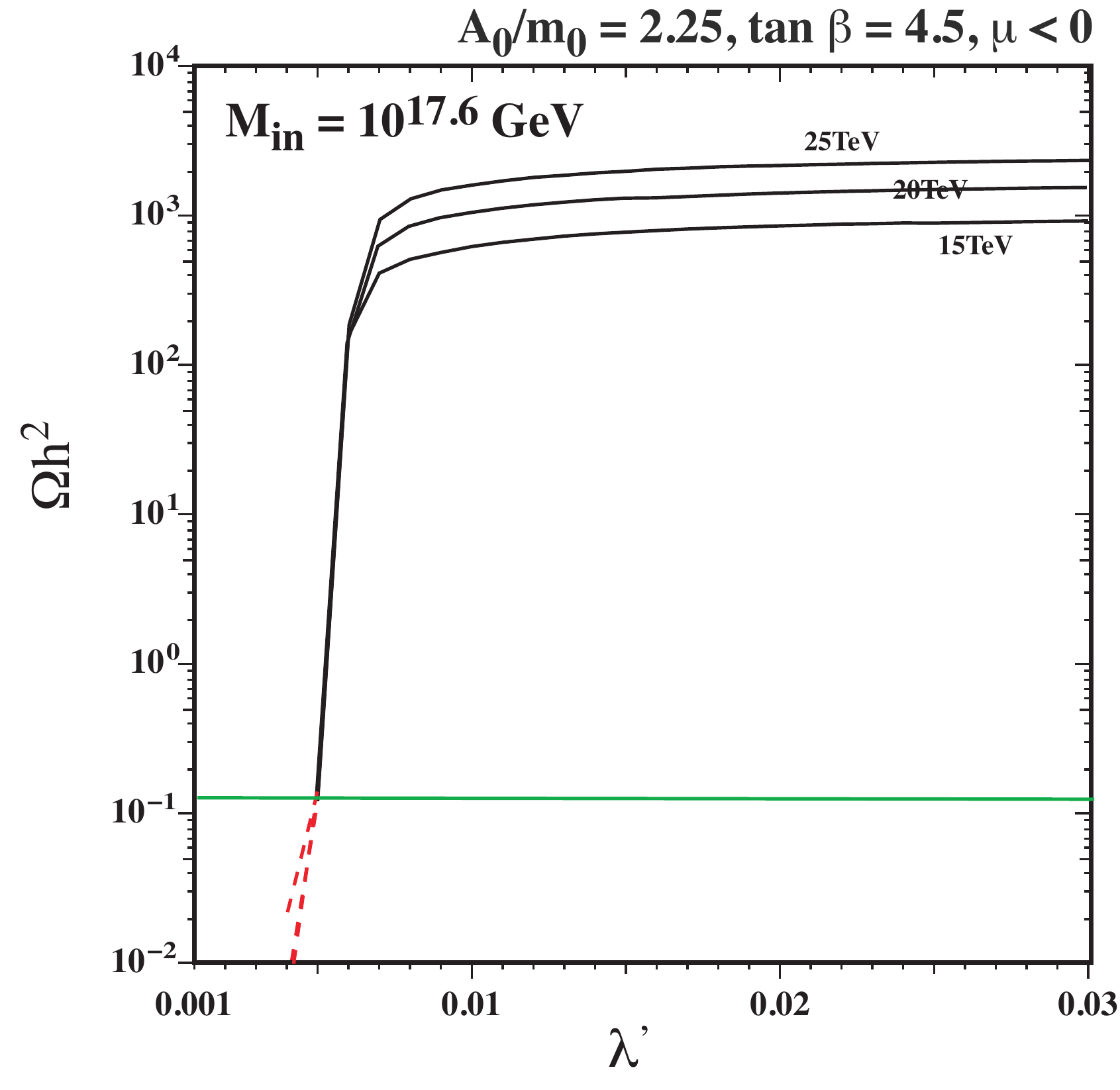}
\includegraphics[width=5.3cm]{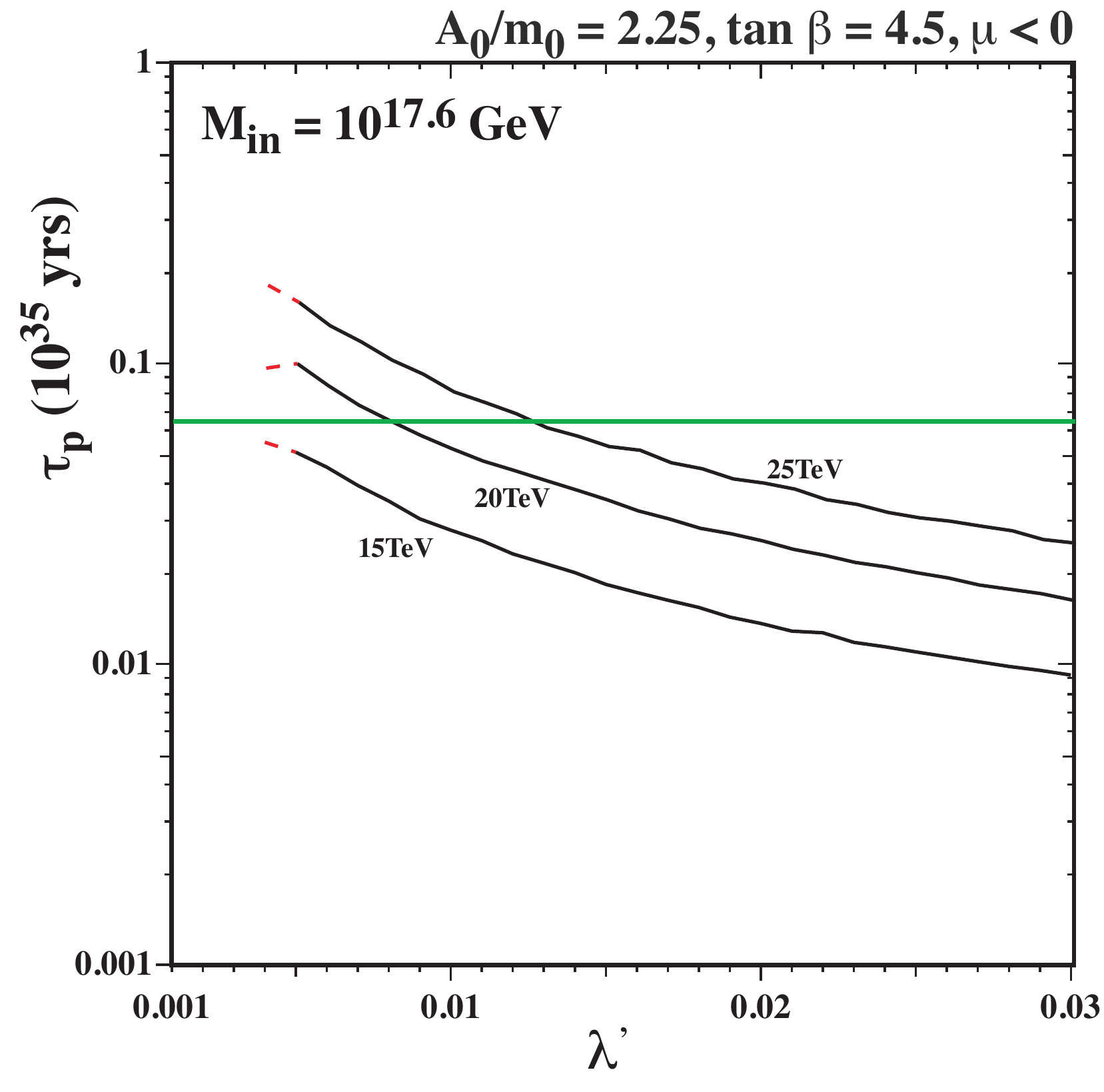}
\includegraphics[width=5.3cm]{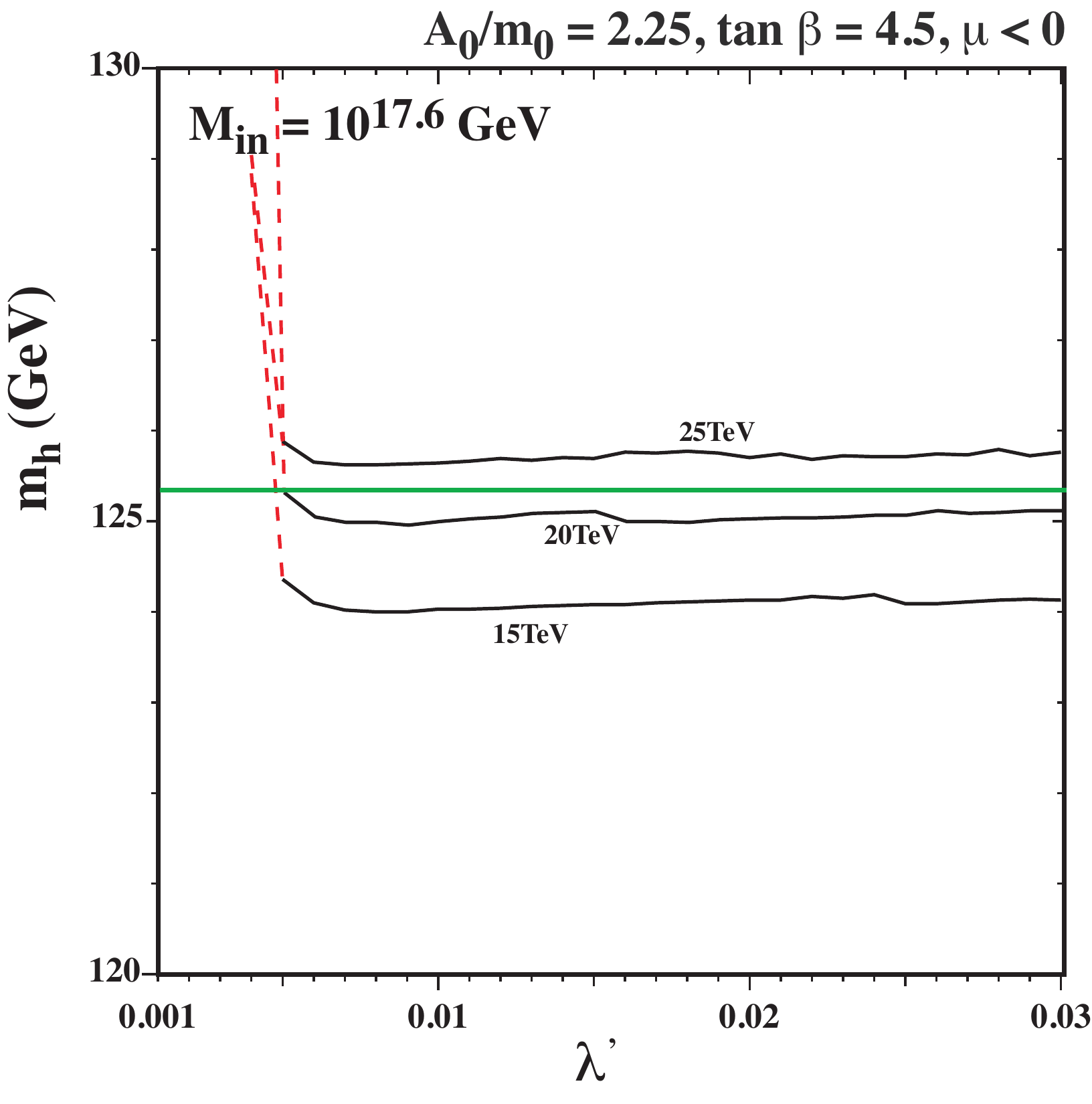}
\caption{\it As in Fig.~\ref{fig:stripprofile}, as functions of $\lambda'$ for
the indicated fixed values of $m_{1/2} \in [15, 25]$~TeV and $m_0$ chosen to obtain the cosmological value of
$\Omega_{\rm LSP} h^2$ for $\lambda_\Theta = 0.005$, with the same values of the other input parameters as in Fig.~\ref{fig:plane1}.}
\label{fig:stripprofile7}
\end{figure}

The conclusion of this analysis is that there is a relatively restricted region of parameter space close to the
default values $\tan \beta = 4.5, M_{\rm in} = 4 \times 10^{17}$~GeV, $M_\Theta = 3 \times 10^{17}$~GeV, 
$\lambda' = 0.005$ and $\lambda_\Theta = 1$, $B_0 = A_0 - m_0$ and $\mu < 0$ that is compatible with all the
experimental constraints. One of the interesting aspects of this conclusion is that $\tau(p \to K^+ \nu)$
is always close to the present experimental lower limit, and hence accessible to the the Hyper-Kamiokande
experiment that is now under construction, and is expected to have 90\% CL exclusion sensitivity to 
$\tau(p \to K^+ \nu)$ at the level of $\sim 5 \times 10^{34}$~yr after 20~yr of operation~\cite{HK}.

\section{Conclusions}

\begin{figure}
\centering
\includegraphics[height=8cm]{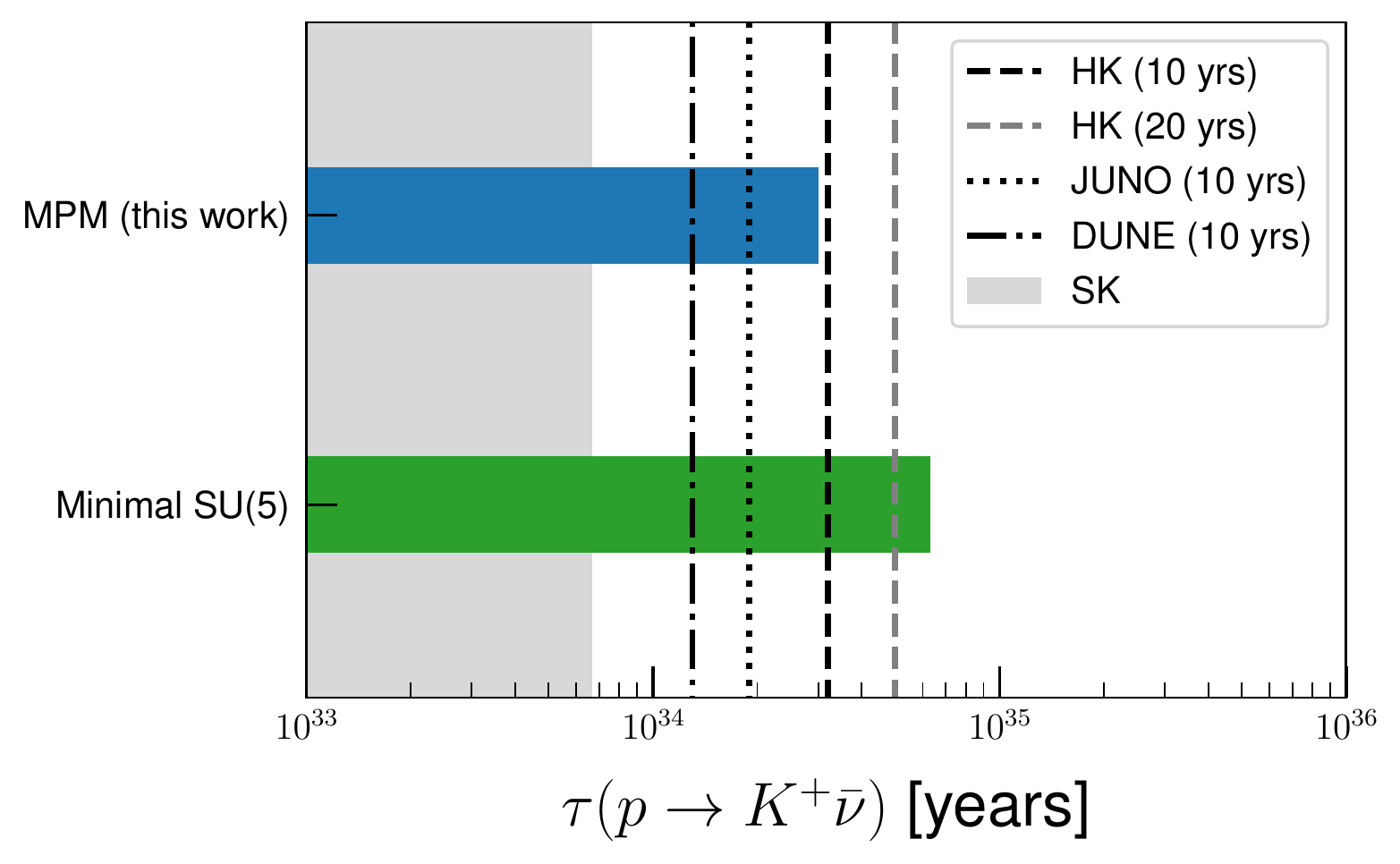}
\caption{
{\it 
The ranges of $p \to K^+ \overline \nu$ lifetimes found here in the CMSSM for the super-GUT MPM (blue band) and in the minimal SU(5) (green band; see Ref.~\cite{Ellis:2019fwf}) compared with the sensitivities of the JUNO~\cite{An:2015jdp}, DUNE~\cite{Abi:2020kei} and Hyper-K~\cite{HK} experiments. The gray shaded area is excluded by the Super-Kamiokande experiment~\cite{Abe:2014mwa}. 
}
}
\label{fig:Summary}
\end{figure}

We have analyzed in this paper the phenomenological viability of the minimal SU(5)
missing-partner super-GUT version of the CMSSM, which contains {\bf 75}, {\bf 50}  and $\mathbf{\overline{50}}$
Higgs representations. The running of the model  parameters above the GUT scale is much faster
than in conventional SU(5), as seen in Fig.~\ref{fig:gaugeunification}, limiting
the ranges above the GUT scale where the RGEs remain perturbative. This imposes constraints
on the ranges of the input scale, $M_{\rm in}$, and the common mass of the {\bf 50}  and $\mathbf{\overline{50}}$
multiplets, $M_\Theta$, limiting them to a few $\times 10^{17}$~GeV. Important phenomenological constraints on the model 
are then imposed by the cosmological relic density, $\Omega_{\rm LSP} h^2$, the proton lifetime,
$\tau(p \to K^+ \nu)$, and the Higgs mass, $m_h$, which we compute using {\tt FeynHiggs~2.18.0}.

The proton lifetime requires the soft supersymmetry-breaking parameters $m_{1/2}$ and $m_0$ to lie
in the multi-TeV range, in which case the relic density constraint forces these parameters to lie along the
stop coannihilation strip, where the MSSM soft supersymmetry-breaking scalar mass $m_0$ is essentially
determined as a function of the gaugino mass $m_{1/2}$. We then find that the Higgs mass prediction is compatible with the relic density
constraint only if the MSSM Higgs mixing parameter $\mu$ is negative. The proton decay and relic density constraints 
set lower and upper limits on $m_{1/2}$ that are compatible with $\Omega_{\rm LSP} h^2$ for only limited ranges of 
$m_{1/2} \sim 15 - 25$~TeV and $\tan \beta \sim 3.5 - 5$. In the allowed range of parameter space we find
that MSSM sparticle masses are typically in the ranges $m_{\rm LSP}, m_{\tilde t_1} \sim 2.5 - 5$~TeV, $m_{\tilde g} \sim 10 - 20$~TeV, 
$m_{\tilde q} \sim 15 - 30$~TeV and $m_{\tilde \ell} \sim 10 - 25$~TeV, beyond the reach of the LHC but potentially within reach
of a 100-TeV proton-proton collider such as FCC-hh or SppC~\cite{Cohen:2013xda}. 

The most promising phenomenological signature
of this model may be proton decay, since we find that $\tau(p \to K^+ \nu) \lesssim 3 \times 10^{34}$~yrs
throughout the allowed range of parameter space. As seen in
Fig.~\ref{fig:Summary}, this range lies within the discovery reaches of searches with the JUNO, DUNE and (in particular) Hyper-Kamiokande experiments,
which are estimated to be $1.9 \times 10^{34}$~yrs~\cite{An:2015jdp}, $1.3 \times 10^{34}$~yrs~\cite{Abi:2020kei}
and $3.2 \times 10^{34}$~yrs~\cite{HK}, respectively, after 10 years of operation.

\section*{Acknowledgements}

The work of J.E. was supported partly by the United Kingdom STFC Grant ST/T000759/1 
and partly by the Estonian Research Council via a Mobilitas Pluss grant. 
The work of N.N. was supported by the Grant-in-Aid for Scientific Research B (No.20H01897), 
Young Scientists B (No.17K14270), and Innovative Areas (No.18H05542).
The work of K.A.O. was supported partly
by the DOE grant DE-SC0011842 at the University of Minnesota.

\end{document}